\documentclass[letterpaper,11pt]{article}
\usepackage{jheppub}
\pdfoutput=1
\usepackage{amsmath,amssymb,amsfonts,bm,amscd,mathtools}
\usepackage{graphicx}
\usepackage{multirow}
\usepackage{verbatim}
\usepackage{appendix}
\usepackage{fancybox}
\usepackage{array}
\usepackage{url}
\usepackage{float}
\usepackage{subfigure}
\usepackage{tikz}

\def\CB{{\mathcal B}}
\def\CC{{\mathcal C}}

\def\CH{{\mathcal H}}

\def\CL{{\mathcal L}}
\def\CM{{\mathcal M}}

\def\CO{{\mathcal O}}

\def\CR{{\mathcal R}}
\def\CS{{\mathcal S}}

\def\CW{{\mathcal W}}

\def\CZ{{\mathcal Z}}

\def\be{\begin{equation}}
\def\ee{\end{equation}}
\def\bea{\begin{eqnarray}}
\def\eea{\end{eqnarray}}

\newcommand{\cU}{{\mathcal U}}

\newcommand{\cH}{{\mathcal H}}
\newcommand{\tq}{{\mathtt q}}

\newcommand{\2}{{\mathtt 2}}
\newcommand{\3}{{\mathtt 3}}
\newcommand{\4}{{\mathtt 4}}
\newcommand{\ta}{{\mathtt a}}
\newcommand{\tb}{{\mathtt b}}
\newcommand{\tc}{{\mathtt c}}

\newcommand{\1}{{1}}

\title{Towards a classification of holographic multi-partite entanglement measures}

\author{Abhijit Gadde, Vineeth Krishna, Trakshu Sharma}
\affiliation{Department of Theoretical Physics \\ 
Tata Institute for Fundamental Research, Mumbai 400005}
\emailAdd{abhijit@theory.tifr.res.in}
\emailAdd{vineeth@theory.tifr.res.in}
\emailAdd{trakshu.sharma@tifr.res.in}

\abstract{In this paper, we systematically study the measures of multi-partite entanglement with the aim of constructing those measures that can be computed in probe approximation in the holographic dual. We classify and count general measures as invariants of local unitary transformations. After formulating these measures in terms of permutation group elements, we derive conditions that a probe measure should satisfy and find a large class of solutions. These solutions are generalizations of the multi-entropy introduced in \cite{Gadde:2022cqi}. We derive their holographic dual with the assumption that the replica symmetry is unbroken in the bulk and check our prescription with explicit computations in $2d$ CFTs. Analogous to the multi-entropy, the holographic dual of these measures is given by the weighted area of the minimal brane-web but with branes having differing tensions. We discuss the replica symmetry assumption and also how the already known entanglement measures, such as entanglement negativity and reflected entropy fit in our framework.}

\begin{document}
\maketitle
\flushbottom

\section{Introduction}
One of the central problems of modern theoretical physics is understanding the quantum nature of gravity. This problem manifests itself most prominently as the black hole information paradox \cite{Hawking:1975vcx, Hawking:1976ra}. One tool to tackle this problem is the AdS/CFT correspondence \cite{Maldacena:1997re, Witten:1998qj, Gubser:1998bc}. The correspondence provides us a definition of the quantum theory of gravity in Anti de-Sitter (AdS) space as a conformal field theory (CFT) living on the boundary. For the dual gravitational theory to be local, the CFT must have certain properties such as a large central charge and a large gap in the spectrum of higher spin operators \cite{Heemskerk:2009pn}. Coupled with quantum information theoretic ideas such as quantum entanglement and quantum error correction, the AdS/CFT correspondence has been effective in advancing our understanding of the black hole information paradox \cite{Almheiri:2019hni, Penington:2019kki}. Quantum information theory ideas have been useful in enriching the correspondence itself via the so-called subregion duality \cite{Dong:2016eik,Harlow:2016vwg,Almheiri:2019hni}. 

The ingredient that has been crucial to making progress in this direction is the  Ryu-Takayanagi (RT) formula \cite{Ryu:2006bv, Hubeny:2007xt} and its quantum generalization \cite{Engelhardt:2014gca}. The RT formula expresses the entanglement entropy of a state in Hilbert spaces associated with two regions of space in terms of a simple geometrical quantity in the dual gravitational theory viz. the area of the minimal surface anchored on the boundaries of the regions in question. Its quantum generalization - the so-called quantum extremal surface formula \cite{Engelhardt:2014gca} - takes into account the contribution of quantum fluctuations of the bulk theory. 

One thing to note is that the progress in understanding the black hole information paradox and the local nature of AdS/CFT correspondence discussed above has stemmed almost exclusively from the ideas related to \emph{bi}-partite entanglement i.e. to entanglement between \emph{two} parties. It is natural to ask how understanding the \emph{multi}-partite entanglement will further this progress. With this motivation, in \cite{Gadde:2022cqi}, we introduced a measure of multi-partite entanglement with a ``probe'' property (described shortly) that is desirable from the holographic point of view. In this paper, we continue to pursue the same line and make progress toward the classification of such probe multi-partite entanglement measures. In addition, we address several points that were left for future work in \cite{Gadde:2022cqi}. 
The somewhat mysterious RT formula for entanglement was explained by Lewkowycz and Maldacena \cite{Lewkowycz:2013nqa} using the replica trick and a certain clever analytic continuation of the bulk geometry. Their work guided us in defining and evaluating the multi-partite measure called multi-entropy in \cite{Gadde:2022cqi}. It continues to do so even in this paper. 

The entanglement in a bi-partite pure state can be measured\footnote{In this paper, we use the word ``measure'' rather liberally without requiring attributes such as positivity and monotonicity under local quantum operations and classical communications etc. that would be demanded by a quantum information theorist.} by Renyi entropies $S_n(\rho)$, defined as 
\begin{align}
    S_n(\rho)= \frac{1}{1-n}{\log}\Big( {\rm Tr} \,\rho^n\Big)
\end{align}
where $\rho$ is the density matrix obtained by tracing one of the two parties. The Renyi entropy is symmetric in both parties even though the definition of Renyi entropy obscures this symmetry. The von-Neumann entropy, also called the entanglement entropy $S$ is defined as $-{\rm Tr} \rho \log \rho$. In many cases, it turns out to be convenient to think of $S$ as the $n\to 1$ limit of $S_n$, analytically continued in $n$. This way of computing the entanglement entropy is known as the replica trick. Once ${\rm Tr}\rho^n$ is analytically continued in $n$,  then $S$ can also be written  as, 
\begin{align}
    S=-\partial_n  \, {\log}\Big( {\rm Tr} \,\rho^n\Big)|_{n=1} = -\partial_n  \, \Big( {\rm Tr} \,\rho^n\Big)|_{n=1}.
\end{align}
In the last equality, we use the fact that the state is normalized i.e. ${\rm Tr} \, \rho = 1$. From a holographic point of view, it is easier to characterize the entanglement entropy compared to the Renyi entropies. This is because the entanglement entropy is given simply by the area of a certain minimal area surface in the dual geometry viz. the RT formula. The Renyi entropies, or rather a certain combination of their derivatives with respect $n$ \cite{Dong:2016fnf}, can also be computed holographically but unlike entanglement entropy, they are not probes of the original dual geometry. Rather their evaluation requires solving the gravitational equation of motion with the same infra-red boundary conditions but with a conical singularity along a certain co-dimension 2 locus. We call the quantities that are computed \emph{in the dual geometry} as opposed to those that are computed in a new solution, the probe quantities. The probe measure of entanglement is the one that is computed by a probe quantity.
With this definition of the probe measure, we can say that the only probe measure in the family of Renyi entropies is the entanglement entropy. 
In this work, we first identify the multi-partite entanglement measures that admit a holographic dual with the same infra-red boundary conditions. These are organized into families where the members are indexed by an integer $n$. We then use a similar sort of $n\to 1$ limit within each family to construct the measures that are probes of the original geometry.

\subsection{Overview of the holographic computation}

In this section, we will give a quick overview of the probe multi-partite measure, named multi-entropy, that was introduced in \cite{Gadde:2022cqi}. Its construction will form the basis of the present work. During this overview, we will also need to refer to \cite{Lewkowycz:2013nqa}, so we will review its relevant aspects as well.

\subsubsection{Bi-partite}
\begin{figure}[t]
    \begin{center}
        \includegraphics[scale=0.7]{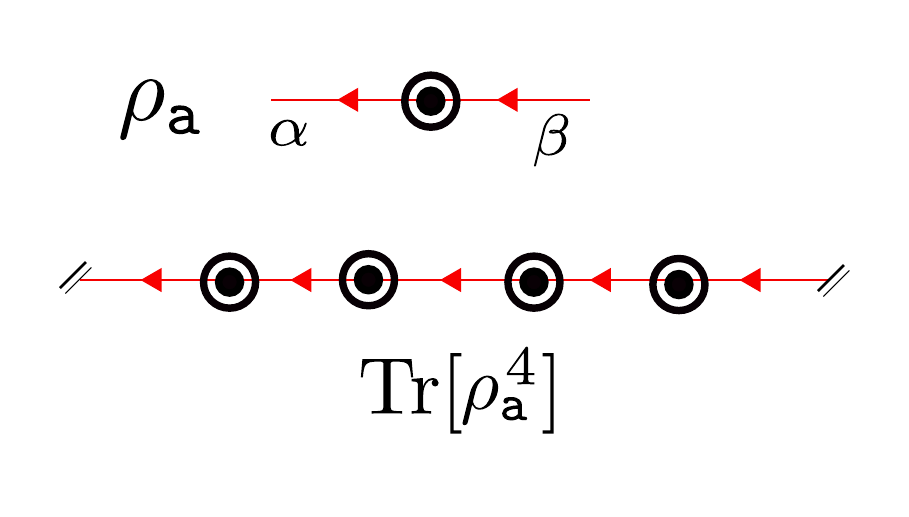}
    \end{center}
    \caption{Graphical notation for the density matrix $\rho_\ta$ and ${\cal E}_n\equiv {\rm Tr} \rho_\ta^n$. Here we have defined the case of $n=4$. The thread with $\rho$'s is periodically identified to make a necklace.}\label{necklace}
\end{figure}

Let us start with the bi-partite entanglement, say between party $\ta$ and party $\tb$. As stated earlier, the pure state of two parties is represented by a density matrix $(\rho_\ta)_\alpha^\beta$, obtained after tracing out, say party $\tb$. So $\rho_\ta$ is an operator acting on ${\cal H}_\ta$. We denote it graphically as in figure \ref{necklace}. With this presentation of $\rho_\ta$, the ${\cal E}_n\equiv {\rm Tr} \rho_\ta^n$ that enters the definition of Renyi entropy is denoted as a necklace with $n$ beads as shown in figure \ref{necklace}. 
This necklace ${\cal E}_n$ has ${\mathbb Z}_n$ rotational symmetry. We call the symmetry of the graph of $\rho$, the replica symmetry. 

\begin{figure}[t]
    \begin{center}
        \includegraphics[scale=0.6]{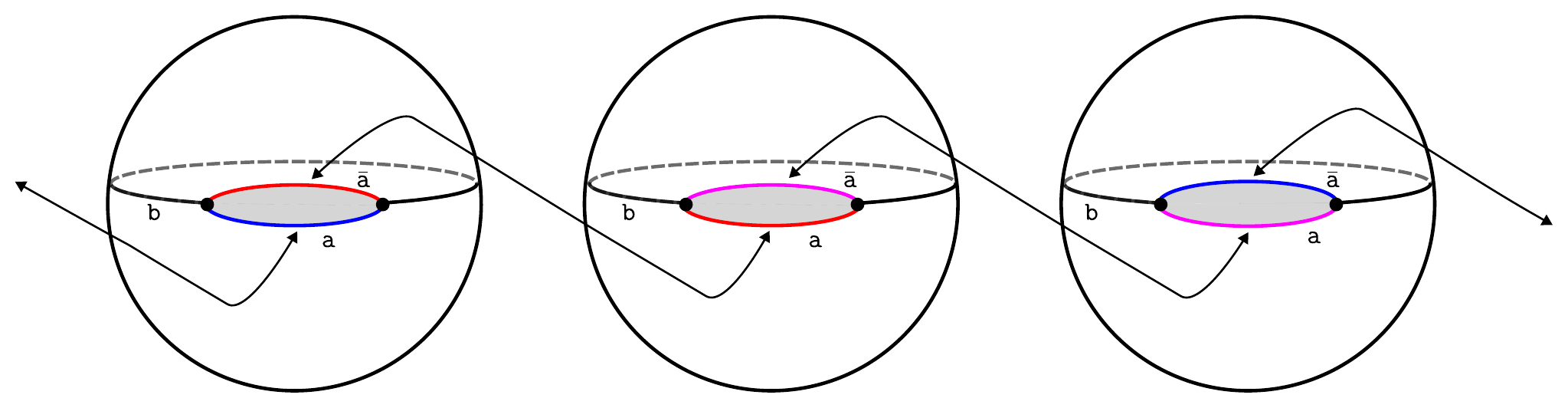}
    \end{center}
    \caption{The path integral that computes ${\cal E}_n$ in two-dimensional quantum field theory}\label{replicaqft}
\end{figure}

The path integral that computes ${\cal E}_n$ in quantum field theory ${\cal Q}$ is constructed by cutting and pasting procedure as shown in figure \ref{replicaqft}. 
For simplicity of drawing,  we have taken the example of a $2d$ theory. 
Even though, in figure \ref{replicaqft} we have taken the party $\ta$ to consist of a single interval, it could also comprise of a union of intervals and similarly for party $\tb$. The state $\psi$ is obtained by a path integral on the southern hemisphere and its conjugate is obtained by a path integral on the northern hemisphere. These two hemispheres are glued along $\tb$ which corresponds to tracing out the party $\tb$ the resulting space is a sphere with slits along the $\ta$ region, which corresponds to the density matrix $\rho_\ta$. The upper lip of the slit corresponds to the ket index and the lower lip corresponds to the bra index of the density matrix. To construct ${\rm Tr} \rho_\ta^n$, we take $n$ copies of $\rho_a$ in a row and glue the upper lip of the $i$-th copy to the lower lip of the $i+1$ copy with periodic identification. The resulting space is a sphere \emph{branched} or \emph{ramified} $n$ times at the boundary points of all intervals. The ${\mathbb Z}_n$ replica symmetry of ${\rm Tr} \rho_\ta^n$ shows up as the symmetry of this space. 
Alternatively, the space obtained by this type of gluing can be thought of as being obtained with the insertion of appropriate twist operators in the direct product of $n$ quantum field theories ${\cal Q}^{\otimes n}$. A general twist operator in  ${\cal Q}^{\otimes n}$ is labeled by an element $\sigma$ of the permutation group $S_n$. If the region on the left (right) of the operator in the replica set of kets is glued with permutation $\sigma_l$ ($\sigma_r$) to the corresponding region in the replica set of bras then the twist operator type is $\sigma_l^{-1}\sigma_r$. 

The quantity ${\rm Tr} \rho_\ta^n$ in CFTs with gravity duals can be computed holographically by computing the action of the dominant bulk solution that fills in the branched cover of the boundary. In \cite{Lewkowycz:2013nqa} this computation was performed with the assumption that the dominant bulk solution preserves the replica symmetry. Orbifolding by the replica symmetry, the authors obtained a geometry with a co-dimension $2$ conical singularity of conical angle $2\pi/n$. As a cone, the geometry admits analytic continuation away from integer $n$ by considering general conical angles not restricted to those of the form $2\pi/n$ with integer $n$. For von-Neumann entropy, we are interested in computing the variation of the gravitational action at $n=1$. In this limit, the conical singularity flattens out and the variation of the gravity action is given by the area of the minimal surface (a geodesic, in the case of $2d$ CFT). See \cite{Lewkowycz:2013nqa} for more details of this argument. 

\subsubsection{Multi-partite}
Here we will only review \cite{Gadde:2022cqi} in the context of the tri-partite multi-entropy. For a more detailed discussion, we refer the reader to \cite{Gadde:2022cqi} and to section \ref{measures}. For a tri-partite pure state on parties $\ta,\tb$ and $\tc$, \cite{Gadde:2022cqi} first constructs the two party density matrix $\rho_{\ta\tb}$ by tracing out party $\tc$. This two-party density matrix $\rho$ with the graphical notation is presented in figure \ref{q3}. The measure ${\cal E}_n^{(\tq)}$ is now defined to be the value of the $n\times n$ square lattice shown in figure \ref{q3} obtained after contracting indices of $\rho_{\ta\tb}$ as shown ($n$ is taken to be $3$ in the figure and parties are taken to be $\1,\2$ and $\3$ with party $\1$ being traced out.). In this case, the replica symmetry is ${\mathbb Z}_n\otimes {\mathbb Z}_n$ where one ${\mathbb Z}_n$ cyclically permutes the rows of the lattice and the other cyclically permutes the columns.  
As in the bi-partite case, the calculation of ${\cal E}_n^{(\tq)}$ is mapped to the calculation of the partition function on the appropriately branched sphere or equivalently as the correlation function of appropriate twist operators. 

A method similar to that of \cite{Lewkowycz:2013nqa} was employed to compute this partition function holographically as the gravitation action of the dominant bulk solution that fills in the branched cover of the boundary sphere. Assuming that the dominant bulk solution preserves replica symmetry, its orbifold with respect to ${\mathbb Z}_n\otimes {\mathbb Z}_n$ was performed. The replica symmetry assumption is certainly true\footnote{This was shown in three-party and three-interval case in \cite{Penington:2022dhr}. The case of four parties and four intervals is discussed in section \ref{replica}.  The case of arbitrary number intervals and parties is treated in \cite{GaddeWIP}.} for $n=2$ and the resulting geometry admits analytic continuation for real values of $n\leq 2$. This geometry has three co-dimension $2$ conical singularities anchored at the three endpoints of the intervals on the boundary that meet up in the bulk at a co-dimension $3$ locus. For $2d$ CFT, this junction is a point in the $3d$ bulk. As in the bi-partite case, in the $n\to 1$ limit, the conical singularity flattens out and the derivative of the gravitational action is given by the area of the minimum area ``soap-film'' that is anchored on the edge of the regions on the boundary. For $2d$ CFTs, this soap film is a trivalent tree of geodesics of minimum total length. Please see \cite{Gadde:2022cqi} for the details of this argument. 
The holographic computation for a more general class of multi-partite entanglement measures is detailed in section \ref{holographic-dual}.

\subsection{Structure of the paper}

The paper is structured as follows. In section \ref{measures}, we consider general entanglement measures as invariants of local unitary transformations. We count them in the limit of large Hilbert space dimension and formulate them in terms of permutation group elements. In section \ref{lm}, we specialize to the entanglement measures in holographic conformal field theories and determine the conditions that a probe measure should satisfy and find a large class of solutions to these constraints. 
These probe conditions are based on the analysis of holographic entanglement entropy by Lewkowycz and Maldacena \cite{Lewkowycz:2013nqa}. In section \ref{holographic-dual}, following \cite{Gadde:2022cqi}, we derive the probe holographic dual of this class of measures with the assumption that the replica symmetry is unbroken in the bulk. We also conjecture the quantum extension of the prescription. 
We discuss the marriage of our multi-partite measures with holographic purification in section \ref{marriage}. In section \ref{2dcft}, we check our proposal for two-dimensional conformal field theories with large central charge and discuss the issue of bulk replica symmetry. We discuss a potential payoff from multi-partite measures as a more refined bulk reconstruction using a form of multi-partite error correction in section \ref{discuss}. In appendix \ref{plethystic}, we give details of the counting of multi-partite measures for large dimensional Hilbert spaces. In appendix \ref{regge}, we discuss the analytic continuation involved in the definition of the probe measure. In particular, we present a numerical approach for the analytic continuation and discuss its validity. This appendix is based on the papers \cite{regge-viano, osti_4065624}. Appendix \ref{monodromy}, gives details of four-point and five-point conformal block computation using the so-called ``monodromy method'' in $2d$ CFTs with large central charge.

\section{Classification of multi-partite entanglement measures}\label{measures}
In quantum information literature, entanglement measures are usually defined for mixed states. They obey several physically motivated properties such as positivity on the ``inseparable'' states and monotonicity under local operations and classical communications (LOCC). See \cite{Horodecki:2009zz} for a detailed review and relevant references. In this paper, we will be interested in the entanglement measures for pure states and we will define them in what is perhaps the most general fashion.
Consider a normalized quantum state $|\Psi\rangle \in \otimes_{\ta=1}^\tq {\cal H}_\ta$ of a $\tq$-party system. Let $d_\ta$ be the dimension of the Hilbert space $\cH_\ta$ and $|\alpha_\ta\rangle, \, \alpha_\ta=1,\ldots, d_\ta$ be its arbitrarily chosen orthonormal basis. In $|\alpha_\ta\rangle$ basis, the state $|\Psi\rangle$ is given as
\begin{align}
    |\Psi\rangle=\sum_{\alpha_1=1}^{d_1}\ldots \sum_{\alpha_\tq=1}^{d_\tq}  \,\,\psi_{\alpha_1\ldots \alpha_\tq}\, \,|\alpha_1\rangle\otimes \ldots \otimes|\alpha_\tq\rangle.
\end{align}
We call $\psi_{\alpha_1\ldots\alpha_\tq}$, the wavefunction.
\begin{itemize}
    \item A q-party entanglement measure $\cal E$ is a function of $\psi$ (and its complex conjugate)
          that is invariant under ``local unitary transformations''.
\end{itemize}
These transformations take the form $\otimes_{\ta=1}^\tq {\cal U}_\ta\in U(d_\ta)$ where ${\cal U}_\ta$ is the unitary transformation that acts on ${\CH_\ta}$.  If we think of the local unitary group $\prod_\ta U(d_\ta)$ as the gauge symmetry and $\psi$'s and $\bar \psi$'s as $\tq$-fundamental and $\tq$-anti-fundamental ``operators'' then the measures are simply the gauge invariant operators. We construct the local unitary invariants by taking multiple copies of $\psi$'s and $\bar \psi$'s and contracting the fundamental indices $\alpha_\ta$ of $\psi$ with anti-fundamental indices\footnote{There may be invariants of the unitary group that are not of this type. For example, one can consider $|\Psi\rangle \in {\cal H}_1 \otimes {\cal H}_2$ such that ${\rm dim}\, {\cal H}_i=d$. In this case, a local unitary invariant can be constructed by taking $d$ copies of only $\psi$ and contracting its party $\1$ as well as party $\2$ indices with the two separate $d$-index $\epsilon$ tensors. Because we are interested in Hilbert spaces of arbitrary dimensions, in particular of infinite dimension, we do not consider such baryon-type unitary invariants. There are even more exotic local unitary invariants such as the so-called tensor eigenvalues \cite{lim2006singular, 10.5555/3240740}. We do not know whether it is possible to express those in terms of simple products of $\psi$'s and $\bar \psi$'s considered here. Restricting only to the class of invariants considered here is sufficient for the purposes of this paper.} $\alpha_\ta$ of $\bar \psi$. As a result, the number of $\psi$'s and the number of $\bar \psi$'s is the same in any entanglement measure. We call this number the replica number $n$. We index the replicas by the superscript $(i)$. The Hilbert space $\CH_\ta$ in the $i$-th replica is denoted as $\CH_\ta^{(i)}$ and its basis as $|\alpha_\ta^{(i)}\rangle$. The wavefunction of the $i$-th replica is then $\psi_{\alpha_1^{(i)}\ldots \alpha_\tq^{(i)}}$ and its conjugate is $\bar \psi^{\alpha_1^{(i)}\ldots \alpha_\tq^{(i)}}$. The analogy with gauge invariance can be pushed further and be made into an equivalence. Consider a $0$-dimensional gauge theory with gauge group $\prod_\ta U(d_\ta)$ coupled with a complex bosonic field $\psi$ that transforms in the fundamental representation with respect to each of the $U(d_\ta)$ factors at weak coupling. The set of local unitary invariants that we are after is precisely the set of gauge invariant operators of this quantum mechanical theory. The Hilbert space of the theory is spanned by these gauge invariant operators. 

\begin{figure}[h]
    \begin{center}
        \includegraphics[scale=1.3]{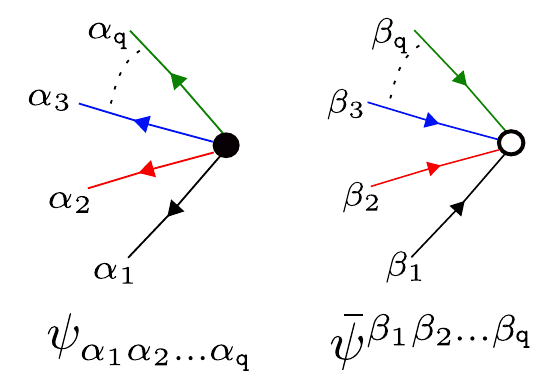}
    \end{center}
    \caption{Graphical notation for the wavefunction $\psi$ and its conjugate $\bar \psi$.}\label{psi-barpsi}
\end{figure}

\begin{figure}[h]
    \begin{center}
        \includegraphics[scale=1.5]{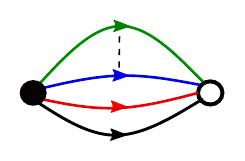}
        \qquad \qquad\qquad
        \includegraphics[scale=1.7]{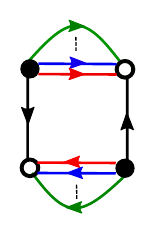}
    \end{center}
    \caption{Examples of multi-partite measures.}\label{examples}
\end{figure}

To classify such measures, it is convenient to introduce a graphical notation in which the wavefunction $\psi$ is denoted as a $\tq$-valent black vertex as in figure \ref{psi-barpsi}.
The vertex has outgoing colored edges as they stand for fundamental indices $\alpha_\ta$ of distinct Hilbert spaces $\CH_\ta$. Similarly $\bar \psi$ is denoted as a $\tq$-valent white vertex with incoming colored edges. The index contraction in $\CH_\ta$ is denoted as joining a white vertex with a black vertex with the appropriately colored edge. An entanglement measure $\cal E$ is then a bi-partite graph made out of these vertices with no dangling edges. Two examples of such measures are given in figure \ref{examples}.
The first of the figures is simply the squared norm of the state and the second is a non-trivial measure.

The set of all entanglement measures forms a commutative ring.  Recall that a ring is a set with two binary operations, addition and multiplication with multiplication being distributive over addition. The addition is commutative and has an inverse. Multiplication need not have an inverse but is associative. In case multiplication is also commutative, the ring is known as a commutative ring.
It is convenient to characterize the ring in terms of its generators. The generating set of a ring is a set of elements that generates the whole ring through inverse and the two binary operations. The simplest example of a commutative ring is the ring of integers. In this case, the generating set can be taken to be the set consisting of a single element $1$. The generating set of the ring of polynomials $P(x)$ with integer coefficients can be taken to be the monomial $x$. When there are no relations between generators then the ring is said to be generated freely. Both the rings mentioned above are freely generated.

The set of invariants described above consists of connected as well as disconnected graphs. The disconnected graphs are obtained by multiplying connected graphs together so it is useful to focus our attention only on the connected graphs. In the quantum mechanical gauge theory, the connected and disconnected graphs correspond to ``single-trace'' and ``multi-trace'' operators respectively. The single traces generate the entire set of invariants by multiplication and hence provide a natural generating set for the ring. For the finite-dimensional case i.e. when all $d_\ta$ are finite, not all elements of this generating set are independent. This is because the total number of components of $\psi$ is finite while the total number of single traces i.e. connected graphs is infinite. It is interesting to understand the independent single trace generators and characterize the relations among them if there are any. 

Let us take the example of a bi-partite state. Tracing over the second party we get the density matrix $\rho$ which transforms in the adjoint representation with respect to $U(d_\1)$. The connected graphs correspond actually to single traces ${\rm Tr} \rho^i$. Thanks to Caley-Hamilton theorem\footnote{Caley-Hamilton theorem states that for any $n\times n$ matrix $A$, $p_A(A)=0$ where $p_A(\lambda)\equiv {\rm det}(\lambda {\mathbb I}-A)$ is the characteristic polynomial of $A$. It expresses $A^n$ in terms of lower matrix powers of $A$. If the rank of the matrix is not maximal then one can construct a lower-order polynomial by removing the factors of $\lambda$. Then $A^r$ is expressed in terms of lower powers of $A$.}, we know that not all such invariants are independent. More precisely, ${\rm Tr}\rho^i$ for $i <  r$ are all independent where $r$ is the rank of the matrix and the rest can be expressed as their linear combinations. We will not be able to characterize the independent generators for a general number of parties but we will be able to count them, along with the relations between them if there are any.

\subsection{Counting}\label{count}
The main tool that we will use to count the number of local unitary invariants is the so-called plethystic program. It first made an appearance in physics in the context of counting gauge invariant operators \cite{Feng:2007ur}. Since then it has been used extensively for computing the superconformal indices of gauge theories. See \cite{Gadde:2020yah} for a pedagogical introduction to the superconformal index and its computation. The counting problem is defined as follows. The number operator $n$ of $\psi$'s  is an operator acting on the vector space $V$ of local invariant operators. As remarked earlier, it is the same as the replica number. The partition function on the measures is defined as
\begin{align}
    \hat Z^{(\tq)}(x)\equiv {\rm Tr}_V x^n = \sum_{n=0}^\infty \, P_n^{(\tq)} \, x^n.
\end{align}
When expanded in the powers of $x$, the coefficient of $x^n$ is the number of local unitary invariants with replica number $n$. While we are at it, let us also define the partition function over vector space $V_C$ of \emph{connected} graphs as $Z(x)\equiv {\rm Tr}_{V_C} x^n$. It is known that these two partition functions are related to each other via the so-called plethystic exponential ({\tt PE}) and plethystic logarithm ({\tt PL}) respectively. 
\begin{align}
    \hat Z(x) & = {\mathtt {PE}}[Z(x)] = \exp\Big( \sum_{k=1}^\infty \frac{1}{k} Z(x^k)\Big )\\
    Z(x) &= {\mathtt {PL}}[\hat Z(x)] = \sum_{k=1}^\infty \frac{\mu(k)}{k} \log \Big(\hat Z (x^k)\Big).
\end{align}
Here the function $\mu(x)$ is the M{\"o}bius function. It is defined as 
\begin{equation}
    \mu_{k} =
      \begin{cases}
        0 & \text{if $k$ has repeated prime factors}\\
        1 & \text{if $k=1$}\\
        (-1)^n & \text{if $k$ is a product of $n$ distinct primes.}
      \end{cases}       
  \end{equation}
The partition function over bosonic multi-particle states is always the {\tt PE} of the single-particle states. Although the formula looks daunting, it can be understood in terms of a simple example. If there is a single particle then its partition function is $Z(x)= x$. Then ${\mathtt {PE}}[Z(x)]= 1/(1-x)$. This is simply the partition function over multi-particle states of identical bosonic particles. So the operation of {\tt PE} can be thought of as ``multi-particling'' the single particle partition function. It essentially implements $x+y+\ldots \to (1-x)^{-1}(1-y)^{-1}\ldots$. 

The partition function over disconnected graphs $\hat Z(x)$ is computed by a matrix integral over unitary groups. In the quantum mechanical gauge theory, we first consider the ``single letter'' partition function over the wavefunction ``field'' $\psi$ and its complex conjugate $\bar \psi$. Recall that $\psi$ transforms in the fundamental representation with respect to all the unitary factors $\prod_{\ta=1}^{\tq} U(d_\ta)$. 
\begin{align}
    z(x,\cU_{\ta})= x \, \prod_{\ta=1}^{\tq} {\rm Tr}(\cU_\ta)+\prod_{\ta=1}^{\tq} {\rm Tr}(\cU_\ta^\dagger).
\end{align}
Here $x$ denotes the contribution from a single replica $\psi$ and the factor $\prod_{\ta=1}^{\tq} {\rm Tr}(\cU_\ta)$ is the character of the fundamental representation of the local unitary group $\prod_{\ta=1}^{\tq} U(d_\ta)$. To compute $\hat Z^{(\tq)}(x)$ we first compute the multi-letter partition function by taking the {\tt PE} of $z(x,\cU_\ta)$ and then projecting onto gauge invariants by integrating over the unitary group $\prod_{\ta=1}^{\tq} U(d_\ta)$ with the Haar measure.
\begin{align}\label{matrix-integral}
    \hat Z^{(\tq)}(x) & =\int \prod_{\ta=1}^{\tq} d\cU_\ta\, {\mathtt {PE}} \,\Big[z(x,\cU_\ta)\Big] =\int \prod_{\ta=1}^{\tq} d\cU_\ta\, \exp\Big(\sum_{k=1}^{\infty}\frac{1}{k}\,z(x^k,\cU_\ta^k)\Big).
\end{align}
This is the partition function over all \emph{disconnected} graphs of $\tq$-party entanglement measures. To compute the partition function $Z^{(\tq)}(x)$ over \emph{connected} graphs we take the {\tt PL} of $\hat Z^{(\tq)}(x)$. 

For $\tq=2$, with $d_\1= d_\2 = d$, 
\begin{align}
    Z^{(\2)}(x)=x+x^2+x^3+\ldots +x^{d-1}.
\end{align}
The term $x^i$ is contributed by ${\rm Tr} \rho^i$. The series truncates because the traces of higher powers of $\rho$ are not independent and are obtained via linear combinations of lower powers.

For higher values of $\tq$ it is difficult to evaluate $Z^{(\tq)}(x)$ for general values of $d_\ta$. Even after fixing $d_\ta$, we can only compute this function as expansion in $x$ up to a few terms. Below we give $Z^{(\tq)}(x)$ for $\tq=3,4$ for the case of qubits i.e. for  $d_\ta=2$.
\begin{align}
    Z^{(\3)}(x) & =x+3x^2+x^3+x^4+x^6-x^{12}. \notag                          \\
    Z^{(\4)}(x) & =x+7x^2+12x^3+ 50x^4+ 111 x^5+ 323x^6+568x^7+{\cal O}(x^8).
\end{align}
We have obtained this result using ${\mathtt {mathematica}}$.  For $Z^{(\4)}(x)$, we have obtained the terms up to $x^{60}$ and unlike  $Z^{(\2)}(x)$ and $Z^{(\3)}(x)$, this series does not  terminate. Let us pause to explain the terms with negative coefficients. As explained earlier, $Z^{(\tq)}(x)$ is a partition function over connected graphs of invariants. When the ring of all invariants is not freely generated by a subset of connected graphs then the {\tt PL} counts not only generators of the ring but also relations among them with a negative sign and relations of relations (if they exist) with a positive sign and so on. That is why the partition function $Z^{(\tq)}(x)$ can have terms with both positive and negative signs.

\begin{figure}[h]
    \begin{center}
        \includegraphics[scale=1.5]{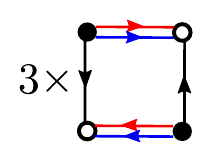}
    \end{center}
    \caption{Three measures that are quadratic in $\psi$. The notation $3\times $ stands for the three measures obtained by permutation of parties i.e. colors.}\label{quadratic}
\end{figure}

\begin{figure}[h]
    \begin{center}
        \includegraphics[scale=1.3]{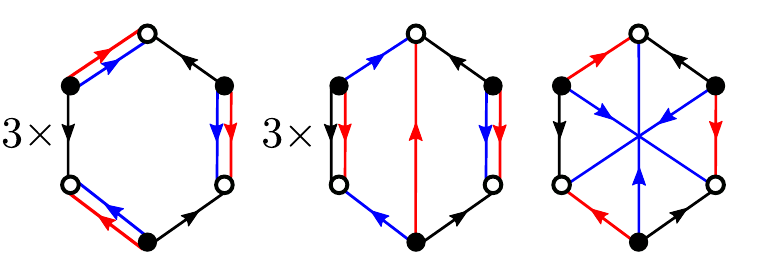}
    \end{center}
    \caption{Seven measures that are cubic in $\psi$. The notation $3\times $ stands for the three measures obtained by permutation of parties i.e. colors.}\label{cubic}
\end{figure}

The ring of invariants is freely generated by connected graphs when all the dimensions $d_\ta$ are taken to be infinity. 
This is the case we will focus on in the rest of the paper because this limit is relevant for counting entanglement measures for quantum field theories where parties are chosen to be distinct spatial subregions. In this case, the matrix integral \eqref{matrix-integral} can be performed using saddle point approximation and a compact expression for $\hat Z^{(\tq)}(x)$ can be obtained for any $\tq$. We have done this computation in appendix \ref{plethystic} and the result is produced below.
\begin{align}\label{Zq}
    \hat Z^{(\tq)}(x) = \prod_{r=1}^\infty \left( \sum_{k=0}^\infty \Big(x^r r^{\tq-2}\Big)^{k} \Big(k!\Big)^{\tq-2} \right).
\end{align}
As a quick sanity check, if we set $\tq=1$, we get $Z^{(\mathtt {1})}(x)\equiv {\mathtt {PL}}[\hat Z^{(\1)}(x)]=x$. This is as expected because the norm is the only local unitary invariant for a single-party system. Also, when we set $\tq=2$, we get the plethystic logarithm $Z^{(\mathtt {2})}(x)\equiv {\mathtt {PL}}[\hat Z^{(\2)}(x)]=x/(1-x) $. This is nothing but the partition function over infinitely many Renyi entropies, except for the first term $x$ which corresponds to the norm of the state. For $\tq=\3$ we get
\begin{align}
    Z^{(\3)}(x)=x + 3x^2 + 7x^3 + 26x^4 + 97x^{5}+\ldots.
\end{align}
The first $3$ non-trivial invariants that are quadratic in $\psi$ are the ones shown in figure \ref{quadratic}.
The next $7$ invariants that are cubic in $\psi$ are shown in figure \ref{cubic}.
One lesson from the expression \eqref{Zq} is that, unlike in the $\tq=2$ case, the number of $\tq \geq 3$ partite measures increases very rapidly with the replica number. It is then a non-trivial task to look for holographically probe measures in this vast zoo. Shortly we will come up with a set of conditions that a measure must obey to have the probe property.

\subsection{Graphs, permutations and replica symmetry}
\begin{figure}[h]
    \begin{center}
        \includegraphics[scale=1.3]{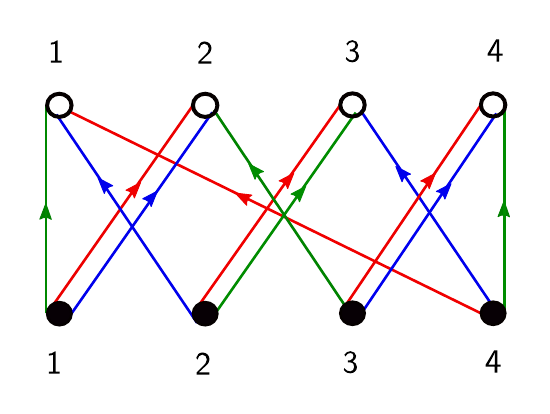}
    \end{center}
    \caption{Describing a multi-partite measure in terms of permutations. The permutations associated with red, blue and green parties, in the cycle notation, are $(1234), (12)(34)$ and $(1)(23)(4)$ respectively}\label{perms}
\end{figure}

In this subsection, we will formulate the entanglement measures in terms of permutations of replicas. A general measure $\cal E$ can be written in terms of contractions of fundamental indices of $n$-replicas of $\psi$ with anti-fundamental indices of $n$-replicas of $\bar \psi$  as follows
\begin{align}
    {\cal E}                                                & =\Big(\psi_{\alpha_1^{(1)}\ldots \alpha_\tq^{(1)}}\ldots \psi_{\alpha_1^{(n)}\ldots \alpha_\tq^{(n)}}\Big)\Big( \bar \psi^{\beta_1^{(1)}\ldots \beta_\tq^{(1)}}\ldots \bar \psi^{\beta_1^{(n)}\ldots \beta_\tq^{(n)}} \Big)\delta^{\vec \alpha_1}_{\sigma_1\cdot\vec\beta_1}\ldots \delta^{\vec \alpha_\tq}_{\sigma_\tq\cdot\vec\beta_\tq}\notag \\
    {\rm where}\quad \delta^{\vec \alpha_\ta}_{\sigma_\ta\cdot\vec\beta_\ta} & \equiv
    \delta^{\alpha_\ta^{(1)}}_{\beta_\ta^{(\sigma_\ta\cdot 1)}}\ldots \delta^{\alpha_\ta^{(n)}}_{\beta_\ta^{(\sigma_\ta\cdot n)}}
\end{align}
If we call the pair of $\psi$ and $\bar \psi$ as a replica then the measure ${\cal E}$ is labeled by $\tq$ permutation elements $\sigma_\ta$ of the permutation group $S_n\equiv G$ acting on the replica set. The element $\sigma_\ta$ indicates how the $n$ fundamental indices of party $\ta$ are contracted with $n$ anti-fundamental indices. Graphical presentation of a measure with $n=4, \tq=3$ is given in figure \ref{perms}.
The red-colored edges of all the replicas of $\psi$ are connected to those of replicas of $\bar \psi$ with a permutation element $\sigma_1$ that is the one cycle $(1234)$. Similarly, the permutation elements for party $2$ (blue colored edges) and $3$ (green colored edges) respectively are $(12)(34)$ and $(1)(23)(4)$. From this discussion, it is also clear that given a $\tq$-colored regular bi-partite graph on $n$ white and $n$ black nodes, one can always read off a set of $\tq$ elements of the permutation group $\{\sigma_1,\ldots, \sigma_\tq\}$ such that the graph is a presentation of the measure ${\cal E}_{\sigma_1\ldots\sigma_\tq}$. 

Let us denote the subgroup of $G$ generated by $\sigma_\ta$ as $K$. 
In special cases, the measure $\cal E$ enjoys a certain symmetry that we suggestively call the replica symmetry. This is the symmetry of the bipartite graph corresponding to $\cal E$. It is generated by the elements of $G$ that commute with all $\sigma_\ta$'s. In other words, it is the commutant of $K$ in $G$, usually denoted as $c_G(K)$.

\section{Towards holographic probe measures}\label{lm}
We are interested in computing multi-partite entanglement measures for holographic CFT states i.e. for CFT states that admit semiclassical dual geometry.
In this section, we will motivate successively stringent conditions on general multi-partite entanglement measures to obtain a measure that for holographic states is computed by a probe of the dual geometry. In deriving these conditions, we will closely follow Lewkowycz and Maldacena's derivation \cite{Lewkowycz:2013nqa} of the Ryu-Takayanagi formula for entanglement entropy.

\subsection{Maximally symmetric measures} 
The CFT path integral that computes any multi-partite measure ${\cal E}(\sigma_1,\ldots,\sigma_\tq)$ is formulated straightforwardly by considering replica number worth of copies of the CFT path integral on ${\cal M}$ and ``cutting and pasting'' along the $\tq$ regions as specified by the permutation elements $\sigma_\ta$. The replicated manifold ${\cal M}_n$ constructed thus has special loci at the boundaries of $\tq$ regions with excess conical angles. Moreover, the background fields enjoy symmetry under the replica symmetry group defined above.

For the bipartite  Renyi-entropy $S^{(2)}_n$, Lewkowycz and Maldacena have computed the CFT path integral using the dual gravitational theory \cite{Lewkowycz:2013nqa}. A crucial part of their derivation is the assumption that the dominant bulk solution ${\cal B}_n$ also preserves the replica symmetry. To derive the Ryu-Takayanagi formula, it is then convenient to orbifold the bulk solution by the replica symmetry group. After orbifolding, the boundary goes back to being the ``un-replicated'' original boundary ${\cal M}$. This is because the replica symmetry group acts freely and transitively on the replica set. So, if we wish to characterize the multi-partite entanglement measures that are computed by dual geometries with the un-replicated original boundary then the replica symmetry group must satisfy the same condition. We formalize this condition as:
\begin{itemize}
    \item {\bf Maximal symmetry condition}: The replica symmetry group $c_G(K)$ contains an abelian subgroup $H$ that acts freely and transitively on the replica set.
\end{itemize}
Strictly speaking, the replica symmetry subgroup $H$ need not be abelian but we consider the abelian case in this paper to make the analysis tractable. It would be interesting to consider non-abelian generalizations in the future.
The measures whose replica symmetry satisfies the maximal symmetry condition are called maximally symmetric measures.
In the rest of the subsection, we will characterize all the maximally symmetric measures.

Recall that transitive means that for every $x,y$ of the replica set there exists $h\in H$ such that $h\cdot x=y$ and free means that no element of $H$ has a fixed point. Transitive abelian subgroups of $G$ have the following properties. See \cite{scott64}, section 10.3:
\begin{itemize}
    \item They are freely acting. So, strictly speaking, the condition of free action in the maximal symmetry condition is not necessary.
    \item They are maximal i.e. they are commutants of themselves. So we have $c_G(H)=H$.
\end{itemize}
The second property will be important for characterizing all maximally symmetric measures.

When a group $H$ acts on a set $X$ then, according to Burnside's lemma,
\begin{align}
    |{X}/{H}|=\frac{1}{|H|}\sum_{h\in H} \, |{X}_h|.
\end{align}
Here the notation $|X|$ stands for the cardinality of the set $X$ and ${X}_h$ is the set of points in ${X}$ left fixed by the action of $h\in {H}$. Then the transitive-ness of the action implies $|{X}/{H}|=1$ and free-ness implies $|{X}_g|=0$ for $g\neq {\rm id}$. So, in particular,  $|{H}|=|{X}|\equiv n$ because ${X}_{\rm id}={X}$. So the subgroup $H$ must contain as many elements as the number of replicas.
In particular, using the fundamental theorem of algebra, $H$ must take the form ${\mathbb Z}_{n_1}\otimes\ldots\otimes{\mathbb Z}_{n_k}$ with $\prod_{k}n_k=n$.
There is a simple way of understanding the action of $H$ on $X$. We take the replica set $X$ as $H$ itself. Then the free and transitive action in question is simply the left (or right) multiplicative action on $H$ onto itself. 

Now we characterize the measures that have $H$ as their replica symmetry group. 
Using the group theoretic property of commutants, if $S_1$ and $S_2$ are two subgroups of $G$ satisfying $S_1 \subseteq S_2$ then $c_G(S_2)\subseteq c_G(S_1)$. Also  for any subgroup $S$ of $G$, $S\subseteq c_G(c_G(S))$. As the subgroup $H\subseteq c_G(K)$, we get $c_G(c_G(K)) \subseteq c_G(H)$. Using $c_G(H)=H$ and  $K\subseteq c_G(c_G(K))$ we get, $K\subseteq H$. As $K$ is the group generated by $\sigma_\ta$, maximally symmetric measures are now completely characterized by explicitly specifying the associated permutation elements $ \sigma_\ta,\, \ta=1,\ldots ,\tq$. We simply pick them arbitrarily from \emph{some} abelian group $H$! The chosen abelian group $H$ is the desired subgroup of the replica symmetry group $c_G(K)$ that acts freely and transitively on the replicas.

\subsubsection*{Examples}
The von Neumann entropy is a maximally symmetric measure. It corresponds to the following choice of permutations for the two parties
\begin{align}
    \sigma_{\1}={\rm id},\qquad \sigma_{\2}=g,
\end{align}
where $g$ is the generator of a ${\mathbb Z}_n$ subgroup. These two permutation elements commute with each other.

Entanglement negativity is also a maximally symmetric measure. It is defined for a tripartite pure state with the following choice of permutation elements.
\begin{align}
    \sigma_{\1}={\rm id},\qquad \sigma_{\2}=g,\qquad \sigma_{\mathtt 3}=g^{-1}.
\end{align}
These elements are mutually commuting.

The other popular measure of tri-partite entanglement, the reflected entropy, is not maximally symmetric because the permutation elements $\sigma_\ta$ are not mutually commuting \cite{Dutta:2019gen}. We will discuss other aspects of the reflected entropy further in section \ref{marriage}.

\subsubsection{Formulation in terms of the density matrix}
We formulated bi-partite graphs in terms of permutation elements $\sigma_\ta$.  
However, the map from the graph to permutation elements is not unique. There are multiple collections of permutation elements $\{\sigma_1,\ldots, \sigma_\tq\}$ which give the same graph. This is due to the freedom of relabeling the vertices.
Relabeling of the $\psi$ vertices ($\bar \psi$ vertices) gives rise to the left (right) multiplication freedom. However, it is more convenient to think of this redundancy as left-multiplication and conjugation as follows.
\begin{itemize}
    \item Left-multiplication: we can relabel only the black vertices keeping the labels of the white vertices fixed. This gives the equivalence of $\cal E$ under left-multiplication by an arbitrary group element $h$ i.e.
          \begin{align}\label{left-gauge}
              {\cal E}(\sigma_1,\ldots,\sigma_\tq)={\cal E}(h\cdot\sigma_1,\ldots,h\cdot\sigma_\tq) \qquad \forall h\in G.
          \end{align}
    \item Conjugation: In addition to this, we can relabel both black and white vertices simultaneously. This results in the equivalence under conjugation by an arbitrary element $g$.
          \begin{align}\label{conj-gauge}
              {\cal E}(\sigma_1,\ldots,\sigma_\tq)={\cal E}(g^{-1}\cdot\sigma_1\cdot g,\ldots,g^{-1}\cdot\sigma_\tq\cdot g) \qquad \forall g\in G.
          \end{align}
\end{itemize}
It is convenient to gauge fix the left multiplication freedom by fixing $\sigma_\1={\rm id}$. Interestingly, it gives a convenient formulation of the measure in terms of the density matrix obtained by tracing out party $\1$. 

\begin{figure}[h]
    \begin{center}
        \includegraphics[scale=0.6]{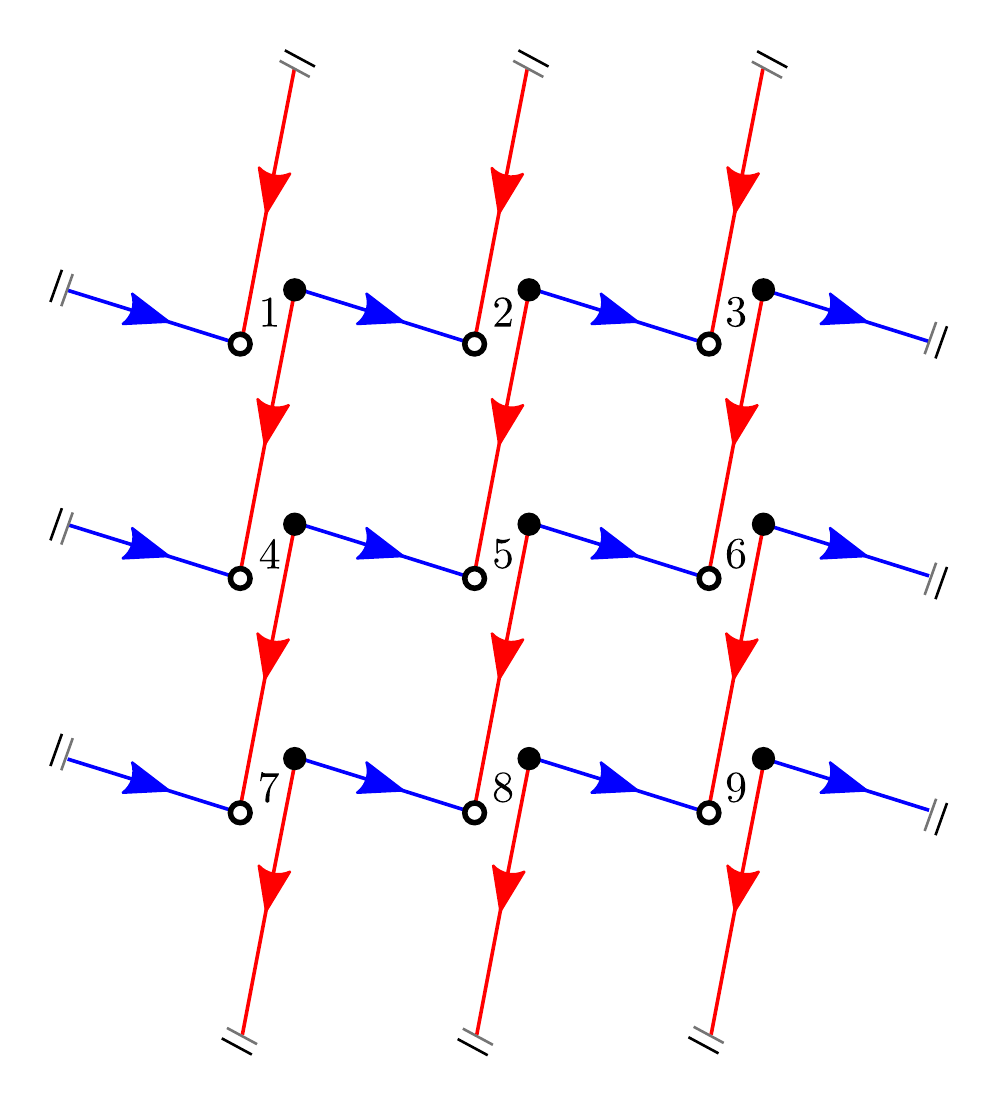}
    \end{center}
    \caption{Measure corresponding to choosing $\sigma_1$ and $\sigma_2$ to be two independent generators of ${\mathbb Z}_3\times {\mathbb Z}_3$. It has three identical connected components. Each of the component corresponds to ${\rm Tr}\rho^3$.}\label{q2}
\end{figure}

Choosing $\sigma_1={\rm id}$ provides a canonical correspondence between $\psi$'s and $\bar \psi$'s. We now take the set of pairs connected by party $\1$ index as the replica set. Because contraction of the party $\1$ index is also the density matrix on the rest of the parties, we effectively treat the set of density matrices as the set of replicas. We have explained this in figure \ref{q2}. 
Consider $H={\mathbb Z}_3\times {\mathbb Z}_3$ and take $\sigma_1=g_1\otimes {\rm id}$ and $\sigma_2= {\rm id} \otimes g_2$ where $g_i$'s are generators of ${\mathbb Z}_3$. The resulting bi-partite graph is shown in figure \ref{q2}. 
As we can see, it consists of three identical disconnected components. If we gauge fix $\sigma_1={\rm id}$, this identifies the $\psi$ and $\bar \psi$ that are connected by the red arrows as a density matrix and the blue arrows give the contractions of the indices corresponding to the second party. Each of the three connected components gives the familiar ${\rm Tr}\rho_{2}^3$. Sometimes it is convenient to describe the measure in this way in terms of the density matrices. 
\begin{figure}[h]
    \begin{center}
        \includegraphics[scale=0.7]{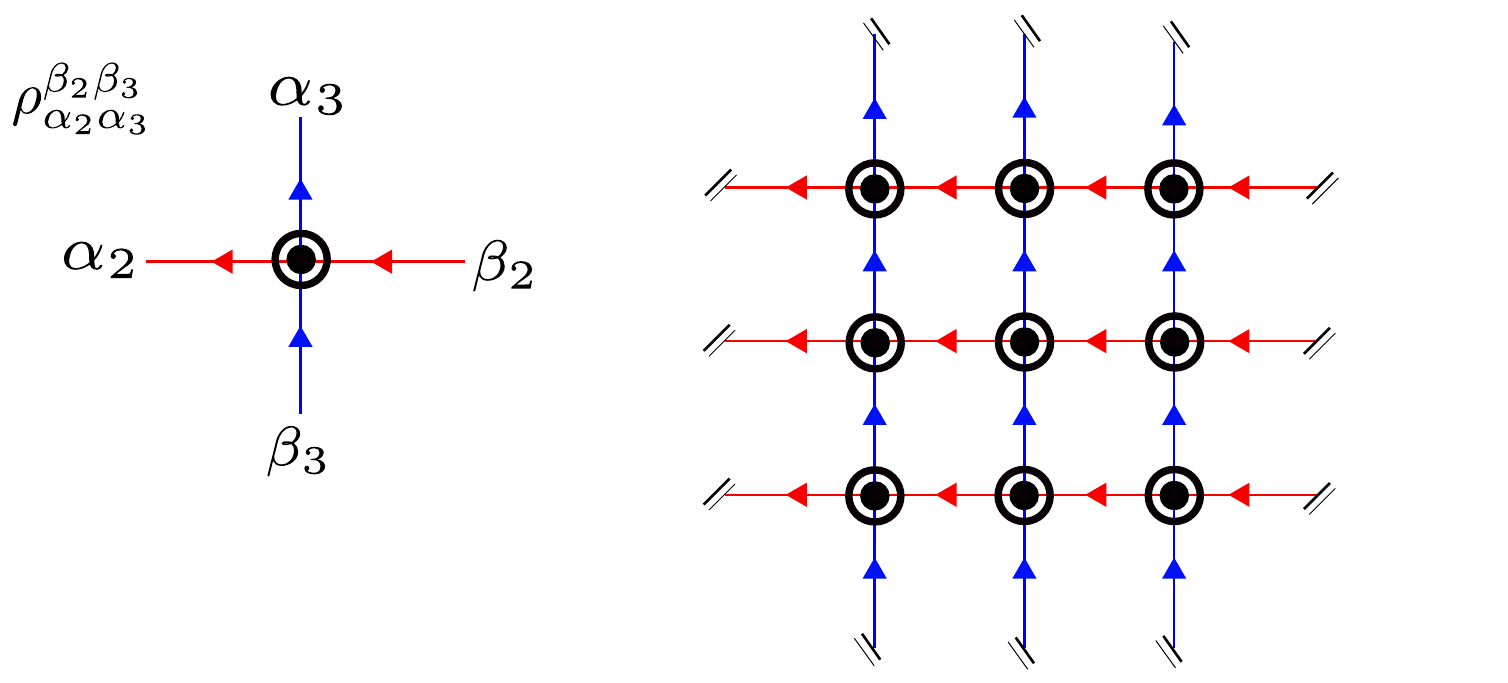}
    \end{center}
    \caption{A single connected component of the measure corresponding to choosing $\sigma_1$, $\sigma_2$ and $\sigma_3$ to be three independent generators of ${\mathbb Z}_3\times {\mathbb Z}_3\times {\mathbb Z}_3$.}\label{q3}
\end{figure}
In figure \ref{q3}, we have shown a single connected component, out of the three identical ones, of the measure obtained with $H={\mathbb Z}_3\times {\mathbb Z}_3\times {\mathbb Z}_3$ with $\sigma_1, \sigma_2$ and $\sigma_3$ being their independent generators respectively.

If we gauge fix some $\sigma_\tb={\rm id}$ instead, the measure might look quite different as a graph constructed out of the density matrices (obtained after tracing out party $\tb$) but it is the same measure as obtained by gauge fixing $\sigma_\1={\rm id}$. This is simply a consequence of the left-multiplication gauge invariance. To see this explicitly, we simply need to observe that the group generated by $\sigma_1^{-1}\sigma_\ta$ is the same as the one generated by $\sigma_\tb^{-1}\sigma_\ta$ for any $\tb$. 

\subsection{Holographic probe family}\label{probe-family}

Let us analyze the conditions that are further required to get a holographic probe measure of entanglement like the Ryu-Takayanagi formula.

In the case of bi-partite Renyi measure $S^{(2)}_n$, the replica symmetric bulk solution ${\cal B}_n$ has certain special co-dimension $2$ loci ${\cal L}_n$ that are fixed under the replica symmetry group ${\mathbb Z}_n$. These loci emanate from points of conical excess on the replicated boundary ${\cal M}_n$.
Orbifolding ${\cal B}_n$ by the replica symmetry group takes the boundary back to ${\cal M}$ and yields a bulk geometry ${\widetilde {\cal B}}_n$ that has conical defects ${\widetilde {\cal L}}_n$. These are essentially images of ${\cal L}_n$ under orbifolding. The conical opening angle at ${\widetilde {\cal L}}_n$ is $2\pi/n$. The path integral for the Renyi-entropy thus gets reduced to the computation of bulk action in the presence of a conical defect. The advantage of this reformulation is that the bulk description admits a natural analytic continuation in $n$. To compute the Renyi measure for a real $n$, we simply consider the singularity with conical opening angle $2\pi/n$ and to obtain von Neumann entropy we take $n\to 1$ limit. In this limit, the conical opening angle approaches $2\pi$ and the defect flattens out. The action for this almost smooth geometry can then be computed in the probe fashion on the bulk solution ${\cal B}$ dual to the CFT state on the un-replicated boundary manifold ${\cal M}$. So the crucial property of the bulk solution is that it is labeled by a single integer $n$ and admits analytic continuation in $n$ such that in the $n\to 1$ limit, the opening angles of all conical defects approach $2\pi$.

Motivated by Lewkowycz and Maldacena's construction, we wish to define a family of multi-partite maximally symmetric measures ${\cal E}_n(\sigma_\ta)$ labeled by a single integer $n$. 
We will consider the condition,
\begin{itemize}
    \item {\bf Probe family condition}: The permutation elements $\sigma_\ta(n)$ are such that the cycle lengths of all the cycles in $\sigma_\ta^{-1}\sigma_\tb$ go to $1$ as $n\to 1$ for all parties $\ta,\tb$.
\end{itemize}
We will call the family of maximally symmetric measures obeying the above condition a holographic probe family or simply, a probe family.
We will show later that conical singularities in the orbifolded geometry associated with the probe family flatten out in the $n\to 1$ limit. As a result, the measure corresponding to the limit $n\to 1$ can be computed by a probe on the original solution ${\cal B}$. We call the $n=1$ element of the family (assuming it exists, see section \ref{analytic}), a probe measure.

\subsubsection*{Examples}
The family of Renyi entropies is a probe family and the associated probe measure is the von Neumann entropy.

For entanglement negativity, the permutation elements $\sigma_1^{-1}\sigma_2, \sigma_1^{-1}\sigma_3$ and $\sigma_3^{-1}\sigma_2$ are $g,g^{-1}$ and $g^2$ respectively. Here $g$ is the generator of a cyclic group. When $|g|$ is even, say $n=2k$ then $g^2$ consists of two cycles of length $k=n/2$ each. So even though $|g|=|g^{-1}|\to 1$ as $n\to 1$, cycle lengths of $g^2$ go to $1/2$. So clearly it does not form a probe measure family. However, if $|g|$ is odd, say $n=2k+1$ then $g^2$ consists of a single cycle of the same length. The cycle lengths of $g,g^{-1}, g^2$ all go to $1$ as $n\to 1$ i.e. $k\to 0$. So entanglement negativity, for odd replica number, does form a probe measure family.

It seems difficult to solve for the probe family condition completely because lengths of cycles in $\sigma_\ta^{-1}\sigma_\tb$ depend on the details of $\sigma_\ta$ and $\sigma_\tb$ in general and not just on their equivalence classes. However, we will soon find a large class of solutions where the equivalence class of $\sigma_\ta^{-1}\sigma_\tb$ can be computed just from the equivalence class of $\sigma_\ta$ and $\sigma_\tb$.

\subsection{Special symmetric measures and special probes}\label{special-probe}

We call the maximally symmetric measures whose permutation elements $\sigma_\ta$  satisfy the following condition,
\begin{itemize}
    \item {\bf Special symmetry condition}:
          \begin{align}
              \langle \sigma_\ta\rangle \cap \langle \sigma_\tb\rangle = {\rm id}\qquad {\rm  for}\qquad  \ta\neq \tb
          \end{align}
\end{itemize}
a special symmetric measure. Here $\langle g\rangle$ is the cyclic group generated by $g$. If the order of $g$ is $|g|$ i.e. if $|g|$ is the smallest integer such that $g^{|g|}=1$ then $\langle g\rangle$ is isomorphic to ${\mathbb Z}_{|g|}$.
The condition $\langle \sigma_\ta\rangle \cap \langle \sigma_\tb\rangle = {\rm id}$ for $\ta\neq \tb$ morally means that the generators $\sigma_\ta$ are linearly independent. In particular, this means $H=K=\otimes_{\ta=1}^{\tq} \langle \sigma_\ta \rangle=\otimes_\ta {\mathbb Z}_{|\sigma_\ta|}$. So essentially, a special symmetric measure is labeled by a choice of $\tq$ integers $m_\ta\equiv |\sigma_\ta|$. We will denote it as ${\cal E}^{\{m_\1,m_\2,\ldots\}}$.

An important property of special symmetric measures is that the order of the element $\sigma_\ta^{-1}\sigma_\tb$ can be characterized in terms of the order of $\sigma_\ta$ and $\sigma_\tb$,
\begin{align}\label{lcm-property}
    |\sigma_\ta^{-1}\sigma_\tb|={\rm lcm}(|\sigma_\ta|,|\sigma_\tb|).
\end{align}
This property lets us straightforwardly construct a holographic probe family. The $\tq$-partite measure constructed in \cite{Gadde:2022cqi}, corresponds to choosing $|\sigma_\ta|=n$ for all $\ta$. Then $|\sigma_\ta^{-1}\sigma_\tb|$ is also $n$. The fact that  $|\sigma_\ta|\to 1$ in the $n\to 1$ limit implies that the length of all the cycles goes to $1$ in the $n\to 1$ limit as required by the holographic probe condition. Explicitly, this probe family is defined as\footnote{This probe family was called Renyi multi-entropy in \cite{Gadde:2022cqi} and was defined without the factor $1/n^{\tq-1}$.}
\begin{align}
    S_n^{(\tq)}\equiv \frac{1}{1-n}\frac{1}{n^{\tq -1}} \log\Big({\cal E}^{\{n,n,\ldots\}}\Big).
\end{align}
The associated probe measure obtained was termed multi-entropy in \cite{Gadde:2022cqi}.

Because we organized the multi-partite measures in this systematic manner, we can quickly recognize that a significantly larger class of holographic probe measures can be constructed.
One obtains a holographic probe measure starting from any special symmetric measure. If $|\sigma_\ta|=m_\ta$ for a special symmetric measure then the family of special symmetric measures with $|\sigma_\ta|=m_\ta^{n-1}$ is a probe measure family. This is  because
\begin{align}
    |\sigma_\ta^{-1}\sigma_\tb|=({\rm lcm}(m_\ta,m_\tb))^{n-1}
\end{align}
and it goes to $1$ as $n\to 1$ for arbitrary integers  $m_\ta$. We will call the probe family obtained from special symmetric measures ${\cal E}^{\{m_\1,m_\2 ,\ldots \}}$ as a special probe family. Explicitly, we define it to be
\begin{align}\label{special-probe-def}
    S_n^{\{m_\1,m_\2,\ldots \}}\equiv \frac{1}{1-n}\frac{1}{(\prod_\ta m_\ta)^{n-1}} \log\Big({\cal E}^{\{m_\1^{n-1},m_\2^{n-1},\ldots\}}\Big).
\end{align}
The special probes are then ${\rm lim}_{n\to 1} S_n^{\{m_\1,m_\2,\ldots \}}$. We will denote them as $S^{\{m_\1,m_\2,\ldots \}}$.

\subsubsection{Number of connected components}
How many connected components does the special symmetric measure have?  This simple-sounding question has an interesting combinatorial answer. The number of replicas in a single connected component is equal to the order of the group $\tilde K$ generated by $\sigma_\1^{-1}\sigma_\ta\equiv \tilde \sigma_\ta$. Note that the group generated  by $\sigma_\tb^{-1}\sigma_\ta$ for all $\ta$ and any fixed $\tb$ is $\tilde K$. The number of connected components of the special symmetric measure is $|H|/|\tilde K|$. Let us discuss the computation of $|\tilde K|$ in terms of in terms of $|\sigma_\ta|=m_\ta$.

For $\tq=2$, the group $\tilde K$ is generated by $\sigma_\1^{-1}\sigma_\2\equiv \tilde \sigma_\2$. It is well known that for independent elements $\sigma_\1$ and $\sigma_\2$, $|\tilde \sigma_\2|={\rm lcm}(m_\1,m_\2)$. 

For $\tq=3$, the group $\tilde K$ is generated by $\sigma_\1^{-1}\sigma_\2\equiv \tilde \sigma_2, \sigma_\1^{-1}\sigma_\3\equiv \tilde \sigma_\3$. Their orders are ${\rm lcm}(m_\1,m_\2)\equiv \tilde m_2$ and ${\rm lcm}(m_\1,m_\3)\equiv \tilde m_\3$ respectively. If $\tilde \sigma_\1$ and $\tilde \sigma_\2$ were independent, $|\tilde K|$ would be product of their orders $\tilde m_1 \tilde m_2$. However, they are not independent since they obey the relation $(\tilde \sigma_\2^{-1} \tilde \sigma_\3)^{{\rm lcm}(m_\2,m_\3)}=(\sigma_\2^{-1} \sigma_\3)^{{\rm lcm}(m_\2,m_\3)}={\rm id}$. In order to compute $|\tilde K|$, we divide $\tilde m_\1\tilde m_\2$ by the order of the group generated by the relation $ (\tilde \sigma_\2^{-1} \tilde \sigma_\3)^{{\rm lcm}(m_\2,m_\3)}$ assuming $\tilde \sigma_\2$ and $\tilde \sigma_\3$ to be independent. For that, we will compute the order of $\tilde \sigma_\2^{{\rm lcm}(m_\2,m_\3)}$ and $\tilde \sigma_\3^{{\rm lcm}(m_\2,m_\3)}$ and the order of the group generated by the relation is the ${\rm lcm}$ of these two orders.
\begin{align}
    |\tilde \sigma_\2^{{\rm lcm}(m_\2,m_\3)}|& = \frac{{\rm lcm}(\tilde m_\2,{\rm lcm}(m_\2,m_\3))}{{\rm lcm}(m_\2,m_\3)}=\frac{{\rm lcm}(m_\1,m_\2,m_\3)}{{\rm lcm}(m_\2,m_\3)}\notag\\
    |\tilde \sigma_\3^{{\rm lcm}(m_\2,m_\3)}|& = \frac{{\rm lcm}(\tilde m_\3,{\rm lcm}(m_\2,m_\3))}{{\rm lcm}(m_\2,m_\3)}=\frac{{\rm lcm}(m_\1,m_\2,m_\3)}{{\rm lcm}(m_\2,m_\3)}\notag\\
    |(\tilde \sigma_\2^{-1} \tilde \sigma_\3)^{{\rm lcm}(m_\2,m_\3)}|&={\rm lcm}(|\tilde \sigma_\2^{{\rm lcm}(m_\2,m_\3)}|, |\tilde \sigma_\3^{{\rm lcm}(m_\2,m_\3)}|)=\frac{{\rm lcm}(m_\1,m_\2,m_\3)}{{\rm lcm}(m_\2,m_\3)}.
\end{align}
This gives,
\begin{align}
    |\tilde K|=\frac{{\rm lcm}(m_\1,m_\2)\cdot {\rm lcm}(m_\2,m_\3)\cdot  {\rm lcm}(m_\3,m_\1)}{{\rm lcm}(m_\1,m_\2,m_\3)}.
\end{align}
This is symmetric between $m_\1,m_\2,m_\3$ as expected.

For higher values of $\tq$, the problem is slightly more involved. It turns out to be useful to think of $|\tilde K|$ as a group generated by $\sigma_\ta^{-1} \sigma_\tb$ for \emph{all} $\ta, \tb$ rather than only by $\sigma_\1^{-1}\sigma_\tb$.\footnote{We thank Arvind Nair for discussion on this point.} To characterize the group, we characterize the relations (and relations among relations, if any and so on) among them. The answer for $|\tilde K|$ will be manifestly symmetric in all the parties with this approach.

Let us again take the case of $\tq=3$. The generators of $\tilde K$ are $\sigma_\1^{-1}\sigma_\2 \equiv h_{\1\2}, \sigma_\2^{-1}\sigma_\3\equiv h_{\2\3}$ and $\sigma_\3^{-1} \sigma_\1\equiv h_{\3\1}$. If these elements were independent, then the group would have been ${\mathbb Z}_{{\rm lcm}(m_\1,m_\2)}\otimes {\mathbb Z}_{{\rm lcm}(m_\2,m_\3)}\otimes {\mathbb Z}_{{\rm lcm}(m_\3,m_\1)}$. But they are not independent and obey the relation $h_{\1\2} h_{\2\3}h_{\3\1}={\rm id}$. The group generated by the relation has the order  ${\rm lcm}(m_\1,m_\2,m_\3)$ because it is the ${\rm lcm}$ of the order of $h_{\1\2}, h_{\2\3}$ and $h_{\3\1}$. The size of $\tilde K$ is then
\begin{align}
    |\tilde K|=\frac{{\rm lcm}(m_\1,m_\2)\cdot {\rm lcm}(m_\1,m_\2)\cdot {\rm lcm}(m_\1,m_\2)}{{\rm lcm}(m_\1,m_\2,m_\3)}.
\end{align}

This analysis generalizes readily to higher values of $\tq$. Let us associate a vertex for every group element $\sigma_\ta$ and denote the generator $\sigma_\ta^{-1}\sigma_\tb\equiv h_{\ta\tb}$ as an edge from vertex $\ta$ to $\tb$. We will have $\,^{q} C_2$ such edges\footnote{Here $^nC_m = n!/(m!(n-m)!)$ is the binomial coefficient. It counts the number of ways $m$ objects can be chosen from the set of $n$ distinct objects.} forming a skeleton of a $\tq$-simplex\footnote{A $\tq$-simplex is a generalization of notion of tetrahedron to $\tq$ dimensions. Examples of a few low-dimensional simplices are the following. A $0$-simplex is a point. $1$-simplex is obtained by connecting a point to the $0$-simplex giving us an interval. A $2$-simplex is obtained by connecting a point to the $1$ simplex with two edges and also filling it in with a face giving us a triangle. A $3$-simplex is obtained by connecting a point to the $2$-simplex by three edges, three triangular faces and a three-dimensional tetrahedral ``face''. In general, a $\tq$-simplex is a $\tq$-dimensional polytope which is the convex hull of its $\tq+1$ vertices, no three of which are collinear.}. The relation among these corresponds to a closed loop and hence a face. For example, in the $\tq=3$ case, the simplex is a triangle and the relation $h_{\1\2} h_{\2\3}h_{\3\1}$ clearly corresponds to the closed loop.

For $\tq=4$, the simplex is the tetrahedron. There are four relations among the $\,^4C_2$ edges corresponding to the $4$ faces. The order of these relations are ${\rm lcm}(m_\1,m_\2,m_\3)$, ${\rm lcm}(m_\1,m_\2,m_\4)$, ${\rm lcm}(m_\1,m_\4,m_\3)$ and ${\rm lcm}(m_\4,m_\2,m_\3)$. 
However, these relations are not independent and have relations among them. These relations of relations correspond to closed two-dimensional surfaces formed out of faces or a 3-simplex. In this case, there is only one. The order of the corresponding element is ${\rm lcm}(m_\1,m_\2,m_\3,m_\4)$ because this is the ${\rm lcm }$ of the above four numbers. All in all, this gives the size of $|\tilde K|$ to be
\begin{align}
    |\tilde K|=\frac{{\rm lcm}(m_\1,m_\2,m_\3,m_\4) \prod_{\ta < \tb} {\rm lcm}(m_\ta, m_\tb)}{\prod_{\ta < \tb < \tc}{\rm lcm}(m_\ta,m_\tb,m_\tc)}.
\end{align}
One can write down the formula for $|\tilde K|$ for higher values of $\tq$ using this ``inclusion-exclusion'' type principle.

\subsection{About analytic continuation}\label{analytic}
The definition of the special probes as well as multi-entropy involves an analytic continuation of the probe family $S^{\{m_\ta\}}_n$, that is defined for integer $n > 1$ to complex $n$ and then taking the $n\to 1$ limit. As we will discuss in section \ref{holographic-dual}, the analytic continuation in $n$ does exist for holographic states if we assume that the dominant bulk saddle preserves replica symmetry.
However, it is \emph{a priori} not obvious that it exists for general quantum states. The problem of analytic continuation can be formulated as a question in complex analysis:
\begin{itemize}
    \item Does there exist an analytic function $f(z)$, that is unique in some sense, that takes given values $f(n)=f_n$ (input data) on positive integers?
\end{itemize}

First of all, the question asked above is ambiguous. We have to specify the sense in which we expect the answer to be unique. 
One solution to this problem is provided by Carlson's theorem. It states the following:
\begin{itemize}
    \item A function $g(z)$ that is analytic for ${\rm Re}(z)>0$ and satisfies
          \begin{align}\label{carlson-cond}
              |g(z)|\leq A e^{c |z|},\quad g(iy)\leq A e^{B y}, \quad B<\pi
          \end{align}
          for some real constants $A$ and $c$ for all $z\in {\mathbb C}$ and $y\in {\mathbb R}$ and vanishes for all non-negative integer must be identically zero.
\end{itemize}
We will call the analyticity and boundedness conditions on $g(z)$ that are required by the theorem as Carlson's conditions. The boundedness condition in the imaginary direction is crucial otherwise $\sin (\pi z)$ forms an immediate counterexample. In particular, this also means that to disallow such a possibility we need to take the bounding exponent $B<\pi$.

The idea is to require that the putative analytic continuation $f(z)$ obeys Carlson's conditions. Thanks to Carlson's theorem this analytic continuation would then be unique. 
For the analytic continuation of $f_n={\rm Tr}\, \rho^n$  imposing Carlson's condition is perfectly suited. This is because $f_n=\sum_i \lambda_i^n$ with\footnote{$\lambda$ can equal $1$ in case of a rank $1$ density matrix. In that case, $f_n={\rm Tr}\rho^n=1$ which analytically continues to $f(z)=1$.} $0\leq\lambda_i<1$ where $\lambda_i$ are (possibly infinitely many) eigenvalues of $\rho$. With this form, it is clear that $f(z)$ defined as $f(z)=\sum_i \lambda_i^z$ obeys Carlson's analyticity and boundedness conditions. This is why Carlson's condition is often used to analytically continue the Renyi entropy $S_n$ away from integer values of $n$.

For multi-partite case, \emph{a priori} it is not clear if the quantity ${\cal E}^{\{m_1^{n-1}, m_2^{n-1},\ldots\}}$ or even ${\cal E}^{\{n,n,\ldots\}}$ is compatible with Carlson's conditions. Let us explain. Let us define the generalized $\tq$-partite GHZ state as 
\begin{align}
    |\Psi\rangle_{\rm GHZ}\equiv \sum_i \lambda_i |e^{1}_i\rangle \otimes |e^{2}_i\rangle \otimes \ldots |e^{\tq}_i\rangle
\end{align}
where all the parties ${\cal H}_\ta$ are take to have the same dimension and $|e_i^{\ta}\rangle $ is a set of orthonormal basis vectors. The measure ${\cal E}^{\{n,n,\ldots\}}$ for this state is given by \cite{Gadde:2022cqi},
\begin{align}\label{ghz}
    {\cal E}^{\{n,n,\ldots\}} = (\sum_i |\lambda_i|^{2n^{\tq-1}})^n.
\end{align}
If we demand that the analytic continuation of ${\cal E}^{\{n,n,\ldots\}}$ is obtained by replacing $n$ by a complex number $z$ then it is clear that the resulting function does not obey Carlson's conditions. Even if we only analytically continue the quantity inside the bracket in this way, this problem persists. It might be that the simple lift of $n$ to $z$ in the formula \eqref{ghz} is not the correct analytic continuation for the generalized GHZ state and that the correct analytic continuation still obeys Carlson's conditions. Here, by correct analytic continuation we mean two things, first is that it should be unique subject to some conditions and second, it should match the analytic continuation for holographic states performed in section \ref{holographic-dual}.
%In the absence of explicit formulas for ${\cal E}^{\{n,n,\ldots\}}$ for general states we can't be sure one way or the other. 
In what follows, we will outline a numerical check of whether the analytic continuation that obeys Carlson's conditions exists. 

This question has been discussed in \cite{regge-viano} in the context of analytic continuation in angular momentum and a necessary condition on the input data $f_n$ has been derived in \cite{osti_4065624}\footnote{We thank Geoff Penington for pointing out this paper to us.}. Below we will review these issues.
It is not difficult to convince oneself that if we wish to uniquely analytically continue a function that is defined on say, even positive integers,  we need to demand Carlson's conditions but with $B <\pi/2$. A generalized version of Carlson's theorem, due to \cite{10.2307/1992882}, implies that if we want uniquely analytically continue a function that is defined on positive points with density $D$ per unit then we should require $B<\pi D$.

From this discussion, we can see that the input data $f_n$ can not be completely independent to have a unique analytic continuation satisfying Carlson's conditions. The argument is as follows. Let us assume a function $f(z)$ satisfying all Carlson's conditions exists. In particular,  $f(iy) \leq e^{C\pi y}, y\in {\mathbb R}$ for some $C<1$. As discussed earlier, we only need points with density $C$ per unit to determine such a function but because the input data is specified for all positive integers i.e. on the set with density $1$, it must not be completely independent.
It is not clear what conditions on the input data are necessary and sufficient for the existence of analytic continuation in the above sense; however in \cite{osti_4065624}, a set of infinitely many necessary conditions is formulated. These conditions express $f_l$ in terms of all the rest of the $f_n$ for all $l$.
We have reproduced the proposed analytic continuation and the consistency conditions it needs to satisfy, in appendix \ref{regge} for the reader's convenience. These conditions of \cite{osti_4065624} are amenable to numerical analysis. If the probe family measures fail these conditions, we need to look for other appropriate conditions that guarantee uniqueness and also match with the analytic continuation performed for the holographic states in section \ref{holographic-dual}. We will not undertake this problem here but continue with the analysis of these measures for holographic states.

\section{Holographic dual of the special probes}\label{holographic-dual}
In this section, we will derive a holographic prescription to compute the special probe measure in CFTs that admit a semiclassical dual. This discussion closely parallels the derivation of the holographic prescription for multi-entropy discussed in \cite{Gadde:2022cqi} which in turn is inspired by the derivation of the Ryu-Takayanagi formula for the entanglement entropy by Lewkowycz and Maldacena \cite{Lewkowycz:2013nqa}.

Let the state  $|\Psi\rangle$ be defined on a time-symmetric Cauchy slice $\CR$ of a $D$-manifold $\CM$. It is given by a Euclidean path integral on the half-space $\CM_\Psi$ such that $\partial \CM_\Psi=\CR$.  The dual bra  $\langle \Psi|$ is constructed by Euclidean path integral on the other half $\CM_{\bar \Psi}$. It is obtained from $\CM_\Psi$ by reflecting about $\CR$. The squared norm of $\Psi$  is the partition function $\CZ_\CM$  on $\CM$. We normalize the state such that $\CZ_\CM=1$.
Let us decompose $\CR$ into  $\tq$ number of disjoint regions $\CR_\ta$,  such that $\cup_\ta \CR_\ta=\CR$. Let the Hilbert space on region $\CR_\ta$ be $\cH_\ta$.
We are interested in computing the special probe measures for the state $|\Psi\rangle$ under the decomposition  $\otimes_\ta \cH_\ta$. For theories that admit a weakly coupled gravity dual, this problem can be addressed holographically.

The probe family measure ${S}_n^{\{m_\ta\}}$ involves working with $|H_n|=\prod_{\ta =1}^{\tq } m_\ta^{n-1}$ number of replica copies of the theory.
The tensor product theory admits co-dimension two twist operators corresponding to the permutation group ${\mathbb S}_{|H_n|}$.
For every pair of regions $(\CR_\ta,\CR_\tb)$ that share a boundary, we insert the twist operator ${\CO}_{\sigma_\ta^{-1}\sigma_{\tb}}$ on the common boundary. The measure $S_n^{\{m_\ta\}}$ is then given by (appropriately normalized) logarithm of the correlation function of these twist operators.
Let us denote the resulting replicated manifold as ${\cal M}_n$. Following \cite{Lewkowycz:2013nqa}, we will proceed to compute this partition function holographically. Let $\CB_n$ be the dominant gravity solution such that $\partial \CB_n=\CM_n$. At leading order in $G_N$,
\begin{align}\label{holo-partition}
    {\log} \,\CZ_{\CM_n}=-\CS_{\rm grav}(\CB_n)+ |H_n| \CS_{\rm grav}(\CB_1).
\end{align}
Here $\CS_{\rm grav}({\cal X})$ is the gravitational action evaluated on the solution ${\cal X}$. The second term on the right-hand side of equation \eqref{holo-partition} serves to normalize the state.

\begin{figure}[h]
    \begin{center}
        \includegraphics[scale=0.25]{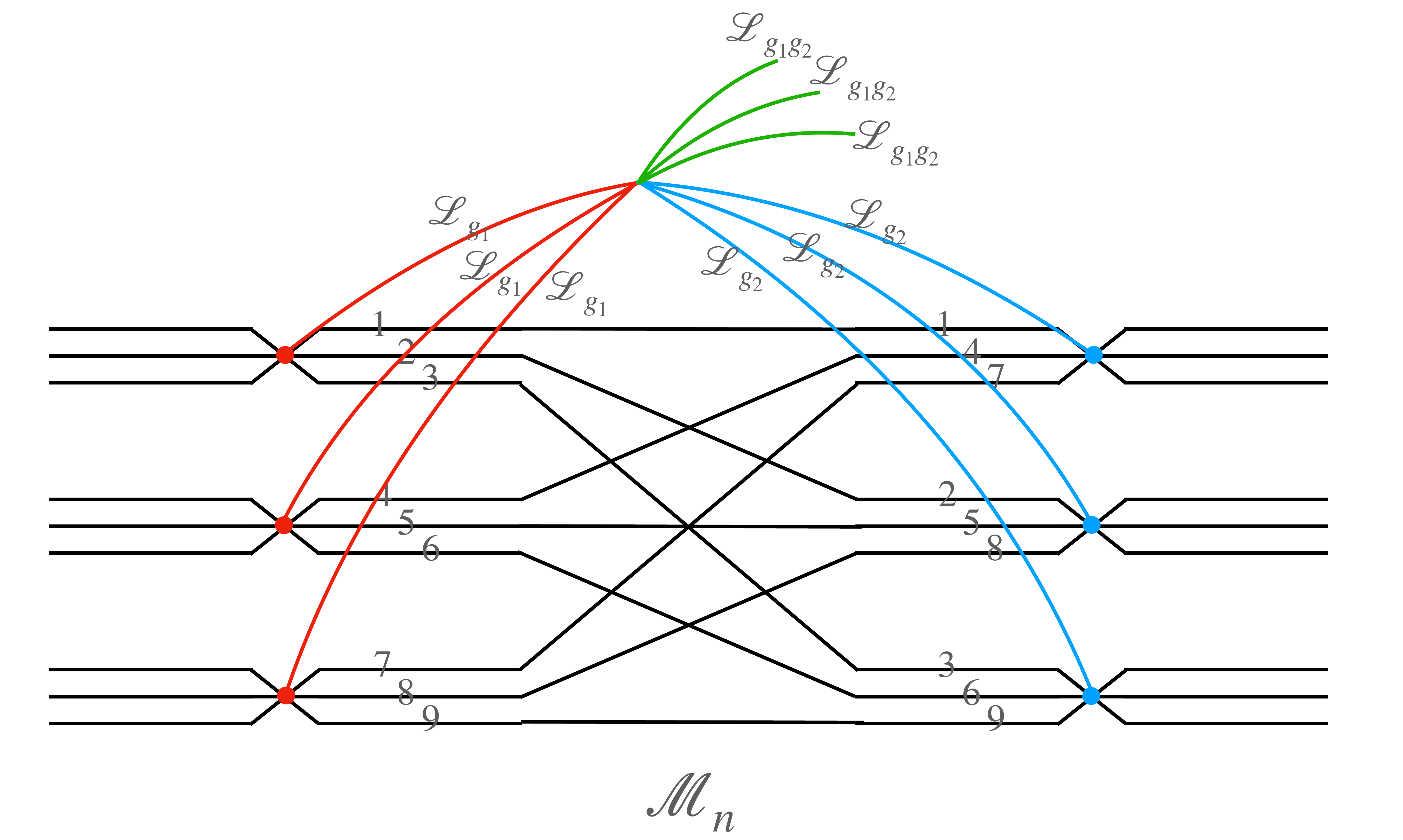}
    \end{center}
    \caption{Branching of the replicated manifold ${\cal M}_n$ and the fixed points. We also show the invariant loci that emanate from the boundary fixed points. They merge according to the rule $\CL_{g_1}\cdot \CL_{g_2}\to \CL_{g_1g_2}$.}\label{branching}
\end{figure}

\begin{figure}[h]
    \begin{center}
        \includegraphics[scale=0.25]{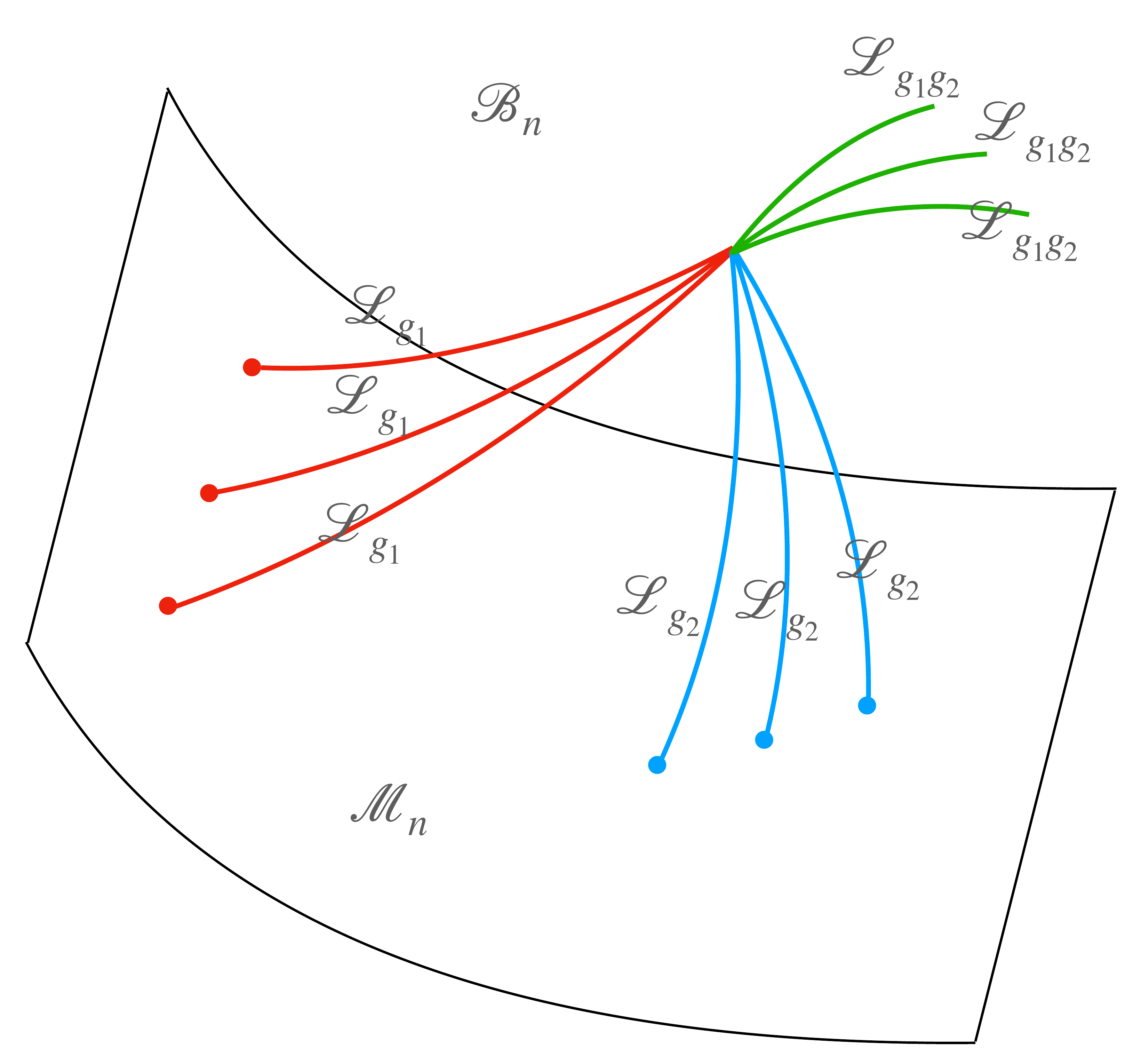}
    \end{center}
    \caption{Families of invariant loci anchored on the boundary. They merge according to the rule $\CL_{g_1}\cdot \CL_{g_2}\to \CL_{g_1g_2}$.}\label{multipleLs}
\end{figure}

The background fields on the manifold $\CM_n$ enjoy a replica symmetry $H_n$. This is the symmetry generated by the permutation elements $\sigma_\ta$ associated with all the twist operators. Following \cite{Lewkowycz:2013nqa}, we will make 
\begin{itemize}
    \item {\bf Replica symmetry assumption:} The dominant bulk solution ${\cal B}_n$ filling in the replicated boundary manifold ${\cal M}_n$ corresponding to the probe family measure preserves replica symmetry.
\end{itemize}
We will offer support for this assumption in section \ref{2dcft} where we compute $S_n^{\{m_\ta\}}$ in $2d$ CFTs with large central charge directly using CFT methods.
The solution $\CB_n$ consists of co-dimension $2$ loci that are invariant under certain subgroups of the replica symmetry group.  Some of these loci, called ``external'', are anchored at the fixed points on the boundary (these are locations of twist operator insertions on $\CM$) while the rest are ``internal''. Let us denote the loci that are anchored at the fixed points corresponding to the twist operator $\CO_g$ as $\CL_g$. They are invariant under the cyclic subgroup of $H_n$ generated by $g$. 
The quotient group $H_n/\langle g \rangle$ acts on one such locus to generate a family of them that is $|H_n/\langle g \rangle|$ in number.
Generically, two $\CL$s can merge to form a different $\CL$. Merging obeys the algebra $\CL_{g_1}\cdot \CL_{g_2}\to \CL_{g_1g_2}$. In fact, because the merging of ${\cal L}_{g_1}$ and ${\cal L}_{g_2}$ to form ${\cal L}_{g_1g_2}$ happens on the locus that is invariant under both $g_1$ and $g_2$, the members of ${\cal L}_{g_1}$ family that are related to each other by ${g_2}$ action merge with members of ${\cal L}_{g_2}$ family that are related to each other by $g_1$ action. The branched structure of the replicated manifold ${\cal M}_n$ along with the invariant loci and their merging is depicted in figure \ref{branching}. Figure \ref{multipleLs} presents a simplified view of the invariant loci and their merging when the boundary manifold ${\cal M}_n$ is unwrapped.
The internal loci come about because of such merging.
Below we will assume that
\begin{enumerate}
    \item Every fixed point locus is of the form $\CL_{ \sigma_\ta^{-1}\sigma_\tb}$.\label{assume-1}
\end{enumerate}

\begin{figure}[h]
    \begin{center}
        \includegraphics[scale=0.25]{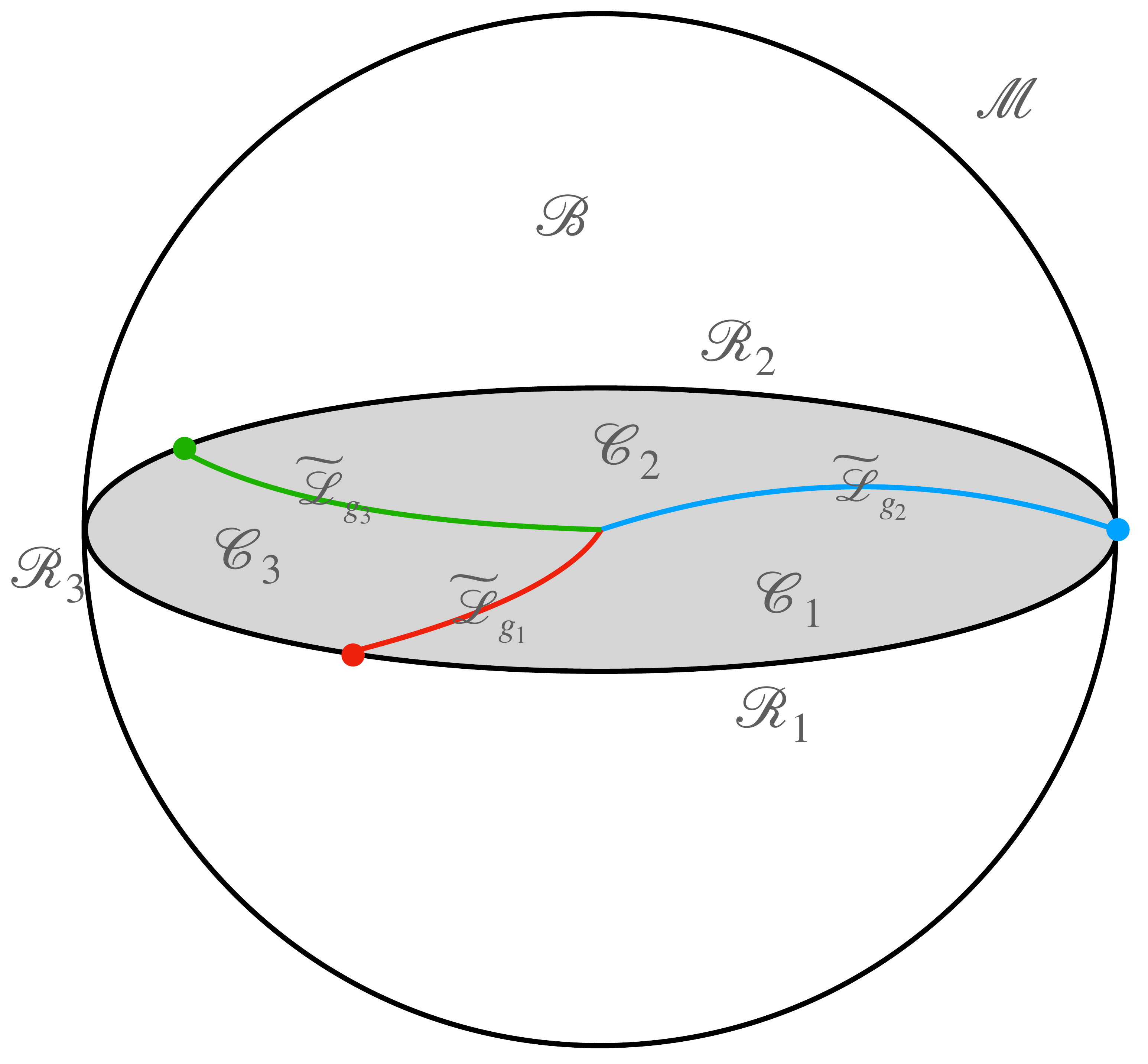}
    \end{center}
    \caption{Here we have denoted the boundary ${\cal M}$ and the regions of interest ${\cal R}_\ta$, the bulk ${\cal B}$ and the chambers ${\cal C}_\ta$. The chambers ${\cal C}_\ta$ and ${\cal C}_\tb$ are separated by walls of the type $\widetilde \CL_{\sigma_\ta^{-1} \sigma_\tb}$.}\label{setup}
\end{figure}

We now make use of the replica symmetry in the bulk to construct the orbifold $\widetilde \CB_n\equiv \CB_n/H_n$. Due to symmetry, the classical gravitational actions on the two spaces are related as
\begin{align}\label{orbifold-action}
    \CS_{\rm grav}(\CB_n)=|H_n|\,\CS_{\rm grav}(\widetilde \CB_n).
\end{align}
The orbifold $\widetilde \CB_n$ has a nice property that $\partial \widetilde \CB_n=\CM$. The family of $\CL_g$ loci discussed above become a \emph{single} conical singularity of opening angle $2\pi/|g|$ in the orbifold $\widetilde \CB_n$. This is because orbifolding by quotient group $H_n/\langle g \rangle$ identifies all the members of the family and orbifolding by the remaining $\langle g\rangle$, which fixes the locus $\CL_g$, gives rise to the conical singularity.
Let us denote this singularity as $\widetilde \CL_g$.
Let us denote the web created by these singularities as $\CW$.
Consider a co-dimension $1$ slice $\CC\in \widetilde \CB_n$ that contains $\CW$ and with the property $\partial \CC=\CR$. There are multiple such slices but the exact choice doesn't matter for the following discussion. Every singularity becomes a co-dimension $1$ wall in $\CC$ and the web $\CW$ yields its chamber decomposition. As we move from $\CR_\ta$ to $\CR_\tb$ with  $\tb \neq \ta$, through $\CC$, we must encounter at least one wall because the permutation elements $\sigma_\ta$ and $\sigma_\tb$ are different.
At this stage, we make an assumption about $\CW$ that
\begin{enumerate}
    \setcounter{enumi}{1}
    \item There is no chamber that lies completely in the interior of $\CC$.\label{assume-2}
\end{enumerate}
As a result, we get a one-to-one map between the chambers and boundary regions. Let us denote the chamber adjacent to $\CR_\ta$ as $\CC_\ta$. It has the property $\partial\CC_a\cap \CM=\CR_\ta$. The web $\CW$ consists of only those walls that separate $\CC_\ta$ and $\CC_\tb$ for some $(\ta,\tb)$. Such a wall must be of the type $\widetilde \CL_{\sigma_\ta^{-1} \sigma_\tb}$. The parent $\CL\in \CB_n$ must also be of the same type. This justifies our assumption \ref{assume-1}. The configuration is summarized in figure \ref{setup}.

To compute the probe family measure using equations \eqref{holo-partition} and \eqref{orbifold-action}, we need to evaluate the gravitational action on the orbifold solution $\widetilde{\CB}_n$. In what follows, we will specialize to the case of Einstein gravity. The orbifold solution is a smooth solution of Einstein's equations with constant negative curvature except at $\widetilde{\CL}_g$'s where it has a conical singularity of opening angle $2\pi/|g|$. The gravitational action for this solution is computed as follows \cite{Lewkowycz:2013nqa,Dong:2013qoa}. To find the solution it is convenient to engineer the conical singularities with the help of a cosmic brane. It is known that a co-dimension two cosmic brane with the action
\begin{align}\label{Sbr}
    \CS_{\rm br}^{(k)}=\frac{k-1}{4kG_{N}}\int dy^{D-1} \sqrt{h}=\frac{k-1}{4kG_{N}}A
\end{align}
gives rise to a conical singularity with the opening angle $2\pi/k$ around its support \cite{Unruh:1989hy, Boisseau:1996bp}. So we consider a brane with the action $S_{\rm br}^{(|g|)}$ supported on every locus ${\widetilde \CL}_g$. In this way, we get a cosmic brane-web $\CW$ in which branes of differing tensions are joined together.  The solution is then computed by solving the equations coming from the action ${\cal S}_{\rm grav}+{\cal S}_{\rm br}(n)$. We then need to evaluate only the action $S_{\rm grav}$ on the solution. This is because the cosmic brane is not actually present at the singularity but merely used as a trick to model the singularity.

\begin{figure}[h]
    \begin{center}
        \includegraphics[scale=.2]{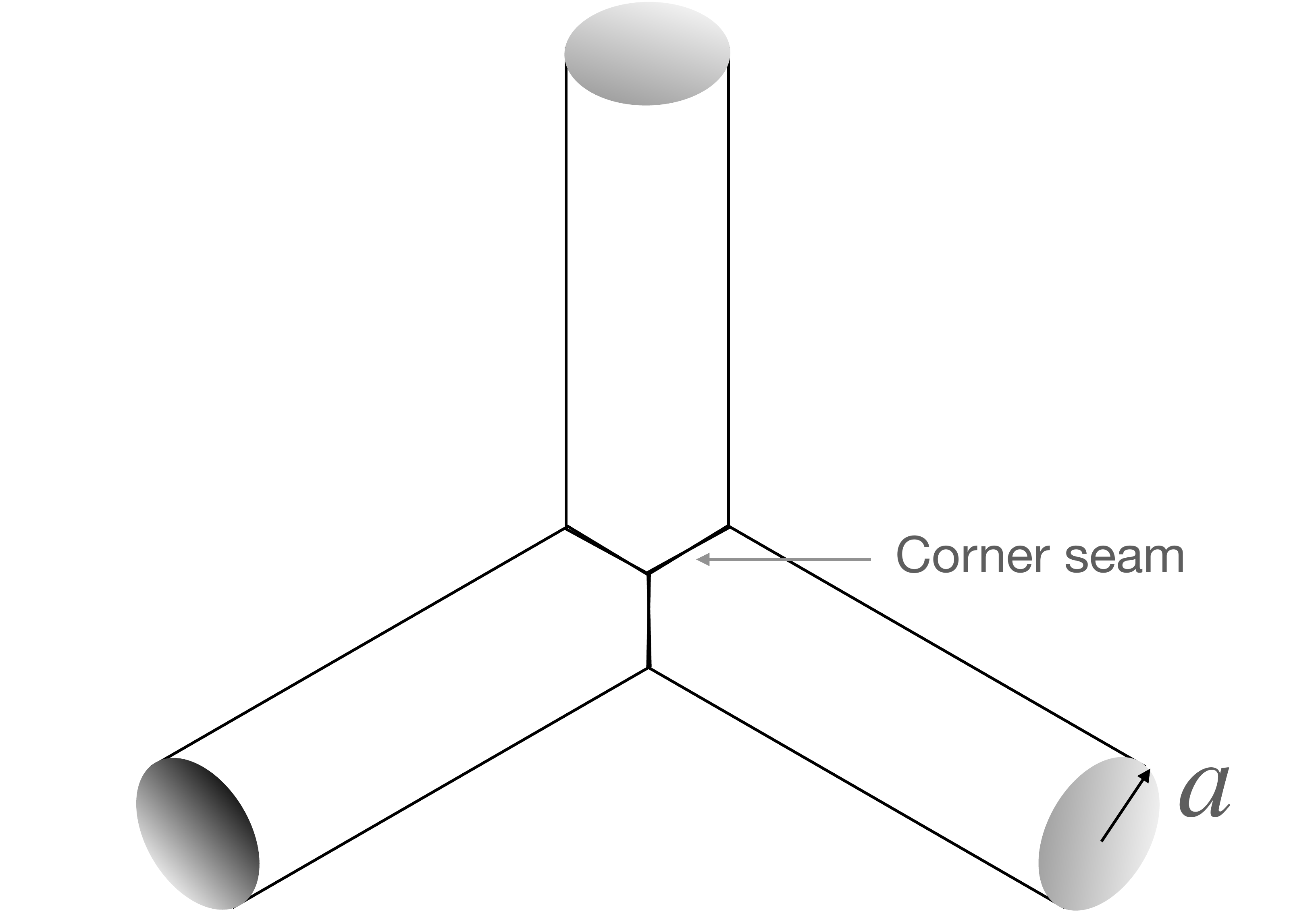}
    \end{center}
    \caption{Top view of the junction of  tubular neighborhoods of three ${\widetilde \CL}'s$.}\label{corner}
\end{figure}

A priori, there can be additional terms in the brane action supported only at the meeting locus. To compute such terms we excise the tubular neighborhood of ${\widetilde \CL}$'s of size $a$ and focus at their co-dimension three junction. As multiple ${\widetilde \CL}$'s meet, their corresponding tubular neighborhoods also meet forming corners as shown in figure \ref{corner}.
For the gravitational variational principle to be well-defined, we need to add the so-called Hayward term at the corner just as the way we add the Gibbons-Hawking-York (GHY) term on the boundary.
\begin{align}\label{hayward}
    {\cal S}_{\rm Hayward}=-\frac{1}{8\pi G_N} \int d\xi^{D-2}\, (\theta-\pi) \sqrt {\gamma}.
\end{align}
Here $\theta$ is the angle subtended by the two boundaries at the corner, $\xi$ is the coordinate along the corner and $\gamma$ is the induced metric on the corner. The contribution of the Hayward term is extensive along the co-dimension $3$ meeting locus and the proportionality constant is computed by integrating the Hayward term along the one-dimensional corner seam of the tubular neighborhood as shown in figure \ref{corner}. This integral is proportional to $a$ and vanishes as we take $a\to 0$. This had to be the case on dimensional grounds because the contribution at the meeting locus must take the form  $\sim \frac{1}{ G_N} \ell^{(3)}_{\rm meeting} \ell_{\rm scale}$.
Here $\ell^{(3)}_{\rm meeting}$ is the length (co-dimension $3$) of the meeting locus in $\CW$ and $\ell_{\rm scale}$ is some length scale needed to obtain a dimensionless answer. The only length scale that could serve this purpose is\footnote{Another length scale present in the problem is the AdS scale $\ell_{\rm AdS}$, however, it is irrelevant to this completely local computation.} $a$ which we take to $0$. This argument also shows that there is no extra contribution to the action even at higher co-dimension meeting loci.

Finding the solution to the theory $\CS_{\rm grav}+\CS_{\rm br}(n)$ for general $n$ is still a daunting task (see \cite{Headrick:2010zt, Hung:2011nu, Dong:2016fnf} for computation of bi-partite Renyi entropy). But to compute the special probe, only the limit $n\to 1$ is relevant. As the only $n$ dependence appears in the coefficients in $\CS_{\rm br}(n)$ analytically through $|\sigma_\ta^{-1}\sigma_\tb|={\rm lcm}(m_\ta,m_\tb)^{n-1}$, the solution can be analytically continued away from $n$ integer. In the limit $n\to 1$, the tensions of all the branes go to zero and the solution can be found in the probe limit as the brane web configuration that extremizes $\CS_{\rm br}$ in the fixed background $\CB_1$. This brane web consists of branes with differing tensions. Every brane wall separating the chambers $\CC_\ta$ and $\CC_\tb$ has a vanishingly small tension proportional to $\log({\rm lcm}(m_\ta,m_\tb))$. 
We are interested in evaluating $\partial_n {\cal S}_{\rm grav}|_{n=1}$. This can be done using the same techniques as \cite{Lewkowycz:2013nqa, Dong:2013qoa, Dong:2016fnf}. We use the geometry that was just discussed near equation \eqref{hayward} where the tubular neighborhood of the singular locus has been excised. The variation of gravitational solution vanishes in the bulk for any $\delta_{g_{\mu\nu}}$. Taking the variation with respect to $n$, we see that the contribution only comes from the GHY term on the boundary. The Hayward term at the corner does not contribute as discussed earlier. The contribution of the GHY term has been evaluated in \cite{Lewkowycz:2013nqa, Dong:2013qoa, Dong:2016fnf}. It produces the area of the singular locus weighted by the tension. Explicitly,
\begin{align}\label{light-branes}
    \partial_n\CS_{\rm grav}|_{n=1} =\frac{1}{4G_N}{\rm min}_{\cal W}\left\{\sum_{\widetilde L_{\sigma_\ta^{-1}\sigma_\tb}\in \CW}\log({\rm lcm}(m_\ta,m_\tb))A_{\ta,\tb}\right\}\equiv \frac{1}{4G_N} A_{\rm min}^{\cal W}.
\end{align}
Here $A_{\ta,\tb}$ is the area of the brane wall that separates the chamber $\CC_\ta$ and $\CC_\tb$ (if there exists such a wall). 
The sum is over all the brane walls and minimization is over all brane webs ${\cal W}$. The weighted sum of areas, after minimization, is denoted as $A^{\cal W}_{\rm min}$. From our discussion it follows that the brane web $\CW$ obeys the topological conditions,
\begin{enumerate}
    \item $\CW$ is anchored at the boundaries of all the regions $\CR_\ta$'s.
    \item $\CW$ contains sub-webs that are homologous to all the regions $\CR_\ta$'s.
\end{enumerate}
The second condition is the reformulation of the statement that between any two chambers $\CC_\ta$ and $\CC_\tb$ there must be at least one wall. As the solution minimizes the area subject to these conditions, it doesn't allow any chamber that lies completely in the interior of $\CC$. This justifies our assumption \ref{assume-2}.

This, along with equations \eqref{special-probe-def} and \eqref{holo-partition}, shows that the special probe measure is given a simple formula
\begin{align}
    S^{\{m_\ta\}}=\frac{1}{4G_N} A_{\rm min}^{\cal W}.
\end{align}
In other words, it is equal to the $1/4G_N$ times the minimal weighted area of the brane web $\CW$.

In the case of $2d$ CFTs, all the loci ${\widetilde {\cal L}_g}$ are one dimensional. Solving the equations of motion for the cosmic brane web means that the individual segments of the brane web are geodesics and that they balance tension at the tri-valent junction. So, for the solution of the tension balance to exist, it is needed that the magnitude of the three tension vectors involved in the junction satisfy the triangle inequality. Happily, for any three positive numbers $(m_\1,m_\2,m_\3)$, the three quantities $\log({\rm lcm}(m_1,m_2)), \log({\rm lcm}(m_2,m_3))$ and $\log({\rm lcm}(m_1,m_3))$ always satisfy the triangle inequality! Hence the solution to the tension balance always exists for the special probes.

If we pick all the integers $m_\ta=m$ then the special probe measure is simply $\log(m)$ times the multi-entropy computed in \cite{Gadde:2022cqi}. In this case, the tensions of all the brane walls are equal and the formula produces the soap film prescription of \cite{Gadde:2022cqi}. In the general case, the soap film prescription is extended to allow the differing tensions to individual soap-films as discussed above.

Interestingly, the brane web corresponding to multi-entropy has appeared already in the discussion of holographic multi-partite entanglement \cite{Harper:2021uuq, Harper:2022sky}. In this paper, the author considers configurations of the so-called hyperthreads that maximize a certain type of flow. The solution to this linear program is provided precisely by the brane webs of multi-entropy. It would be interesting to explore the connection between multi-entropy and hyperthreads further. 

\subsection{Quantum correction}
The discussion in \cite{Gadde:2022cqi} of quantum correction to multi-entropy generalizes straightforwardly to the more general case of special probe measures. The quantum corrections to the special probe measure are obtained using the replica trick in the bulk. The web ${\cal W}$ gives a decomposition of the bulk Cauchy slice into regions, called chambers above, ${\cal C}_\ta$'s. In the un-orbifolded geometry ${\cal B}_n$, the bulk regions ${\cal C}_\ta$'s get glued to each other precisely in the same way boundary regions ${\cal R}_\ta$'s were glued. This evaluates nothing but the same special probe measure for the bulk quantum fields. If this replica trick was performed in the bulk geometry $\CB$ along ${\cal W}$, it would have given, in the $n\to 1$ limit, the bulk probe measure $S^{\{m_\ta\}}_{\rm bulk}(\CW)$ corresponding to the chamber decomposition $\CC_a$ directly, however, because the replica trick was performed on $\widetilde \CB_n$, it is not immediately clear that what we get is $S^{\{m_\ta\}}_{\rm bulk}(\CW)$. 
This situation is similar to the bi-partite case \cite{Faulkner:2013ana} (see also \cite{Barrella:2013wja}). There, the difference between the two quantities is captured by changing the classical solution by $\CO(G_N)$ to account for the one-loop expectation value of the stress tensor. This changes the area and hence the entanglement entropy by $\CO(1)$. We expect a similar formula to give the sub-leading correction to the special probe measure.
\begin{align}\label{quantum-web}
    S^{\{m_\ta\}}=\sum_{\widetilde L_{\sigma_\ta^{-1}\sigma_\tb}\in \CW}\log({\rm lcm}(m_\ta,m_\tb))\frac{\langle \hat A_{\ta,\tb}\rangle }{4G_N}+S^{\{m_\ta\}}_{\rm bulk}(\CW)+{\rm c.t.}.
\end{align}
Here $\hat A_{\ta,\tb}$ is the area operator of the $(\ta,\tb)$ wall of the minimal web ${\cal W}$ and ${\rm c.t.}$ are the counter-terms that render $S_{\rm bulk}^{(\tq)}(\CW)$ finite. 

Following \cite{Engelhardt:2014gca}, \cite{Dong:2017xht} we conjecture a formula that is valid to all orders in $1/G_N$ perturbation theory: special probe measure $S^{\{m_\ta\}}$ is given by the above formula but $\CW$ is not the ordinary area minimizing web but rather the ``quantum extremal web'' i.e. the web that minimizes the combination 
\begin{align}
    \sum_{\widetilde L_{\sigma_\ta^{-1}\sigma_\tb}\in \CW}\log({\rm lcm}(m_\ta,m_\tb))\frac{ A_{\ta,\tb} }{4G_N}+S^{\{m_\ta\}}_{\rm bulk}(\CW).
\end{align}

For general gravitational theories, we expect that the special probe measure is given by Wald-type corrections as computed in  \cite{Dong:2013qoa, Hung:2011xb} for the bi-partite case.

\section{Marriage with holographic purification}\label{marriage}
Perhaps the most well-known multi-partite measure that has been studied in the holographic context is the so-called reflected entropy. 
It is defined for a tri-partite pure state $|\psi_{ABC}\rangle$ or equivalently a bi-partite mixed state $\rho_{AB}$.

Before we review the definition of reflected entropy, let us review the notion of canonical purification. Consider a density matrix $\rho_A\in {\rm End}\,H_{A}$ on party $A$. It admits the Schmidt decomposition
\begin{align}
    \rho_A= \sum_i \lambda_i |i\rangle \langle i|
\end{align}
with positive $\lambda_i$'s and mutually orthonormal basis vectors $|i\rangle$. The canonical purification of $\rho_A$ is the following pure state  ``$|\sqrt {\rho_A}\rangle$'' in $H_A\otimes H_{A^\star}$ where $H_{A^\star}$ is the isomorphic ``reflection'' of $H_A$.
\begin{align}
    |\sqrt{\rho_A}\rangle\equiv \sum_i \sqrt{\lambda_i} |i\rangle |i^\star\rangle.
\end{align}
Here $|i^\star\rangle$ is the basis of $H_{A^\star}$.
Now the reflected entropy between $A$ and $B$ in the mixed state $\rho_{AB}$ can be defined straightforwardly. It is simply the von Neumann entropy of $AA^{\star}$ (or equivalently of $BB^{\star}$) in the canonical purification $|\sqrt{\rho_{AB}}\rangle\in H_A\otimes H_B\otimes H_{A^\star}\otimes H_{B^\star}$.

\begin{figure}[h]
    \begin{center}
        \includegraphics[scale=0.2]{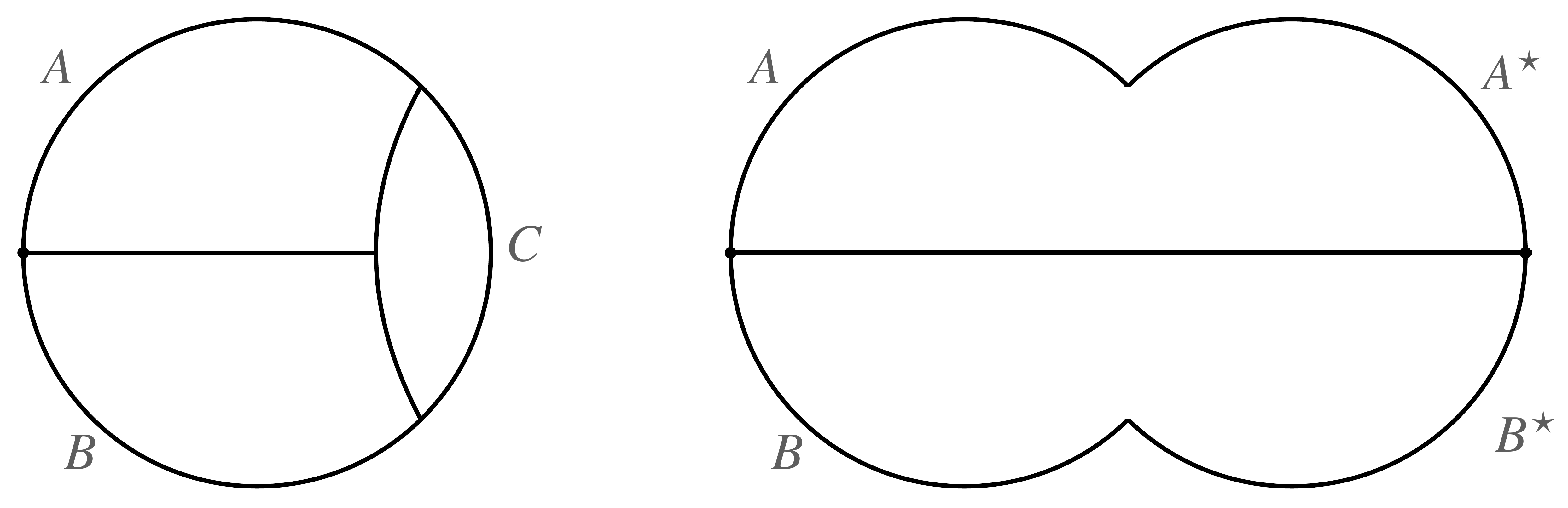}
    \end{center}
    \caption{The reflected geometry dual to the canonically purified state $|\sqrt{\rho_{AB}}\rangle$ is denoted in the right figure. The entanglement entropy of $AA^\star$ is twice the entanglement wedge cut as shown in the left figure.}\label{reflected}
\end{figure}

Following \cite{Engelhardt:2017aux, Engelhardt:2018kcs, Faulkner:2018faa}, Dutta and Faulkner \cite{Dutta:2019gen} gave a prescription for constructing the holographic dual of the state $|\sqrt{\rho_{AB}}\rangle$ starting from the holographic dual of the original pure state $|\psi_{ABC}\rangle$. For simplicity, let us assume that all three parties have been defined on the special section at the moment of time symmetry. Then the holographic dual of $|\sqrt{\rho_{AB}}\rangle$ is obtained by cutting the dual slice of $|\psi_{ABC}\rangle$ at the Ryu-Takayanagi surface of $C$ and gluing it to its reflected copy along the cut. As the Ryu-Takayanagi surface is of vanishing extrinsic curvature, this procedure does yield a solution to Einstein's equations. Once the dual geometry to $|\sqrt{\rho_{AB}}\rangle$ is obtained, entanglement entropy of $AA^{\star}$ is computed using the Ryu-Takayanagi prescription on the new geometry. This gives the answer that is two times the minimal cut of the $AB$ entanglement wedge in the original geometry. This is graphically denoted in figure \ref{reflected}.

Even though the reflected entropy is defined for a tri-partite pure state, the idea can be extended to define multi-partite entanglement measures. We think of canonical purification ${\cal CP}$ as a map from a $\tq$-partite pure state with a distinguished party to a $2(\tq-1)$-partite pure state. 
\begin{align}
    {\cal CP} [|\psi_{A_\1,\ldots,A_\tq}, A_i\rangle ] = |\sqrt{\rho_{A_1,\ldots, A_\tq}}\rangle 
\end{align}
Here $A_i$ is absent from the subscript set of $\rho$. The pure state on the right-hand side is on $\otimes_{j\neq i}H_{A_j}\otimes H_{A_j^\star}$. The map can be applied once again on $|\sqrt{\rho_{A_1,\ldots, A_\tq}}\rangle$ with the choice of a distinguished party and so on. Note that, while doing so, $|\sqrt{\rho_{A_1,\ldots, A_\tq}}\rangle$ can be thought of as either $2(\tq-1)$-partite state or we can re-partition the parties to obtain a $k$-partite state with $k<2(\tq-1)$.  One can obtain a  rich class of multi-partite measures of the original state $|\psi_{A_\1,\ldots,A_\tq}\rangle$ by evaluating a general multi-partite measure ${\cal E}$ on the pure state obtained after successive application of ${\cal CP}$ on $|\psi_{A_\1,\ldots,A_\tq}\rangle$. Even the entanglement entropy corresponding to any bipartition would be a multi-partite entanglement measure of the original state. The reflected entropy is just one such measure.

\begin{figure}[h]
    \begin{center}
        \includegraphics[scale=0.2]{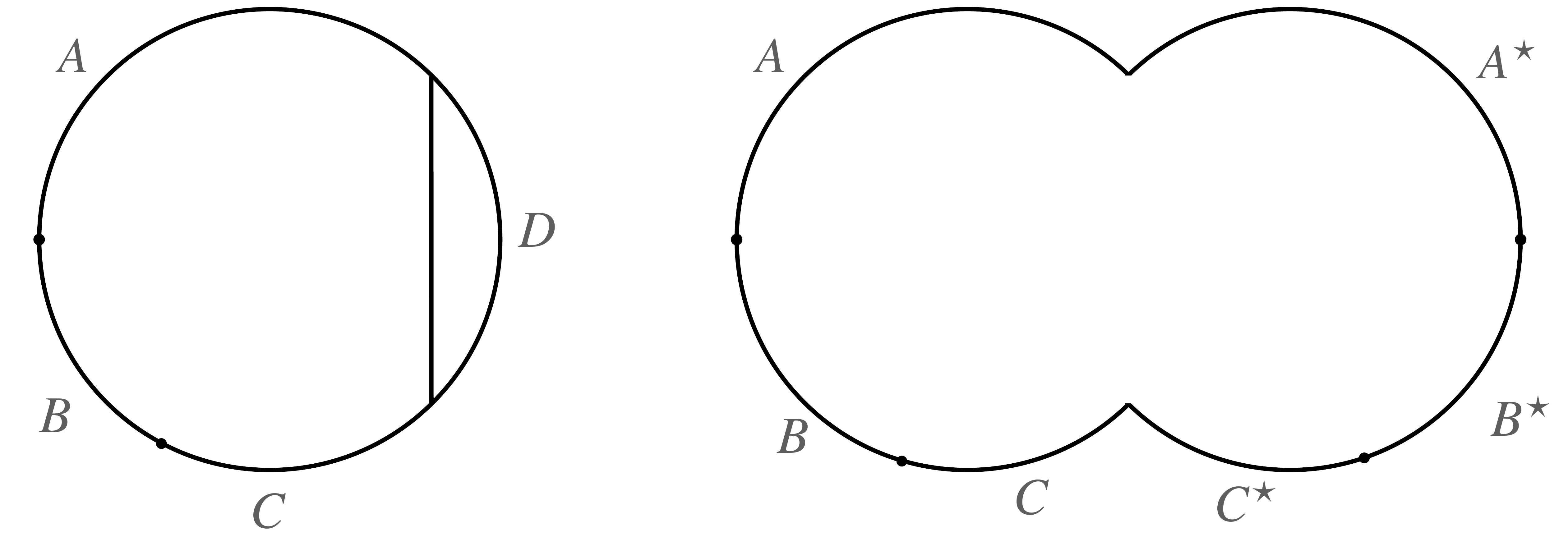}\\
        \qquad 
        \includegraphics[scale=0.2]{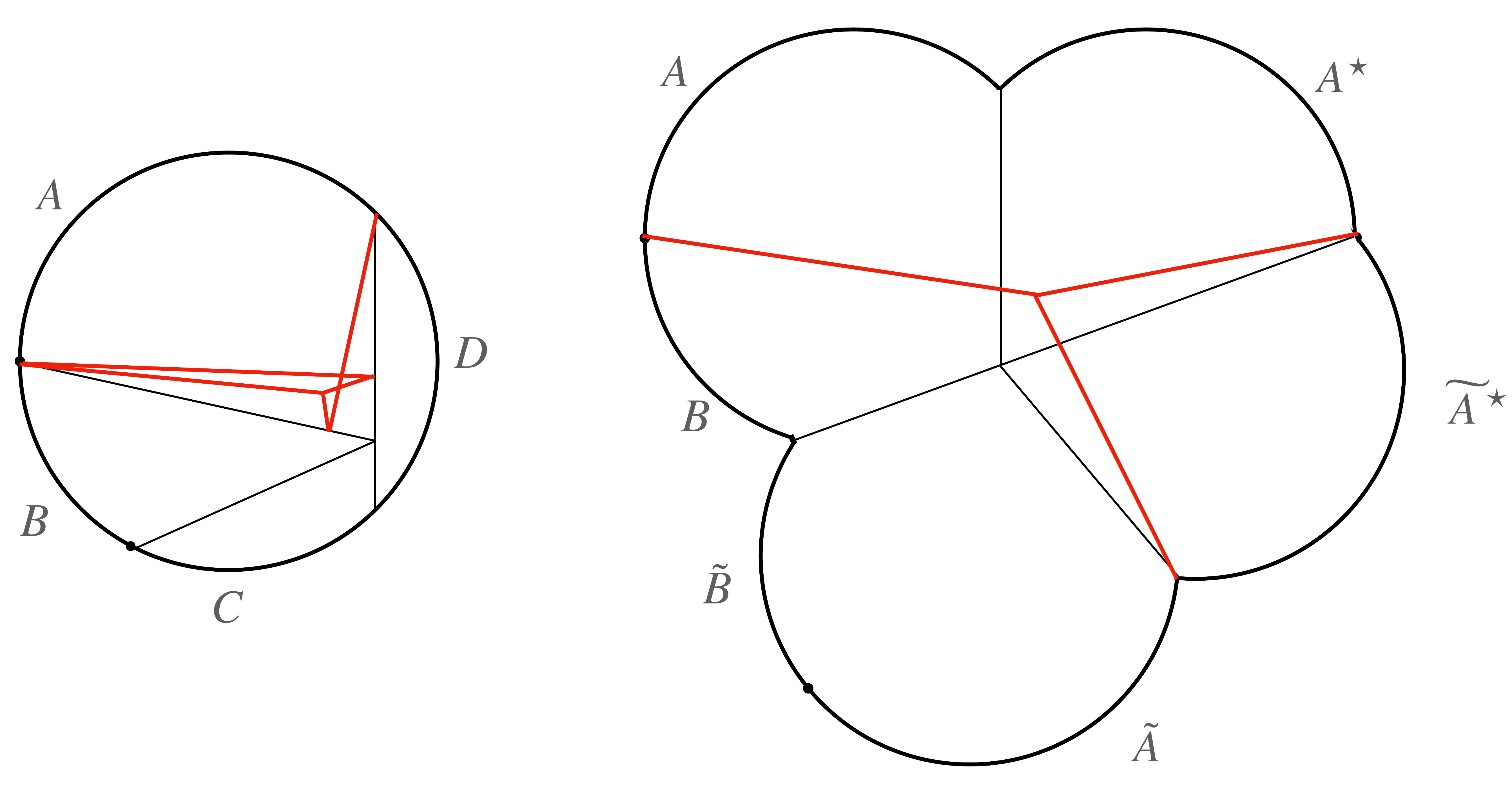}
    \end{center}
    \caption{The first figure shows the geometry dual to the pure state on $ABCD$. The second figure shows the reflected geometry dual to the canonical purification of the density matrix on $ABC$. The bottom right shows the geometry dual to another purification, this time of the density matrix on $AA^\star B$. We have also denoted the tri-partite probe measure $S^{\{m_\ta\}}(AA^\star, {\widetilde {A^\star}}, B\tilde B \tilde A)$ in red. The resulting brane web is described in the original geometry through reflections across certain minimal surfaces. For simplicity, all the minimal length loci have been drawn as straight lines.}\label{multiple-refl}
\end{figure}

The canonical purification procedure of \cite{Dutta:2019gen} is precisely the map ${\cal CP}$. Remarkably it is closed on holographic states. Applying this multiple times constructs a bulk solution with multiple boundary regions. Any probe measure can then be evaluated in the new geometry. As argued above, this measure is then a multi-partite measure also of the original holographic state. In figure \ref{multiple-refl}, we show two applications of ${\cal CP}$ on a holographic state and tri-partite probe measures evaluated on the resulting bulk geometry. 
All in all, we have shown how the class of computable multi-partite measures of a holographic state can be enlarged by marrying the holographic canonical purification of \cite{Dutta:2019gen} with our probe measures defined in section \ref{lm}.

The idea of canonically purifying a given holographic state multiple times has previously appeared in \cite{Umemoto:2018jpc, Bao:2018gck, Bao:2019zqc, Chu:2019etd, Bhattacharya:2020ymw, Harper:2020wad}.

\section{Computations in $AdS_3/CFT_2$}\label{2dcft}

In this section, we will check our holographic prescription in $2d$ conformal field theories. To avoid clutter, we will focus on the case of
special probe measures with all the integers $m_\ta=m$  taken equal\footnote{The discussion applies directly to the case of multi-entropy after removing all factors of $\log m$.}. For the case of three parties and three intervals, when the results are relatively less complicated, we will also give results for the case of unequal $m_\ta$'s. 

\subsection{Using holographic prescription}
In this subsection, we will compute the special probe measure between regions of a particular Cauchy slice $\CR$ of a $2d$ holographic CFT. We will take $\CR$ to be circular and at the moment of time symmetry. We extend it into the bulk so that the bulk slice $\CC$ also has the moment of time symmetry. $\CC$ is a two-dimensional hyperbolic disk bounded by the circular conformal boundary $\CR$. For our computation, we only need to focus on this $2d$ bulk-$1d$ boundary system in this time slice. The special probe measure is expected to be covariant under the conformal symmetry of the boundary. For the calculation at hand, it is sufficient to impose the subgroup of conformal symmetry that preserves $\CR$. It is $SO(2,1)$, also the isometry of $\CC$.
To impose conformal constraints, it is convenient to carry out the calculations in the so-called embedding space. It is ${\mathbb R}^{2,1}$ on which $SO(2,1)$ acts linearly.  The boundary coordinates $P$ obey $P^2=0$ with the projective identification $P\sim \lambda P$ and the bulk coordinates $X$ obey $X^2=-1$ (we have set the AdS radius to $1$). In what follows, we will fix the projective identification $P\sim \lambda P$ by picking the slice $P_3=1$. In this gauge, the boundary is a circle of radius $1$.

\subsection*{$\tq=3$}
\begin{figure}[h]
    \begin{center}
        \includegraphics[scale=0.5]{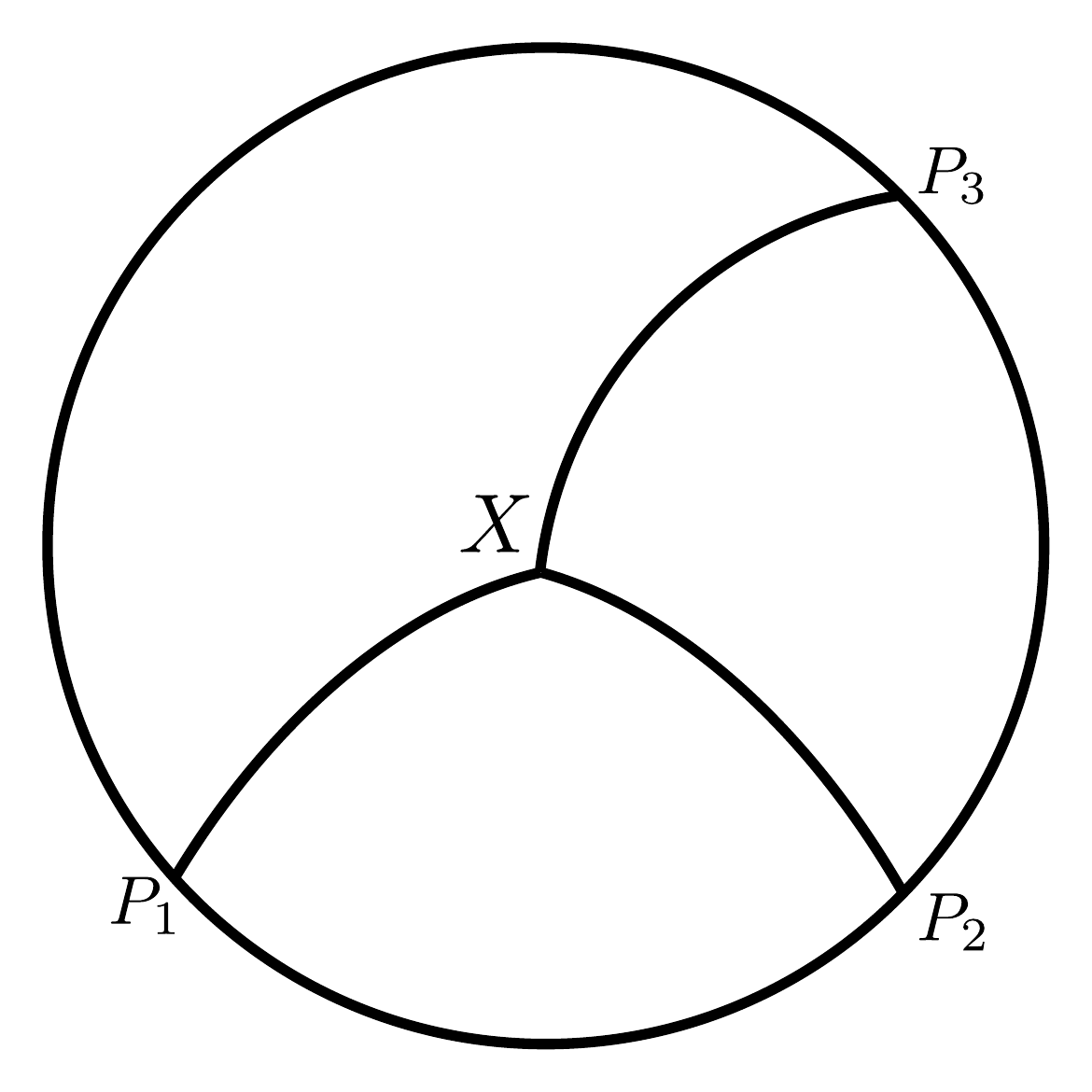}
    \end{center}
    \caption{Geodesic network to compute the tri-partite special probe measure.}\label{3-pt-WL}
\end{figure}

Let us first compute the probe measure $S^{\{m,m,m\}}$  between the three connected regions defined by points $P_1,P_2,P_3$ on $\CR$.
According to our holographic prescription for the special probe measure,
\begin{align}
    S^{\{m,m,m\}}=\frac{\log  m}{4G_N}{\rm min}_X\Big(\sum_i \ell (P_i,X))\Big).
\end{align}
Here $X$ is some bulk point and $\ell(P,X)$ denotes the geodesic distance between the points $P$ and $X$, ${\rm min}_X$ means that the expression is to be minimized over the bulk point $X$. See figure \ref{3-pt-WL}.
In this dimension, the soap-film prescription is more usefully thought of as a minimal geodesic network prescription. The geodesic distance between a bulk point from the boundary is infinite. It is regularized as follows. We think of the boundary point $P$ as a limit $\alpha\to \infty$ of the bulk point $Z=\alpha P+Y$ with $P\cdot Y=0, Y^2=-1$.
\begin{align}
    \ell(Z,X) & =\cosh^{-1}(|Z\cdot X|)=\cosh^{-1}(|(\alpha P+Y)\cdot X|) \notag \\
              & =\log (2|P\cdot X|)+\log (\alpha).
\end{align}
We will regularize the length $\ell(P,X)$ by subtracting the fixed infinity, $\log(\alpha)$, for all boundary points $P$. Then
\begin{align}\label{tri-2d}
    S^{\{m,m,m\}}=\frac{\log  m}{4G_N}{\rm min}\Big(\sum_i \log (2|P_i\cdot X|))\Big)
\end{align}
subject to the constraint $X^2= -1$. This is done by introducing a Lagrange multiplier $\lambda$ that multiplies $(X^2+1)$ and solving for $P_i\cdot X$. We give the solution for $X$ below.
\begin{align}
    X=\frac{1}{\sqrt 3}\Big(\sqrt{\frac{P_{23}}{P_{12}P_{13}}}P_1+\sqrt{\frac{P_{13}}{P_{12}P_{23}}}P_2+\sqrt{\frac{P_{12}}{P_{13}P_{23}}}P_3\Big).
\end{align}
Here $P_{ij}\equiv -2P_i\cdot P_j$. This can be deduced purely from conformal symmetry as $X$ is the only $SO(2,1)$ vector invariant under $P_i\sim \lambda_i P_i$ with $X^2=-1$  that is symmetric in all $P_i$'s.
Substituting this in equation \eqref{tri-2d}, we get
\begin{align}\label{q3holography}
    S^{\{m,m,m\}}=\frac{\log  m}{4G_N}\Big[3\log\Big(\frac{2}{\sqrt 3}\Big) +\frac12 \log\Big(P_{12}P_{23}P_{31}\Big)\Big].
\end{align}
Note that
\begin{align}
    e^{-\frac{4G_N}{\log  m} S^{\{m,m,m\}}}=\Big(\frac{\sqrt 3}{2}\Big)^{3} (P_{12}P_{23}P_{31})^{-1/2}.
\end{align}
has the same conformal transformation properties as the three-point function of local operators of dimension $1$ inserted at $P_i$.

Let us now consider the case where $m_\ta$ are unequal. In this case, the tension of the three segments $P_1\cdot X, P_2 \cdot X$ and $P_3\cdot X$ is unequal and is given by $\log ({\rm lcm}(m_1,m_2)), \log ({\rm lcm}(m_2,m_3))$ and $\log ({\rm lcm}(m_3,m_1))$. Let us label these tensions $t_3, t_1$ and $t_2$ respectively. The special probe measure
\begin{align}\label{unequal}
    S^{\{m_1,m_2,m_3\}}= \frac{1}{4G_n}{\rm min}\Big(\sum_i t_i \log(2|P_i\cdot X|)\Big)
\end{align}
The minimization over length is performed with respect $X$ satisfying $X^2=-1$. This is achieved with the help of Lagrange multiplier $\lambda$. We define 
\begin{align}
    g(\lambda, X)\equiv \sum_i t_i \log(2|P_i\cdot X|) + \lambda(X^2+1)
\end{align}
and vary with respect to both $\lambda$ and $X$. 
The solutions are,
\begin{align}\label{saddle_sol_3pt}
    \lambda &= \frac{t_{1}+t_{2}+t_{3}}{2} \notag \\
    P_{1} \cdot X &= t_{1} \sqrt{\frac{(-t_{1}+t_{2}+t_{3})}{(t_{1}-t_{2}+t_{3})(t_{1}+t_{2}-t_{3})(t_{1}+t_{2}+t_{3})}} \sqrt{\frac{P_{12} P_{13}}{P_{23}}} \notag \\
    P_{2} \cdot X &= t_{2} \sqrt{\frac{(t_{1}-t_{2}+t_{3})}{(-t_{1}+t_{2}+t_{3})(t_{1}+t_{2}-t_{3})(t_{1}+t_{2}+t_{3})}} \sqrt{\frac{P_{12} P_{23}}{P_{13}}} \notag \\
    P_{3} \cdot X &= t_{3} \sqrt{\frac{(t_{1}+t_{2}-t_{3})}{(-t_{1}+t_{2}+t_{3})(t_{1}-t_{2}+t_{3})(t_{1}+t_{2}+t_{3})}} \sqrt{\frac{P_{13} P_{23}}{P_{12}}}
    \end{align} 
Substituting in equation \eqref{unequal}, we get
\begin{align}
    4G_N \,S^{\{m_1,m_2,m_3\}}&= \log \Big( \frac{\chi(2t_{1}) \chi(2t_{2}) \chi(2t_{3})}{ \chi(-t_{1}+t_{2}+t_{3}) \chi(t_{1}-t_{2}+t_{3}) \chi(t_{1}+t_{2}-t_{3}) \chi(t_{1}+t_{2}+t_{3}) } \Big) \notag \\
    & + \log \Big( P_{12}^{\frac{(t_{1}+t_{2}-t_{3})}{2}} P_{23}^{\frac{(-t_{1}+t_{2}+t_{3})}{2}} P_{13}^{\frac{(t_{1}-t_{2}+t_{3})}{2}} \Big), \qquad \qquad \chi(x) = x^{\frac{x}{2}}.
\end{align}

\subsection*{$\tq=4$}
\begin{figure}[h]
    \begin{center}
        \includegraphics[scale=0.4]{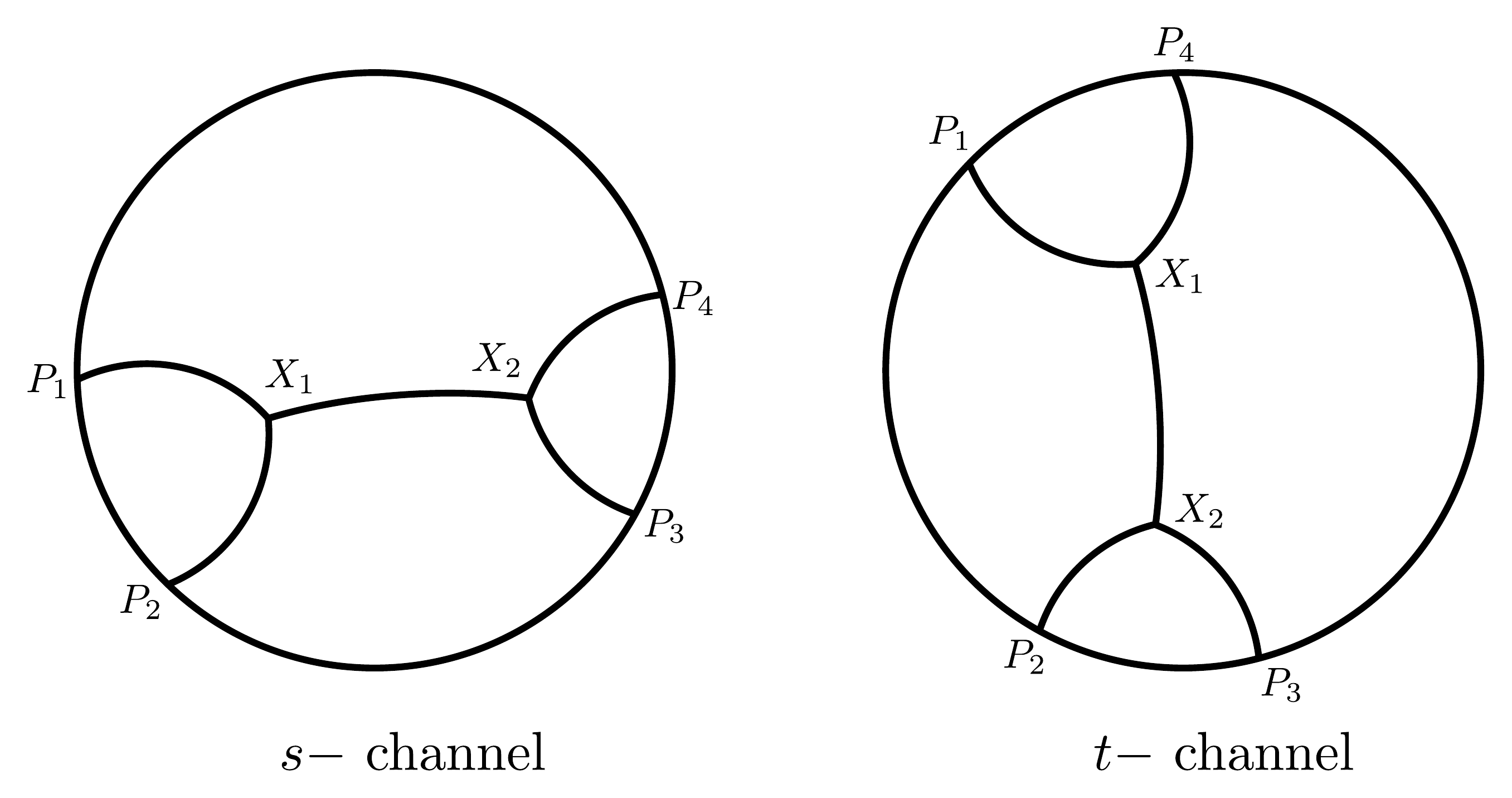}
    \end{center}
    \caption{Geodesic network to compute the 4-partite special probe measure.}\label{4-pt-WL}
\end{figure}

\begin{figure}[h]
    \begin{center}
        \includegraphics[scale=0.5]{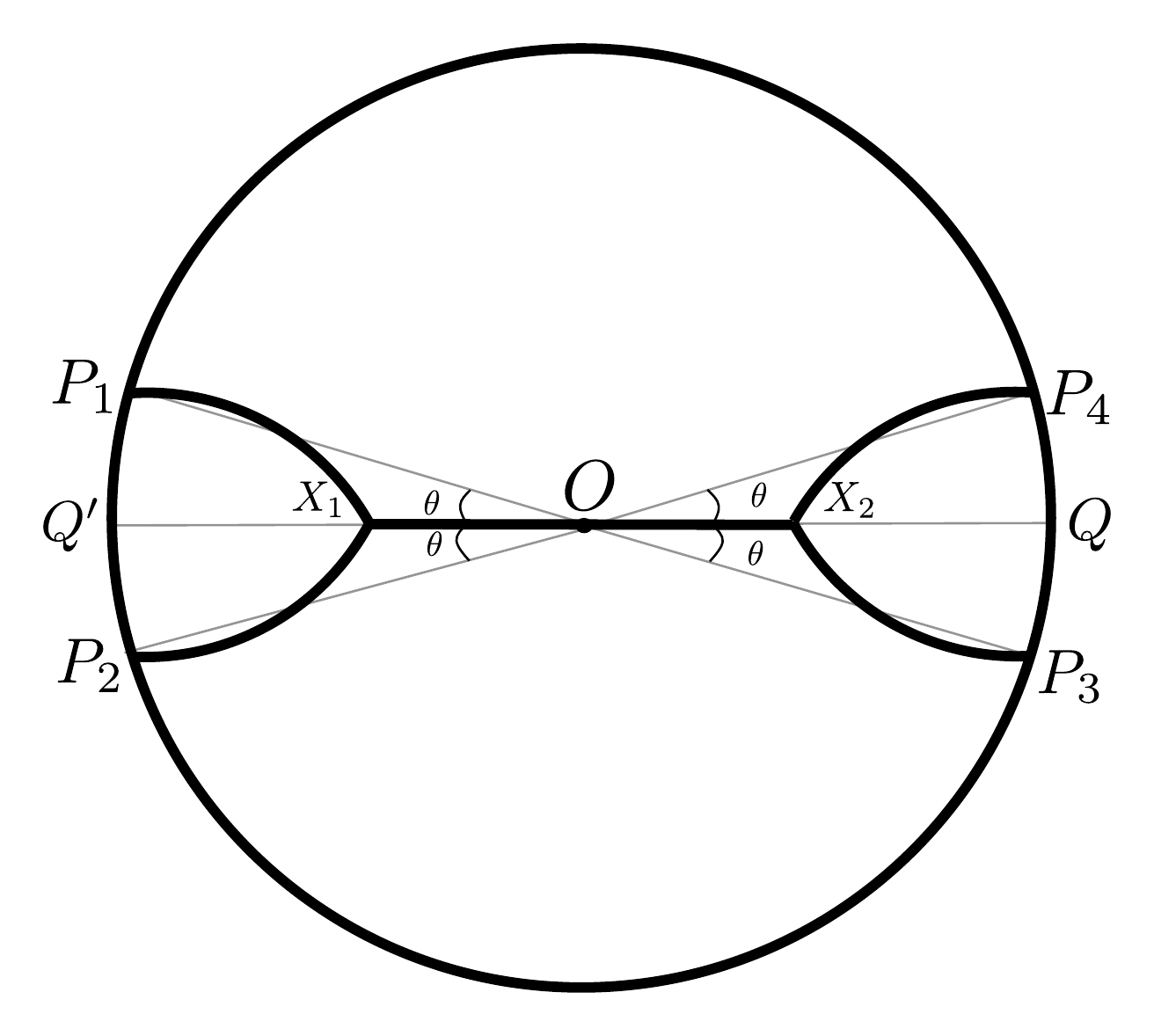}
    \end{center}
    \caption{Geodesic network to compute the 4-partite special probe measure.}\label{symm-4}
\end{figure}

Let us look at the special probe measure of four connected regions defined by the points $P_1,\ldots,P_4$. It is given by the minimum of the two extremal geodesic networks given in figure \ref{4-pt-WL}.
We will refer to them suggestively as s-channel and t-channel geodesic networks.
From conformal symmetry, we expect it to take the form
\begin{align}
    \frac{4G_N}{\log m} S^{\{m,m,m,m\}}=\log(P_{12}P_{34})+ f(z),\quad z\equiv\Big(\frac{P_{12}P_{34}}{P_{13}P_{24}}\Big)^{\frac12}.
\end{align}
The coefficient of the first term $\log(P_{12}P_{23})$ is determined by requiring $e^{-\frac{4G_N}{\log m} S{\{m,m,m,m\}}}$ transform as a conformal correlation function of operators with dimension $1$.
Here $z$ is the usual cross-ratio of four points. It is clear that near $z=0$, $S^{\{m,m,m,m\}}$ is given by the s-channel network and near $z=1$, it is given by t-channel network, with phase transition occurring at an intermediate value of $z$. For concreteness let us take $z$ to be finite but close to $0$.
We can compute $S^{\{m,m,m,m\}}$ without actually solving the extremization problem but rather by using symmetry and $\tq=3$ computation.
To compute $f(z)$, we use conformal transformation to arrange the point symmetrically as shown in figure \ref{symm-4}.
The parametrization of $P_i$ in terms of $\theta$ is
\begin{align}
     & P_1=(-\cos\theta,\sin\theta,1), \quad P_2=(-\cos\theta,-\sin\theta,1)\notag \\
     & P_3=(\cos\theta,-\sin\theta,1),\quad P_4=(\cos\theta,\sin\theta,1)
\end{align}
This gives $z=\sin^2\theta$.  Now realize that the length of the minimal four-point geodesic network is the sum of lengths of minimal three-point networks $(P_1,P_2,Q)$ and $(P_3,P_4,Q')$ minus the ``diameter'' $(Q,Q')$. We get,
\begin{align}\label{4partyhol}
    S^{\{m,m,m,m\}}=\frac{\log m}{4G_N}\Big[6\log\Big(\frac{2}{\sqrt 3}\Big)+\log\Big(\frac{1+\sqrt{1-z}}{2z}\Big)+\log(P_{12}P_{34})\Big].
\end{align} 
When $z$ is closer to $1$, the t-channel network dominates and $S^{\{m,m,m,m\}}$ is obtained from the above answer by permuting $P_1$ and $P_3$. 
%Clearly, this phase transition occurs at $\theta=\pi/4$ i.e. at $z=1/2$.

\subsection*{$\tq=5$}\label{holo5section}
\begin{figure}[h]
    \begin{center}
        \includegraphics[scale=0.8]{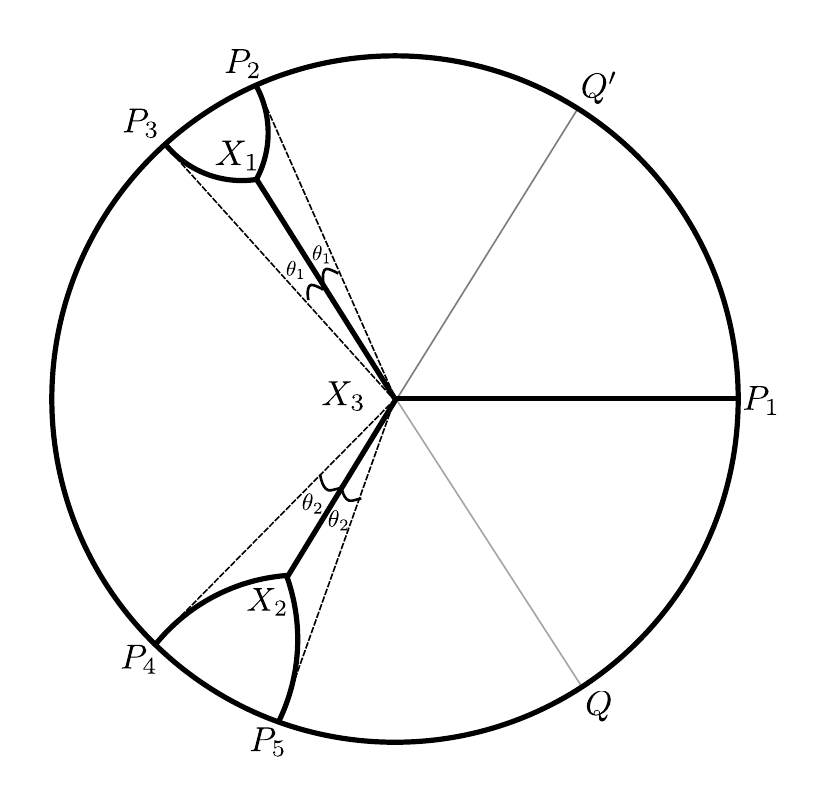}
    \end{center}
    \caption{Symmetric configuration of the geodesic network to compute the general 5-partite special probe measure.}\label{symm-5}
\end{figure}

To compute $S^{\{m,m,m,m,m\}}$ for five connected regions the same trick can be used. The number of cross-ratios for five points in $1d$ is two. Unlike in the $\tq=4$ case, there is no canonical choice of cross-ratios, we simply parametrize this space with two angles $\theta_1,\theta_2$ labeling the cyclically symmetric configuration of points shown in figure \ref{symm-5}.
The points are $P_1=(1,0,1)$ and
\begin{align}
    P_{2,3} & =(\cos(\frac{2\pi}{3}\mp\theta_1),\sin(\frac{2\pi}{3}\mp\theta_1),1),\notag \\
    P_{4,5} & =(\cos(\frac{2\pi}{3}\mp\theta_2),\sin(\frac{2\pi}{3}\mp\theta_2),1).
\end{align}
The length of the minimal geodesic network is just the sum of lengths of three-point networks $(P_2,P_3,Q)$ and $(P_4,P_5,Q')$ minus half the diameter. This can be easily evaluated. We get
\begin{align}\label{holo5}
    S^{\{m,m,m,m,m\}} & =\frac{\log m}{4G_N}\left[9\log\Big(\frac{2}{\sqrt 3}\Big)+\frac12 \log(P_{12}P_{23}P_{34}P_{45}P_{51})\right.                                                                            \\
            & \left.+\frac12\log\left(\left(\frac{3}{4}\right)^3\frac{(1+\cos\theta_1)^2(1+\cos\theta_2)^2}{(1+\cos(\frac{\pi}{3}+\theta_1))(1+\cos(\frac{\pi}{3}+\theta_1))(1+\cos(\frac{\pi}{3}+\theta_1+\theta_2))}\right)\right]\notag.
\end{align}
Here we have made a cyclically symmetric choice of the ``prefactor'' $(P_{12}P_{23}P_{34}P_{45}P_{51})^{1/2}$ in $e^{-\frac{4G_N}{\log m} S^{\{m,m,m,m,m\}}}$ to saturate the unit conformal dimensions of all the operators.

\subsection{CFT computation}
Now we move on to compute special probe measures with $m_\ta=m$ for $3$, $4$ and $5$ for connected regions directly from CFT. We do this by first computing the twist operator correlation function ${\cal E}^{\{m^{n-1},m^{n-1},\ldots\}}$ using CFT techniques. The special probe measure  is then computed using the formula
\begin{align}\label{special-probe-def1}
    S^{\{m,m,\ldots \}}=\lim_{n\to 1} \frac{1}{1-n} \log\Big({\cal E}^{\{m^{n-1},m^{n-1},\ldots\}}\Big).
\end{align}

As a warm-up consider the bi-partite case $\tq=2$.
\begin{align}
    {\cal E}^{\{m^{n-1},m^{n-1}\}} & =\langle {\CO}_{\sigma_1^{-1}\sigma_2}(P_1){\CO}_{\sigma_2^{-1}\sigma_1}(P_2)\rangle
\end{align}
where $\sigma_1$ and $\sigma_2$ are two independent ${\mathbb Z}_{m^{n-1}}$ generators. 
This correlation function has been computed in \cite{Lunin:2000yv} using the Liouville action for uniformization for general conformal field theories. It only depends on the conjugacy class of the permutation element associated with the twist operator.
\begin{align}
    {\cal E}^{\{m^{n-1},m^{n-1}\}}=(P_{12})^{-\Delta_{\sigma_1^{-1}\sigma_2}}.
\end{align}
For a permutation element of the conjugacy class $\{p_k\}$ ($p_k$ is the number of $k$ cycles), we have
\begin{align}
    \Delta= \frac{c}{12}\sum_k p_k\Big(k-\frac1k\Big).
\end{align}
For twist operators of special probes for $\tq$ parties, only $p_{m^{n-1}}= (m^{n-1})^{\tq-1}$ and the rest are $0$. so $\Delta_{\hat \sigma_1^{-1}\hat\sigma_2}=\frac{c}{12}(m^{n-1})^{\tq-2}(m^{2(n-1)}-1)\equiv \Delta^{(\tq)}_n(m)$. Using equation \eqref{special-probe-def1} it follows that
\begin{align}
    S^{\{m,m\}}=\frac{c \, \log m}{6}\log(P_{12}).
\end{align}
Using the holographic dictionary $cG_N=3/2$, we see that $e^{-\frac{4G_N}{\log m} S^{\{m,m\}}}$ indeed transforms like a two-point function of operators with dimension $1$ as expected.

\subsection*{$\tq=3$}
In this case,
\begin{align}\label{3party}
    {\cal E}^{\{m^{n-1},m^{n-1},m^{n-1}\}} & =\langle{\CO}_{\sigma_1^{-1}\sigma_2}(P_1){\CO}_{\sigma_2^{-1}\sigma_3}(P_2){\CO}_{\sigma_3^{-1}\sigma_1}(P_3)\rangle \\
                 & =C_n(m)\,(P_{12}P_{23}P_{31})^{-\frac{1}{2} \Delta_n^{(3)}(m)}.
\end{align}
All the twist operators in this correlation function have the same conformal dimension $\Delta_n^{(\tq)}(m)$.
The OPE coefficient $C_n(m)$ can not be fixed using conformal symmetry. In the cases where the replicated manifold turns out to be a sphere the OPE coefficient can be computed using uniformization, see \cite{Lunin:2000yv,Lunin:2001pw}. Unfortunately, this approach seems difficult in our case since the genus of the replicated manifold increases with $n$. Pressing on,
\begin{align}\label{kappa-def}
    S^{\{m,m,m\}}                  & =\frac{c \log m}{6}\Big[\kappa(m)+\frac12\log(P_{12}P_{23}P_{31})\Big]\notag \\
    {\rm where}\qquad \kappa(m) & =\frac{1}{\log m}\log\Big(\partial_n\Big(\frac{C_n(m)}{C_1^n(m)}\Big)_{n=1}\Big).
\end{align}
We can't compute $\kappa(m)$ from CFT as we don't know $C_n(m)$.  If our holographic prescription is correct then comparing the above expression to the one obtained using holographic prescription \eqref{q3holography}, we arrive at the conjecture
\begin{align}\label{kappa}
    \kappa(m)=3 \log\Big(\frac{2}{\sqrt 3}\Big).
\end{align}

\subsection*{$\tq=4$}
In this case, we need the four-point function of twist operators.
\begin{align}
    {\cal E}^{\{m^{n-1},m^{n-1},m^{n-1},m^{n-1}\}} & =\langle{\CO}_{\sigma_1^{-1}\sigma_2}{\CO}_{\sigma_2^{-1}\sigma_3}{\CO}_{\sigma_3^{-1}\sigma_4}{\CO}_{\sigma_4^{-1}\sigma_1}\rangle.
\end{align}
The operators are inserted at $P_1,\ldots, P_4$ respectively.
When the cross-ratio $z$ is in a finite neighborhood of $0$ (of $1$), it can be shown that for a large $c$ conformal field theory, the correlator is dominated by the Virasoro block of the lightest operator in the s-channel (t-channel). This operator is the twist operator ${\CO}_{\sigma_1^{-1}\sigma_3}$ (${\CO}_{\sigma_2^{-1}\sigma_4}$).
\begin{align}
    {\cal E}^{\{m^{n-1},m^{n-1},m^{n-1},m^{n-1}\}}=\tilde C_n(m) {\tilde C'}_n(m) (P_{12}P_{34})^{-\Delta_{n}^{(4)}(m)} F_{\Delta^{(4)}_n(m)}(z,\bar z).
\end{align}
Here $F_{\Delta}(z,\bar z)$ is the Virasoro block of an operator of dimension $\Delta$. The OPE coefficients $\tilde C_n(m)$ and $\tilde C'_n(m)$ are the three point function coefficients in $ \langle{\CO}_{\hat \sigma_1^{-1}\hat\sigma_2}{\CO}_{\hat \sigma_2^{-1}\hat\sigma_3}{\CO}_{\hat \sigma_3^{-1}\hat\sigma_1}\rangle$ and $\langle{\CO}_{\hat \sigma_1^{-1}\hat\sigma_3}{\CO}_{\hat \sigma_3^{-1}\hat\sigma_4}{\CO}_{\hat \sigma_4^{-1}\hat\sigma_1}\rangle$ respectively. As our permutation elements are symmetric under the exchange of parties, both these three-point functions are the same and are equal to $C_n(m)$ in equation \eqref{3party}. This gives
\begin{align}
    S^{\{m,m,m,m\}} =\frac{c\log m}{6}\Big[2\kappa(m)+\log(P_{12}P_{34})-\partial_n\log F_{\Delta_n^{(4)}(m)}|_{n=1}\Big].
\end{align}
The $\log F_{\Delta}$ can be computed for large $c$ CFT in a perturbative expansion in $\Delta$ using the so-called monodromy method. This method is reviewed in appendix \ref{monodromy}. Because $\Delta\to 0$ in the $n\to 1$ limit, the special probe $S^{\{m,m,m,m\}}$ can be computed just from the first term in the perturbative expansion. We have done this computation in appendix \ref{4-pt-monodromy} and have found the result to agree with the holographic expectation \eqref{4partyhol}. Note that the conformal symmetry only guarantees that the holographic probe measure depends on the cross-ratio, the fact that it is the same function of cross-ratio that we get from the CFT (up to a constant) serves as a check of our holographic prescription.

\subsection*{$\tq=5$}
We will parametrize the five points in the same way as in section \ref{holo5section}. For simplicity, we will only consider the channel shown in figure \ref{symm-5}. The correlation function is dominated by the lightest Virasoro block in this channel when $\theta_1$ and $\theta_2$ are small but finite. The OPE coefficients appearing at all three vertices are the same, $C_n(m)$. So we have,
\begin{align}
    S^{\{m,m,m,m,m\}} & =\frac{c\log m}{6}\Big[3\kappa(m)+\frac12\log(P_{12}P_{23}P_{34}P_{45}P_{51})\notag        \\
            & -\partial_n\log F_{\Delta_n^{(5)}(m),\Delta_n^{(5)}(m)}(\theta_1,\theta_2)|_{n=1}\Big].
\end{align}
Here $F_{\Delta_1,\Delta_2}(\theta_1,\theta_2)$ is the conformal block in the channel displayed in figure \ref{symm-5}.
We again use the monodromy method for the case of five points and we compute the leading term in $\log F_{\Delta_n^{(5)},\Delta_n^{(5)}}$ in appendix \ref{5-pt}. The resulting $S^{\{m,m,m,m,m\}}$ agrees with the holographic prediction \eqref{holo5}. Again, \emph{a priori} this agreement is not guaranteed so this match does strengthen the check of our holographic prescription.

\subsubsection{General argument}
A beautiful argument - using the $SL(2,C)$ Chern-Simons theory description of gravity in AdS$_3$ - has been given in \cite{Hijano:2015rla} that shows in full generality that the logarithm of a general point conformal block, at leading order in $G_N$, is given by the length of the minimal geodesic network up to an overall constant. The constant is fixed by taking the OPE limit. We will reproduce this argument here for the reader's convenience. For details, please see \cite{Hijano:2015rla}.

Consider the Fefferman-Graham expansion of the locally $AdS_3$ metric in the presence of a massive particle
\begin{align}
    ds^2=d\rho^2+e^{2\rho} g^{(0)}_{\mu\nu} dx^\mu dx^\nu+g^{(2)}_{\mu\nu}dx^\mu dx^\nu+\ldots
\end{align}
with $g^{(0)}_{\mu\nu} dx^\mu dx^\nu=dzd\bar z$. Let the world-line of the massive particle, i.e. with the action $S_{\rm particle}=2M\int d\lambda \sqrt{h}$ where $h$ is the induced metric on the world-line, pierce the boundary at $z_0$. We wish to solve Einstein's equations near this point. Expanding the Einstein equations at leading order in large $\rho$, the only non-vanishing equations are
\begin{align}
    g_{z\bar z}^{(2)}                                                       & =2\pi M  \delta^{(2)}(z-z_0)\notag     \\
    \partial_{\bar z} g_{zz}^{(2)}-\partial_{z}g_{z\bar z}^{(2)}            & =-4\pi p_z\delta^{(2)}(z-z_0)\notag   \\
    \partial_{z} g_{\bar z \bar z}^{(2)}-\partial_{\bar z}g_{z\bar z}^{(2)} & =-4\pi p_{\bar z}\delta^{(2)}(z-z_0).
\end{align}
Here $p_\mu$ is the momentum defined through, $\frac{c}{6}p^\mu=2M \,dx^\mu/d\lambda$ with $c$ being the Brown-Henneaux central charge $c=3/2G$. When it is large, the mass is related to the conformal dimension as $M=\Delta\equiv \delta c/6$. Using the fact that boundary stress tensor is $T_{\mu\nu}=g^{(2)}_{\mu\nu}$, singular terms in the holomorphic stress tensor can be computed using the above expression.
\begin{align}\label{T-massive}
    T(z)=\frac{\delta}{(z-z_0)^2}+\frac{2p_z}{z-z_0}+\ldots
\end{align}
As the momentum $p_z$ is canonical conjugate to the coordinate $z$, we have $p_z=d S_{\rm particle}/dz$. In terms of $T(z)$, the metric takes the form
\begin{align}
    ds^2=d\rho^2-Tdz^2-\bar T d {\bar z}^2+(e^{2\rho}+T\bar T e^{-2\rho})dzd\bar z.
\end{align}

At this point, it is useful to pass to the Chern-Simons description of $AdS_3$ gravity. The $SL(2)\times SL(2)$ connection that corresponds to the above metric is
\begin{align}
    A=\begin{pmatrix} \frac12 d\rho & e^{-\rho} T dz \\ -e^{\rho} dz  & -\frac12 d\rho \end{pmatrix},\qquad  \bar A=\begin{pmatrix} -\frac12 d\rho & e^{\rho}  dz \\ -e^{-\rho} \bar T dz  & \frac12 d\rho \end{pmatrix}.
\end{align}
In the absence of matter, $T(z)$ obeys the equation $\partial_{\bar z} T(z)=0$ and hence the holonomy of this connection around a closed contour would be zero. However, when the contour encloses a massive particle, the holonomy of $A$ is nontrivial due to the conical singularity. Computing this holonomy is equivalent to computing the monodromy of a two-component vector function $\psi(z)$ that satisfies the equation
\begin{align}
    \frac{d\psi}{dz}=A \psi.
\end{align}
The bottom component of $\psi$ then obeys,
\begin{align}
    \psi_2''+T(z)\psi_2=0.
\end{align}
This is exactly the differential equation that appears in the monodromy method \eqref{diff-con}. Moreover, the function $T(z)$ given in equation \eqref{T-massive} is exactly equal to the one appearing in \eqref{Tform} with $p_z$ playing the role of the accessory parameter $c_\ta(z)$. When the $\psi_2$ is taken around a group of external sources for the massive operators, its path is homologous to a closed loop around the world-line of the massive particle appearing in this channel. The holonomy of $A$ around such a loop is such that it produces the monodromy of $\psi_2$ that matches the one required in the monodromy method of appendix \ref{monodromy}.

This argument shows in full generality that computation of the action of a world-line network of massive particles is equivalent to the computation of the corresponding Virasoro conformal block in the CFT with large central charge $c$ (at leading order in $c$). Note that, although the actual computation of either is possible only in perturbation theory in $\delta$ where $\Delta=\delta c/6$ i.e. in the limit $n\to 1$, the equivalence of the two is shown exactly in $n$.

The minimal geodesic network in a hyperbolic disc is also called a Steiner tree. The connection between Virasoro conformal blocks and Steiner trees has been analyzed in detail in \cite{Alkalaev:2018nik}. Some of our analysis in section \ref{2dcft} overlaps with theirs.

\subsubsection{Limitation of the check}
As remarked earlier, using the CFT methods we are only able to compute the dominant conformal block in the twist operator correlation function. To compute the correlation function itself in the large $c$ limit, we also need to compute the three-point function coefficient $C_n(m)$ of twist operators at large $c$. This computation is difficult to do as it maps to the computation of the CFT partition function on a higher genus Riemann surface which could potentially be non-universal. This does not allow the uniformization methods of \cite{Lunin:2000yv, Lunin:2001pw} to be applicable. The holographic prescription gives a conjecture 
\begin{align}\label{conj}
    \kappa(m)=3\log\Big(\frac{2}{\sqrt 3}\Big)
\end{align}
where $\kappa(m)$ is defined in the equation \eqref{kappa-def}. It would be nice to check the same from the CFT. One possible way to do so could be along the lines of \cite{Cardy:2017qhl}. In this paper, the authors use the modular invariance of genus $2$ partition function of 2d CFT to compute the average of the three-point coefficient squared for heavy operators. If one believes, that the twist operators considered here are generic enough that their three-point function should match with the average, then methods of \cite{Cardy:2017qhl} could be used to test the conjecture \eqref{conj}. We will not have anything more to say about this.

Even after fixing this value of $\kappa(m)$, there is yet another check. The prediction from the CFT is that $\tq$-partite special probe measure for $\tq$ connected regions must take the form,
\begin{align}
    S^{\{m,m,\ldots \}}=(\tq-2)\kappa(m) + \partial_n\log F^{(\tq)}_n|_{n=1}
\end{align}
where $F^{(\tq)}$ is the Virasoro conformal block in some channel. This is because there are $\tq-2$ trivalent vertices that appear in a $\tq$-point conformal block.
The argument outlined in this section shows that  $ S^{\{m,m,\ldots \}}_{\rm holographic}-\partial_n\log F^{(\tq)}_n|_{n=1}$ is a constant, say $\kappa^{(\tq)}$. It is then nontrivial that the  $\tq$ dependance of $\kappa^{(\tq)}$ is such that  $\kappa^{(\tq)}=(\tq -2)\kappa$. We have checked this for $\tq=4,5$. It would be nice to see that this continues to hold for higher points and hopefully produce an argument that works for all higher values of $\tq$.

\subsection{Comment on bulk replica symmetry}\label{replica}
In this section, we will discuss the status of the replica symmetry assumption. 
We will discuss this assumption only for the case of the probe family $S^{(\tq)}_n$ that arises from the special symmetric measure ${\cal E}^{\{n,n,\ldots\}}$ in $2d$ CFTs. Just like any other measure, this measure is computed from the path integral on the ramified manifold obtained by cutting and gluing replicas along given spatial regions in the fashion specified by the permutations. The genus of the ramified boundary is easily computed using the Riemann-Hurwitz formula
\begin{align}
    g_{(\tq,n)}=1+\frac{n^{\tq-2}}{2}(n(\tq-2)-\tq).
\end{align}
As expected, for $\tq=2$ we get $g=0$ for all $n$. This is simply the statement that the replicated manifold corresponding to the bi-partite Renyi entropy between an interval and its complement is topologically a sphere. For $(\tq,n) =(3,2)$ also we get $g=0$. This case has been studied in \cite{Penington:2022dhr}. For $(\tq,n)=(3,3),(4,2)$, the genus is $1$ and for every other case $g\geq 3$. The case of $(\tq,n)=(3,3)$ was also considered in \cite{Penington:2022dhr} and it was argued that the dominant bulk solution does not preserve the replica symmetry ${\mathbb Z}_3 \times {\mathbb Z}_3$. The argument is as follows.\footnote{We thank Geoff Penington for the discussion on this issue.} 

Configuration of three points on a sphere does not admit any modulus so the complex structure of the ramified manifold is fixed. It is fixed to the ${\mathbb Z}_3$ symmetric point $\tau=e^{i\pi/3}$. As a result, the torus has three shortest cycles and they are transformed into each other by the action of a subgroup of the replica symmetry. The dominant bulk solution must correspond to the torus handle-body where one of the three shortest cycles becomes contractible in the bulk. This bulk solution breaks the symmetry between the three cycles and hence also a part of the replica symmetry. This is an example where the assumption that the dominant bulk saddle is replica symmetry preserving is invalid. Does this mean that the replica symmetry assumption is incorrect even for other values of $(\tq, n)$? We have verified the correctness of our prescription, which followed from the replica symmetry assumption, for $\tq\geq 4$ using conformal field theory methods. However, it is possible that even in the absence of full replica symmetry of the dominant bulk solution, the correct prescription may coincide with the one obtained from CFT methods. Therefore, it would be useful to have a direct bulk argument that supports the replica symmetry assumption at least for $\tq \geq 4$. In what follows, we will give such an argument for $(\tq,n)=(4,2)$. We will explicitly construct the dominant bulk solution and show that it is replica symmetric. 

\begin{figure}[h]
    \begin{center}
        \includegraphics[scale=0.17]{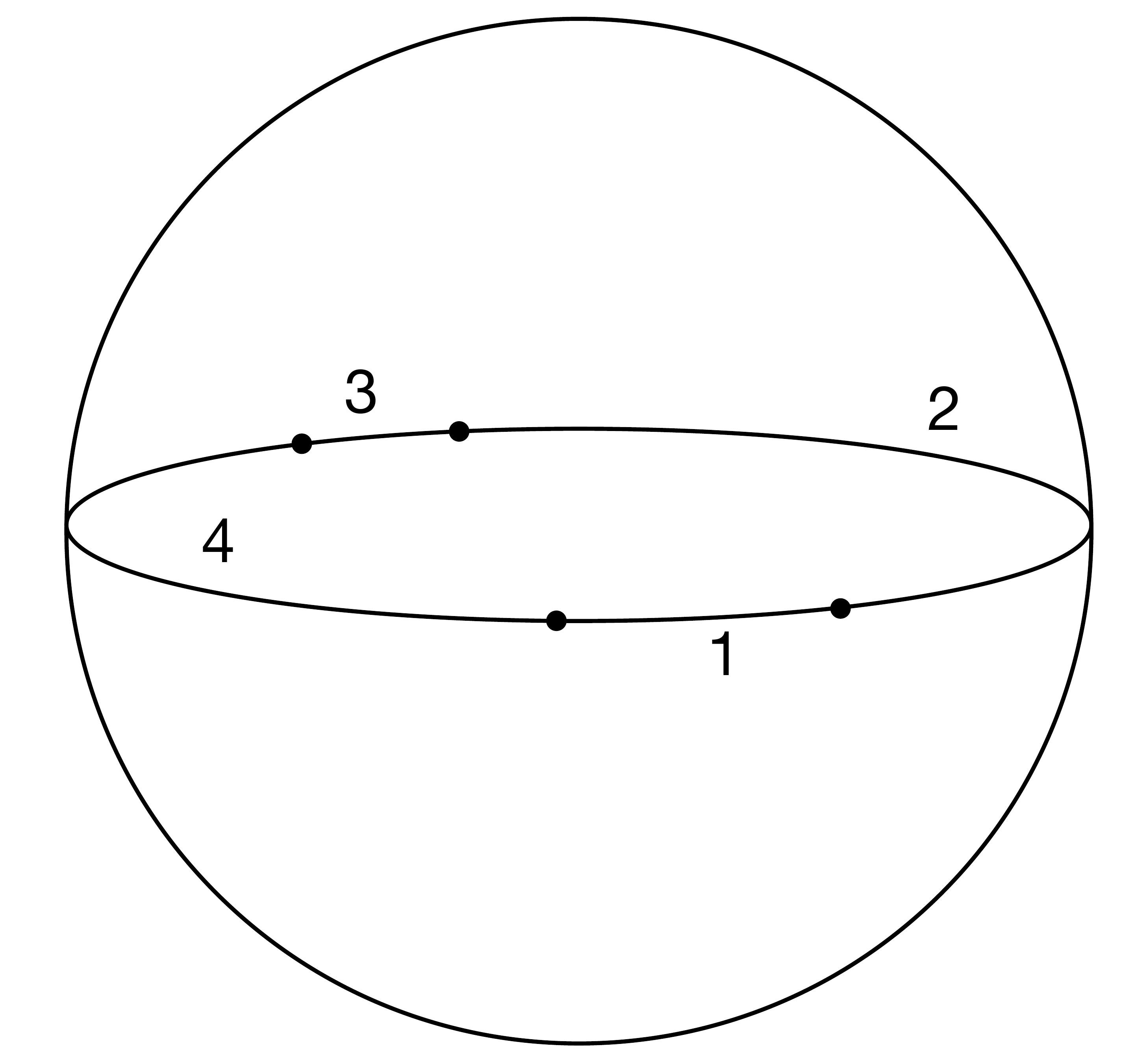}
    \end{center}
    \caption{Configuration of twist operators on the equator.}\label{4ptsphere}
\end{figure}

\begin{figure}[h]
    \begin{center}
        \includegraphics[scale=0.22]{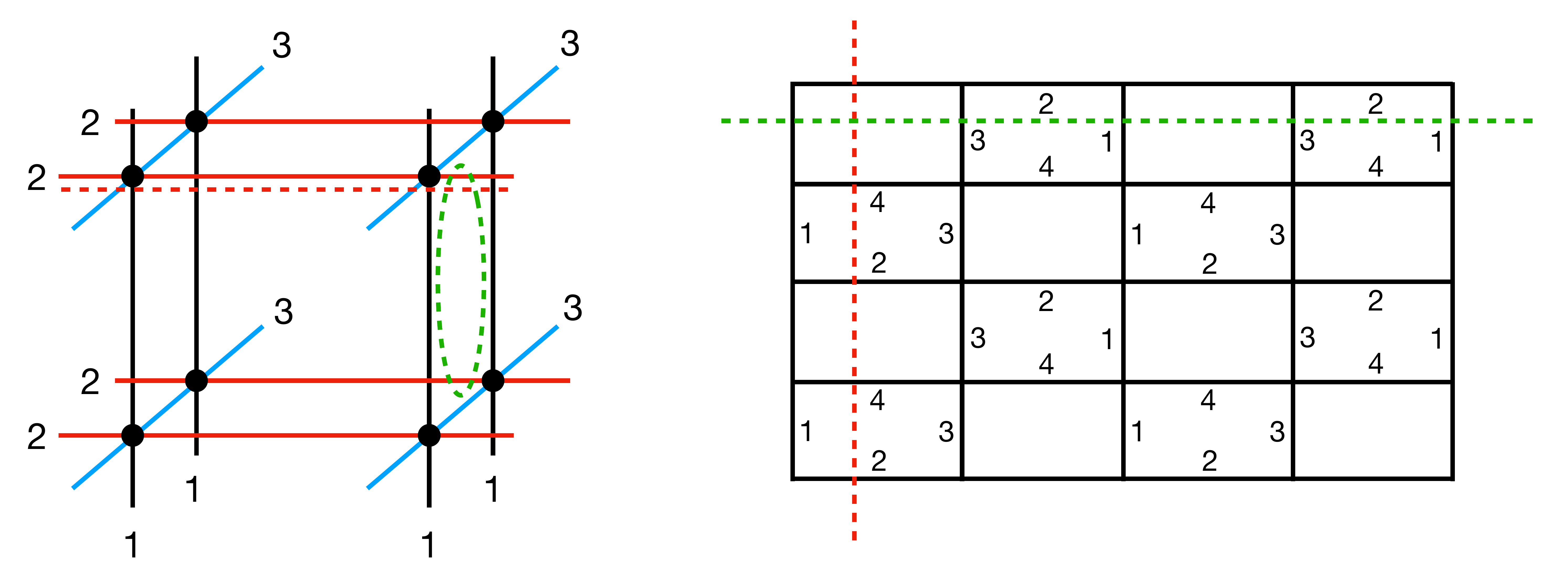}
    \end{center}
    \caption{The first figure gives the gluing pattern of all the replicas used to construct the measure. The second figure shows the ramified manifold viz. torus tessellated by the replicas. In the second picture, we identify the two homologically independent cycles. They are denoted by the red and green dotted lines. These cycles are then pulled back to the figure.}\label{ramified}
\end{figure}

Consider the measure $S_2^{(4)}$ where the parties correspond to four connected regions on the equator of the sphere as shown in figure \ref{4ptsphere}.
For simplicity, we have taken the regions $1$ and $3$ to be smaller than $2$ and $4$. The equatorial configuration of twist operators maps to a rectangular torus. The gluing pattern of all the regions in eight replicas is given in the first diagram of figure \ref{ramified}. This is the graph with the density matrix on parties $1,2,3$ as vertices. 
In the second diagram in figure \ref{ramified}, we have shown how the individual replicas glue together to form the torus. The two homologically independent cycles of the torus are denoted by the red and green dotted lines. 
Their pull back to the graph is also shown in the same way. The action of the replica group ${\mathbb Z}_2\times {\mathbb Z}_2 \times {\mathbb Z}_2$ is clear on the graph. The independent cycles are not mixed under the replica group action. The shortest cycle of the torus is the one shown by a red dotted line. The dominant bulk saddle fills out this cycle to form a hyperbolic genus-$1$ handle-body. Because the cycles are invariant under the replica symmetry group the dominant bulk saddle is also invariant. 

This construction is possible because $g_{(4,2)}=1$. For $n=2$ and higher values of $\tq$, the genus is larger and hence the analysis becomes cumbersome but not completely intractable. Currently, we are in the process of computing replica symmetric solutions for the case of $(\tq, 2)$ for general $\tq$ \cite{GaddeWIP}. We don't have anything definitive to say about the status of replica symmetric solutions for higher values of $n$.

\section{Discussion}\label{discuss}

\subsection*{Usefulness to Quantum information theory}
The measures that we have constructed are genuinely multi-partite. Let us elaborate. 
One can find pairs of multi-partite states, known as isospectral states, with the property that every density matrix constructed out of them by partial trace has the same spectrum and even then they are not equivalent to each other up to local unitary transformations. As a result, any combination of bi-partite Renyi entropies and Von Neumann entropies yields identical results. A very simple example is furnished by the following pair \cite{Nielsen_2001, Gadde:2022cqi}.
\begin{align}
    |\psi_1\rangle= \frac{1}{\sqrt{3}}|000\rangle+\frac{\sqrt{2}}{\sqrt{3}}|111\rangle, \quad |\psi_2\rangle = \frac{1}{\sqrt{3}}(|100\rangle+|010\rangle+|001\rangle).
\end{align}
However, as emphasized in \cite{Gadde:2022cqi}, the multi-partite measures introduced in this paper, do distinguish them. This illustrates their usefulness.

Another useful property that a multi-partite entanglement measure should satisfy is that it should be monotonic under coarse-graining \cite{Hein_2004} i.e. if ${\cal M}^{(\tq)}$ is a $\tq$-partite measure evaluated on a $\tq$-partite state $|\psi\rangle$ and ${\cal M}^{\tq-1}_{[\ta,\tb]}$ is a $(\tq-1)$-partite measure evaluated on a $(\tq-1)$-partite state obtained from $|\psi\rangle$ by identifying a pair of parties $\ta,\tb$ and thinking of their tensor product as a single party then it is desirable that the measure (or more appropriately the family of measures ${\cal M}^{(\tq)}$) obey
\begin{align}\label{coarse}
    {\cal M}^{(\tq)}-{\cal M}^{(\tq-1)}_{[\ta,\tb]}\geq 0.
\end{align}
This is intuitive because by identifying a pair of parties, the measure allows for a scrambling unitary operator that mixes the two parties. It is expected that the entanglement should reduce under such scrambling. We are currently exploring this inequality for the measures discussed in this paper. Another viewpoint on the inequality \eqref{coarse} can be obtained by thinking of the left-hand side as a sort of ``multi-partite'' mutual information ${\cal I}^{(\tq)}_{\ta,\tb}$ between the parties $\ta$ and $\tb$ so that the inequality \eqref{coarse} is a statement of its positivity. To be clear, this is not the usual mutual information but a novel one that has to do with the entanglement of parties $\ta$ and $\tb$ with themselves directly as well as ``through'' other parties. It is then useful study conditions that lead to the vanishing of this ``multi-partite'' mutual information ${\cal I}^{(\tq)}_{\ta,\tb}$. The advantage of this point of view will become clear in this next discussion point.

\subsection*{Application to bulk reconstruction}
Arguably the most important application of information-theoretic ideas to AdS/CFT correspondence has been the so-called entanglement wedge reconstruction \cite{Dong:2016eik, Harlow:2016vwg}. This idea relies on formulating the AdS/CFT correspondence as an error-correcting code \cite{Almheiri:2014lwa} and on the observation that boundary modular flow is equal to the bulk modular flow \cite{Jafferis:2015del}. It was also shown that a version of the quantum-corrected Ryu-Takayanagi formula holds for any error-correcting code \cite{Harlow:2016vwg}. Let's review the statement of entanglement wedge reconstruction in the simplest context of the vacuum state. 

Let the AdS/CFT map be an isometry $W$ from the bulk Hilbert space $H_{\rm bulk}$ into the boundary Hilbert space $H_{\rm boundary}$. Decomposition of the boundary into regions $A$ and its complement $\bar A$ corresponds to the tensor decomposition 
\begin{align}
    H_{\rm boundary}=H_A\otimes H_{\bar A}.
\end{align}
The Ryu-Takayanagi surface for a given boundary region $A$ divides the bulk into the entanglement wedge $a$ of $A$ and its complement. This gives the tensor decomposition of the bulk Hilbert space 
\begin{align}
    H_{\rm bulk}=H_{a}\otimes H_{\bar a}.
\end{align}
We can now ask refined questions about the isometry for the above tensor product decomposition. The entanglement wedge reconstruction is the statement that bulk operator $O_{\rm bulk}$ acting on $H_{a}$ can be reconstructed on the boundary as an operator $O_{\rm boundary}$ that only acts on $H_A$. A similar statement can also be made for the region $\bar A$ and its entanglement wedge $\bar a$. Alternatively, the reconstruction can also be phrased in the language of error correcting codes viz. the encoding of $H_a$ is robust against arbitrary errors acting on $H_{\bar A}$. In simpler language, any quantum operation on $\bar A$ will not destroy ``information'' contained in $a$. The idea of quantum error correction (QEC) is used heavily in arguing for entanglement wedge reconstruction. Let us briefly outline this idea. Let us consider a maximally entangled state $|\psi\rangle$ in $H_a\otimes H_{\rm ref}$ where is $H_{\rm ref}$ is a reference Hilbert space introduced only to diagnose the quantum errors. Applying $W$ to $|\psi\rangle\otimes |e_0\rangle, |e_0\rangle \in H_{\bar a}$, we get a state in $H_A \otimes H_{\bar A}\otimes H_{\rm ref}$. One of the fundamental theorems of QEC, known as the decoupling principle, states that if the mutual information between parties $H_{\rm ref} \otimes H_{\bar A}$ in the state $W(|\psi\rangle\otimes |e_0\rangle)$ is zero then the errors on $H_{\bar A}$ are reconstructible. Here we want to highlight the role played by the vanishing of mutual information in QEC. 

\begin{figure}[h]
    \begin{center}
        \includegraphics[scale=0.17]{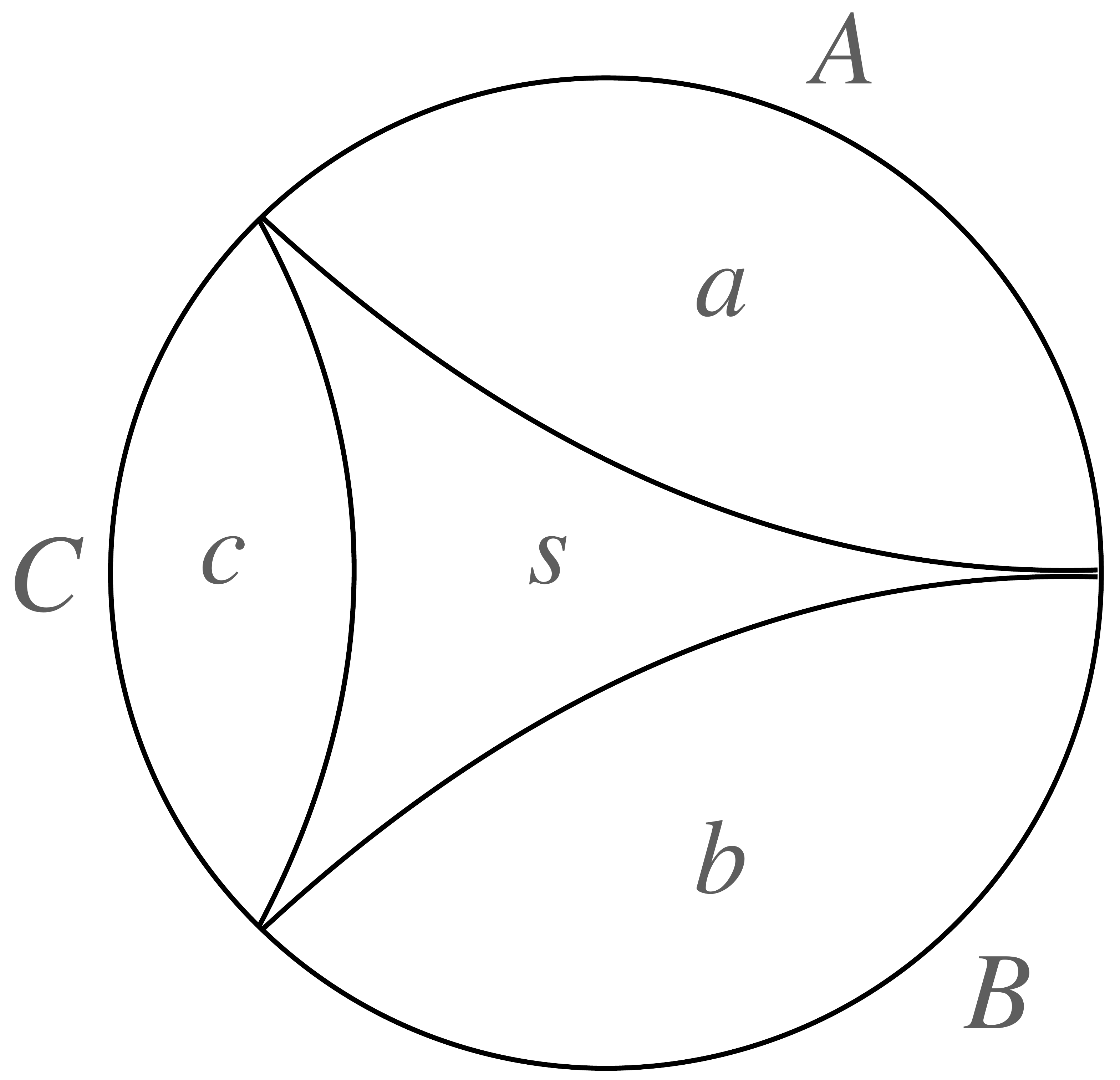}
    \end{center}
    \caption{The regions $A,B$ and $C$ on the boundary and their entanglement wedges $a,b$ and $c$ respectively in the bulk. The so-called entanglement shadow is labeled by $s$.}\label{shadow}
\end{figure}

Now consider the decomposition of the boundary into three regions $A,B$ and $C$ as shown in figure \ref{shadow}. 
We have also shown their respective entanglement wedges $a,b$ and $c$ in the bulk. There is a region in the bulk, labeled $s$, that is not included in any of these entanglement wedges. An operator acting on $H_s$ can not be reconstructed as an operator acting on a single interval, either $H_A, H_B$ or $H_C$ on the boundary. However, it can be reconstructed as an operator acting on either $H_{A\cup B}, H_{B\cup C}$ or $H_{C\cup A}$ because $s$ is part of the entanglement wedge of any one of them. Alternatively, we could say that the encoding of $H_s$ is not robust against \emph{arbitrary} errors acting on either regions $A\cup B, A\cup C$ or $B\cup C$. It is robust, however, against arbitrary errors acting on any one of the regions $A,B,C$.

\begin{figure}[h]
    \begin{center}
        \includegraphics[scale=0.17]{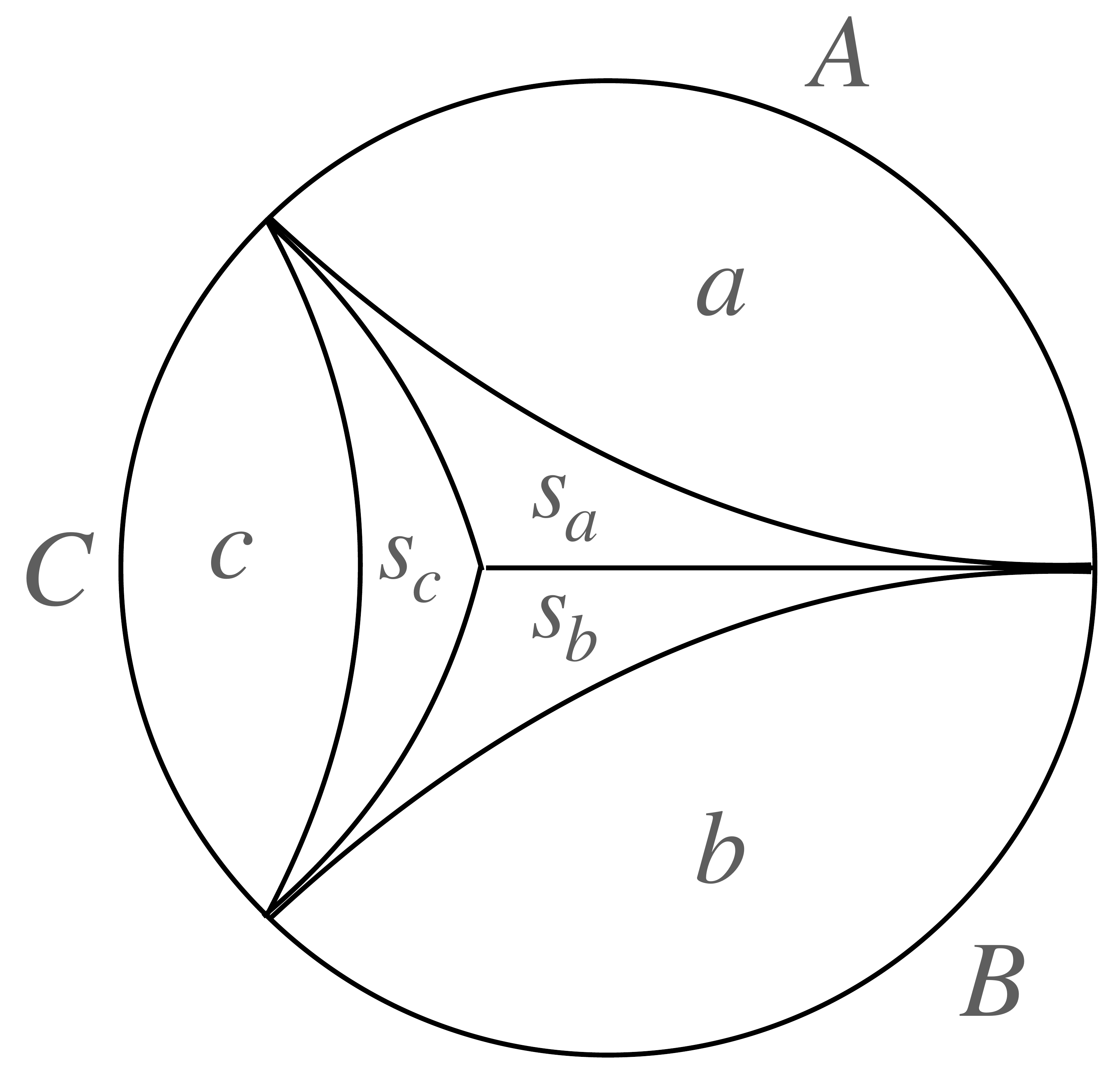}
    \end{center}
    \caption{Decomposition of bulk due to a tripartite probe measure. In particular, the region $s$ is partitioned into three regions $s_a, s_b$ and $s_c$.}\label{3-error}
\end{figure}

\begin{figure}[h]
    \begin{center}
        \includegraphics[scale=0.17]{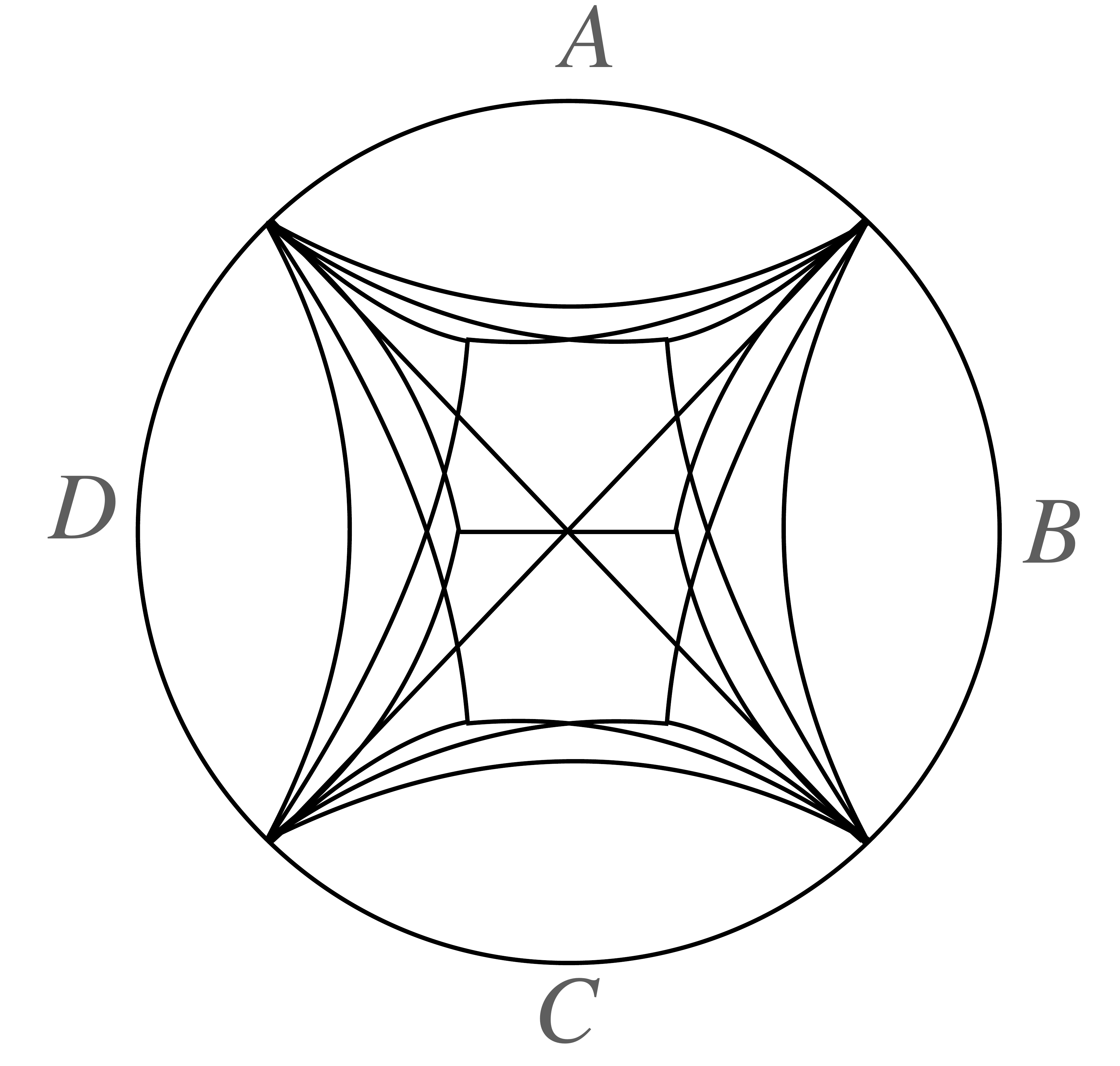}
    \end{center}
    \caption{The bulk Cauchy slice getting chopped up into fine pieces due to brane-webs corresponding to various multi-party probe measures.}\label{4-chop}
\end{figure}

Any of the probe measures discussed in this paper yields a decomposition of the bulk, in particular of the region $s$ into three regions $s_a,s_b$ and $s_c$ as shown in figure \ref{3-error}.
It is natural to expect that the quantum probe measure formula \eqref{quantum-web} leads to a statement of error correction and hence of reconstruction that distinguishes the regions $s_a,s_b$ and $s_c$. It would be very interesting to find this new notion of multi-partite quantum error correction. Given the role played by the mutual information in the usual bi-partite QEC, we expect a novel notion of multi-partite mutual information could be useful in, first of all, phrasing the statement of multi-partite QEC and then diagnosing it. A quantity similar to ${\cal I}^{(\tq)}_{\ta,\tb}$ discussed above could be useful for this purpose.  The multi-partite QEC would enrich our understanding of the local properties of the AdS/CFT correspondence as we refine the boundary region into multiple parties. For example consider four parties $A,B,C$ and $D$ on the boundary as shown in figure \ref{4-chop}. 
With this decomposition of the boundary we can compute $^4C_2=6$ of tri-partite measures $S^{(3)}(A,B,CD), S^{(3)}(A,C,BD), \ldots$, one four-partite measure $S^{(4)}(A,B,C,D)$ in addition to six bi-partite ones. Almost all of these measures give a different brane-web in the bulk. If we overlay them, then the bulk gets chopped into many fine pieces and error-correcting properties of Hilbert space associated with every such piece would be distinct. As we consider more and more regions on the boundary, the bulk gets chopped up into finer and finer pieces, hopefully leading to a very local AdS/CFT correspondence. 

\section*{Acknowledgements}
We would like to thank Xi Dong, Jonathan Harper, Matthew Headrick, Gautam Mandal, Shiraz Minwalla, Arvind Nair, Pratik Rath, Pranab Sen, Piyush Shrivastava, Douglas Stanford, Sandip Trivedi for interesting discussions. We are particularly indebted to Onkar Parrikar for stimulating discussions and insightful comments. We would also like to thank Shraiyance Jain and Harshal Kulkarni for their collaboration on related projects. 
This work is supported by the Infosys Endowment for the study of the Quantum Structure of Spacetime and by the SERB Ramanujan fellowship.  We acknowledge the support of the Department of Atomic Energy, Government of India, under Project Identification No. RTI 4002. 
AG would like to thank the hospitality of KITP during the program ``Bootstrapping quantum gravity'' where this work was partially carried out.
Finally, we acknowledge our debt to the people of India for their steady support to the study of the basic sciences.

\appendix
\section{Counting measures for large dimensional systems}\label{plethystic}
In this appendix, we will perform the integral \eqref{matrix-integral} in the $d_i\to \infty$ limit in the saddle point approximation.
\begin{align}
    \hat Z^{(\tq)}(x) & =\int \prod_{i=1}^{\tq} d\cU_i\, \exp\Big(\sum_{k=1}^{\infty}\frac{1}{k}\,z(x^k,\cU_i^k)\Big)\notag                                                                     \\
                      & =\int \prod_{i=1}^{\tq} d\cU_i\, \exp\Big(\sum_{k=1}^{\infty}\frac{1}{k}\,x^k \, \prod_{i=1}^{\tq} {\rm Tr}(\cU_i^k)+\prod_{i=1}^{\tq} {\rm Tr}(\cU_i^{\dagger k})\Big)
\end{align}
We first gauge fix the integral to the Cartan subgroup of every $U(d_i)$. The integral over $\cU_i$ is reduced to $e^{i\theta_{\alpha_{i}}}$ where $\alpha_i=1,\ldots, d_i$. The we define the eigenvalue density $\rho_i(\theta)\equiv \sum_{\alpha_i=1}^{d_i} \delta(\theta- \theta_{\alpha_{i}})$. We then rewrite the integral as integral over modes of $\rho_i(\theta)$. These modes are defined as $\rho_{i,r}=\int \rho_i(\theta) e^{ir \theta} d\theta=\sum_{\alpha_i=1}^{d_i} e^{ir\theta_{\alpha_i}}={\rm Tr}(\cU_i^r)$.

In these variables, the Haar measure takes the form
\begin{align}
    d\cU_i=\prod_{r=1}^{\infty} d\rho_{i,r}d\bar \rho_{i,r} \,\exp\Big(-\frac{|\rho_{i,r}|^2}{r}\Big).
\end{align}
The partition function integral is simplified to
\begin{align}\label{fr}
    \hat Z^{(\tq)}(x) & =\prod_{r=1}^{\infty} \int  \Big(\prod_{i=1}^{\tq} d\rho_{i,r}d\bar \rho_{i,r}\Big)\,\exp\Big(-\sum_{i=1}^{\tq} \frac{|\rho_{i,r}|^2}{r}+\frac{x^r}{r} \prod_{i=1}^{\tq} \rho_{i,r}+ \prod_{i=1}^{\tq} \bar \rho_{i,r}\Big)\notag \\
                      & \equiv \prod_{r=1}^{\infty} F_r(x).
\end{align}
Using change of variables in the integral $F_r(x)$, it is easy to show that
\begin{align}\label{frf1}
    F_r(x)=F_1(x^r\,r^{\tq-2}).
\end{align}
Hence, to compute $\hat Z^{(\tq)}(x)$, we need only to do a single integral $F_{1}(x)$. After expanding in powers of $x$, this integral reduces to a sum of Gaussian integrals. We get,
\begin{align}
    F_{1}(x)=\sum_{k=1}^\infty x^k (k!)^{\tq-2}.
\end{align}
Using the equations \eqref{fr} and \eqref{frf1}, we get
\begin{align}
    \hat Z^{(\tq)}(x) = \prod_{r=1}^\infty \left( \sum_{k=0}^\infty \Big(x^r r^{\tq-2}\Big)^{k} \Big(k!\Big)^{\tq-2} \right).
\end{align}
This is exactly the result quoted in equation \eqref{Zq}.

\section{Analytic interpolation problem}\label{regge}
In this section, we will summarize the results of the paper \cite{regge-viano} without going into its derivation. We will also discuss the consistency conditions for the proposed analytic continuation that were derived in \cite{osti_4065624}.

Let $f(z)$ be a function that is analytic in the right half plane that obeys the following conditions,
\begin{align}
     & \lim_{\rho\to \infty} \frac{\log (f(|\rho e^{i\theta}|))}{\rho}\leq B \sin \theta \leq \pi \label{bounded} \\
     & f(n+\lambda)=f_n\qquad  n\in \{0,1,2,\ldots\}, \lambda>0. \label{values}
\end{align}
Here $f_n$ are some fixed value, sometimes referred to as input data. The function $f(z)$ with these properties was constructed in \cite{regge-viano} just from the input data. The solution is as follows.

\begin{align}\label{interpolate}
    f(z+\lambda)= \frac{\sin \pi z}{\pi}\frac{\Gamma(z+1)\Gamma(2\lambda)}{\Gamma(z+2\lambda)}\Big(\sum_{n=0}^{\infty} (-1)^{n}\frac{\Gamma(n+2\lambda)}{\Gamma(n+1)\Gamma(2\lambda)}\frac{f_n}{z-n}-2^{1-2\lambda}\sum_{k=0}^{\infty} a_k Q_k^{\lambda}(-i(z+\lambda))\Big).
\end{align}
Let us take a moment to explain the second term on the right-hand side. The coefficients $a_k$ are defined as
\begin{align}
    f(iy)=\sum_{k=0}^{\infty} a_k P_k^\lambda(y), \qquad y\in {\mathbb R}.
\end{align}
The functions $P_k^{\lambda}(y)$ are the so called Pollaczek polynomials. The coefficients $a_k$ can be evaluated using the orthogonality of $P_k^{\lambda}(y)$.
\begin{align}
    a_k=2^{2\lambda} \sum_{n=0}^{\infty} (-1)^n f_n \frac{\Gamma(n+2\lambda)\Gamma(k+1)}{\Gamma(k+2\lambda)\Gamma(n+1)}P_k^{\lambda}(-i(n+\lambda)).
\end{align}
The functions $Q_k^\lambda(z)$ are known as the associated Pollaczek functions,
\begin{align}
    Q_k^\lambda(z)=\frac{i}{2}\int_{-\infty}^{\infty}\frac{P_k^\lambda(x)}{x-z}\Big(\frac{2^{2\lambda-1}\Gamma(\lambda+i x)\Gamma(\lambda-ix)}{\pi \Gamma(2\lambda)}\Big).
\end{align}
It is easy to check that if we set $z$ to be a non-negative integer, the formula \eqref{interpolate} becomes a tautology. 

For the analytic continuation of the probe family, we have the data defined at $n=2,3,\ldots$ and we are interested in analytically continuing to $n=1$. Let us shift all these values to the left by one unit so the input data is specified at $n=1,2,\ldots$ and we analytically continue it to $n=0$. It is convenient to set $\lambda =1$ in the above formulas and define $\tilde f_n\equiv  f(n)=f_{n-1}, \,\tilde a_k \equiv  a_{k-1}$ and $q_k(z)\equiv Q_{k-1}^1(-iz)/i^{k-1}, \,p_k(z)\equiv P_{k-1}^1(-iz)/(i^{k-1}k)$. The interpolation formula \eqref{interpolate} simplifies to 
\begin{align}
    f(z)= \frac{\sin \pi z}{\pi}\frac{1}{z}\Big(\sum_{n=1}^{\infty} (-1)^{n}\frac{ n \tilde f_n}{z-n}+2\sum_{k=1}^{\infty} (-1)^{k}q_k(z) \sum_{n=1}^\infty (-1)^n  n \tilde f_n\, p_k(n) \Big).
\end{align}
Explicitly,
\begin{align}
    p_k(z)&= \,\,_2F_1(1-k, 1+z, 2; 2),\notag\\
    q_k(z)&= 2^{-z} \frac{\Gamma(k+1) \Gamma(1 + z)}{\Gamma(1+k+z)} \,_2F_1(z, 1 + z, 1 + k + z; 1/2).
\end{align}
Evaluating the formula in the $z\to 0$ limit,
\begin{align}
    f(0)=\sum_{n=1}^{\infty} (-1)^{n+1}\tilde f_n+2\sum_{k=1}^{\infty} (-1)^{k} \sum_{n=1}^\infty (-1)^n  n \tilde f_n\, p_k(n).
\end{align}
This is the analytically continued value of the function at $z=0$ if the analytic continuation exists. The existence is not always guaranteed. In \cite{osti_4065624}, a set of necessary conditions was found. It follows from taking the input data to be $h_n^{(l)}=(n-l) \tilde f_n$ and requiring that they yield the function $h^{(l)}(z)=(z-l)f(z)$. Differentiating this function at $z=l$, we get $\tilde f_l$. This is then expressed in terms of other input values $\tilde f_n$ for $n\neq l$. 
\begin{align}
    \tilde f_l= \frac{1}{l}\Big(\sum_{n=1,n\neq l}^{\infty} (-1)^{n+l+1}n \tilde f_n+2\sum_{k=1}^{\infty} (-1)^{k} q_k(l) \sum_{n=1}^\infty (-1)^{n+l}  n (n-l) \tilde f_n\, p_k(n) \Big).
\end{align}
This is a nontrivial condition on the input data that needs to be satisfied if there were to exist the analytic continuation satisfying Carlson's conditions given below \eqref{carlson-cond}. We are currently exploring these conditions for the special probe family $S_n^{(\tq)}$ numerically.

\section{$2d$ conformal blocks at large $c$} \label{monodromy}
In this section, we will discuss the computation of Virasoro conformal blocks for $2d$ CFTs with large central charge $c$ using the so-called ``monodromy method''. We will first outline this method in the most general case i.e. in the case of a $\tq$ point conformal block where $\tq$ of the external operators are taken to have distinct conformal dimension $\Delta_\ta\equiv \delta_\ta c/6$ and the $\tq-3$ of the internal operators also have distinct conformal dimensions $\Delta_\alpha\equiv \delta_\alpha c/6$. Then we will specialize this method to the four-point conformal block and evaluate it in the limit all $\delta_\ta$'s and $\delta_\alpha$'s are taken to be identical and small. We will also consider five-point conformal block in this limit numerically.

The trick is to consider a correlation function $\Psi(\xi,z_\ta)$ of $\tq$ heavy operators $\CO_\ta(z_\ta)$ and one particular degenerate operator $\psi(\xi)$ in the Liouville theory. Constrain this correlator using the null state condition obeyed by $\psi$ and project this constraint on the particular conformal families that appear in the OPE limit that defines the channel of the conformal block. The degenerate operator $\psi$ obeys,
\begin{align}
    {\cal L}_{-2}-\frac{c}{6}{\cal L}_{-1}^2=0.
\end{align}
This constraint translates into a differential constraint on the correlator $\Psi$
\begin{align}\label{diff-con}
    \partial_\xi^2 \Psi+T(\xi,z_\ta) \Psi=0
\end{align}
where
\begin{align}\label{Tform}
    T(\xi,z_\ta)=\sum_\ta\Big(\frac{\delta_\ta}{(\xi-z_\ta)^2}-\frac{c_\ta}{(\xi-z_\ta)}\Big), \qquad \quad c_\ta\equiv -\frac{6}{c}\partial_{z_\ta} \log \Psi.
\end{align}
At this stage, we will project on the conformal family of the specified internal operators in the specified OPE channel. If we denote the resulting conformal block as $e^{-(c/6)f}$ then the ``accessory parameters'' $c_\ta$ become $c_\ta=\partial_{z_\ta}f$. We demand that $T(\xi,z_\ta)$ vanishes as $\xi^{-4}$ at infinity. This reduces the number of free accessory parameters to $\tq-3$. The projection on a given conformal family in a specified channel where a given group of $z_\ta$'s come together implies that $\Psi$ has a particular monodromy as $\xi$ is taken around the said group of $z_\ta$'s. This monodromy depends on the conformal dimension of the internal operator appearing in that channel. This monodromy constraint, in principle, is sufficient to fix the $\tq-3$ unfixed  $c_\ta$'s. Because $c_\ta$'s are derivatives of $f$, this gives $f$ up to an additive constant. This constant is fixed using consistency with the OPE limit.

\subsection{Four-point block}\label{4-pt-monodromy}
Now we will carry out this method for $\tq=4$.  We take the dimensions of all the external operators to be equal $\Delta=\delta c/6$ and gauge fix their positions to be $z_1=0,z_2=z,z_3=1$ and $z_4=\infty$. The cross-ratio for this configuration is $z$. We also solve for $c_1,c_3,c_4$ in terms of $c_2$ using the condition that $T$ must vanish as $\xi^{-4}$ at infinity. The resulting expression for $T(\xi, z)$ is,
\begin{align}\label{4pt-T}
    T(\xi)=\frac{\delta}{\xi^2}+\frac{\delta}{(\xi-z)^2}+\frac{\delta}{(1-\xi)^2}+\frac{2\delta}{\xi(1-\xi)}-\frac{c_2 z(1-z)}{\xi(\xi-z)(1-\xi)}.
\end{align}
For $s$-channel conformal block, $c_2$ is computed by requiring that the solution $\psi(\xi)$ undergoes a fixed monodromy as $\xi$ is taken around a contour that encloses $z_1=0$ and $z_2=z$. If we take the internal operator to be of dimension $\delta_\alpha$, the monodromy matrix has the eigenvalues $\lambda_{\pm}=e^{\pi i (1\pm \sqrt{1-4\delta_\alpha})}$. In principle, this constraint is enough to fix $c_2$. To compute $c_2$ explicitly however, we need to take $\delta$ and $\delta_s$ to be small. This is the so-called  ``light operator approximation''. Now we solve equation \eqref{diff-con} perturbatively in $\delta$.
\begin{align}
    O(\delta^0): & \quad \Psi^{(0)}(\xi)''=0.\notag                       \\
    O(\delta^1): & \quad \Psi^{(1)}(\xi)''+T^{(1)}(\xi)\Psi^{(0)}(\xi)=0.
\end{align}
The solution of the first equation is $\Psi^{(0)}=a+b\xi$. We choose the two linearly independent ones
\begin{align}
    \Psi_1^{(0)}=1,\qquad \Psi_2^{(0)}=1-\xi.
\end{align}
Choosing a different pair of solutions modifies the monodromy matrix by a similarity transformation. Since we are interested only in the eigenvalues of monodromy, this freedom doesn't matter.

The solution to the second equation can be written in terms of $\Psi^{(0)}$ as follows,
\begin{align}\label{order-2}
    \Psi_i^{(1)}=\Psi^{(0)}_2 \int  \Psi_1^{(0)}T^{(1)}\Psi_i^{(0)} d\xi-\Psi^{(0)}_1 \int  \Psi_2^{(0)}T^{(1)}\Psi_i^{(0)} d\xi, \qquad i=1,2.
\end{align}
Note that, in the integrands of equation \eqref{order-2} only the simple pole terms contribute to non-trivial monodromy because they give logarithmic terms when integrated. Such terms have a monodromy of $2\pi i$ times the residue of the integrand at the points around which the monodromy contour is drawn.
For the case of four point block, since we are interested in $s$-channel computation, the monodromy contour is taken around the points $z_1 = 0$ and $z_2=z$. Doing a residue analysis of equation \eqref{4pt-T}, we get the following monodromy matrix at $O(\delta^1)$,
\begin{equation}
    M^{(1)} = 2\pi i \begin{pmatrix} 0 & 2\delta-c_2 z \\ (2\delta-c_2 z)(1 - z)  & 0 \end{pmatrix}
\end{equation}
At $O(\delta^0)$ the monodromy matrix $M^{(0)}$ is of course the identity. The complete monodromy matrix is then, $M = M^{(0)} +  M^{(1)}$. Since the trace is $2$ even at the first order, the non-trivial monodromy condition comes from the determinant of the matrix,
\begin{equation}
    \det[M] = 1 + 4\pi^2 \delta_\alpha^2
\end{equation}
where $\Delta_\alpha=\delta_\alpha c/6$ is the conformal dimension of the exchanged operator.
Solving this quadratic equation for $c_2$ gives us two solutions, out of which we choose the one with the right OPE limit $(f^{(1)} \rightarrow (2\delta - \delta_\alpha)\log(z)$ as $z \rightarrow 0)$ and integrate it to give the conformal partial wave,
\begin{equation}
    2\delta \log(z) + 2\delta_\alpha \log \left( \frac{1+\sqrt{1-z}}{2\sqrt{z}} \right).
\end{equation}
Here we have also fixed the integration constant by requiring the correct OPE limit. We recognize the first term proportional to the scaling dimension of the external operators to be the coordinate-dependent pre-factor of the conformal partial wave. After removing that term we are left with only the part which depends on the cross-ratio,
The conformal block for four heavy perturbative operators, at leading order in $\delta$, is
\begin{equation}
    f^{(1)}(z|\delta,\delta_\alpha) = 2\delta_\alpha\log\left(\frac{1+\sqrt{1-z}}{2\sqrt{z}} \right).
\end{equation}

\subsection{Five-point block}\label{5-pt}
Now we will carry out this method for $\tq=5$.  We take the dimensions of all the external operators to be equal $\Delta=\delta c/6$ and gauge fix their positions to be $z_1=2,z_2=1,z_5=0$, the positions $z_3$ and $z_4$ are left unfixed and thought of as cross-ratios. We also solve for $c_1,c_2,c_5$ in terms of $c_3$ and $c_4$ using the condition that $T$ must vanish as $\xi^{-4}$ at infinity. The resulting expression for $T(\xi, z)$ is,
\begin{align}
    T(\xi |z_3,z_4) &= \frac{\delta}{(\xi-z_3)^2} + \frac{\delta}{(\xi-z_4)^2} + \frac{\delta}{(\xi-2)^2} \\
    &+ \frac{\delta}{(\xi-1)^2} + \frac{\delta}{\xi^2}+ \delta\frac{9(1-\xi) + 2(\xi-z_3) + 2(\xi-z_4)}{\xi(\xi-1)(\xi-2)} \notag\\
        & - \frac{1}{\xi(\xi-1)(\xi-2)} \left( c_3 \frac{z_3(z_3-1)(z_3-2)}{\xi-z_3} + c_4 \frac{z_4(z_4-1)(z_4-2)}{\xi-z_4}\right)\notag
\end{align}
We compute the conformal block in a channel that corresponds to the pair of points $z_2,z_3$ and $z_4,z_5$ coming close. This is achieved by fixing the monodromy matrices $M_{23}$ and $M_{45}$ of $\psi (\xi)$ as $\xi$ encircles pair of points $z_2,z_3$ and $z_4,z_5$ respectively.  We have reproduced the expression for $M_{23}^{(1)}$ below. The matrix $M_{45}^{(1)}$ takes somewhat cumbersome form.
\begin{align}
    M_{23}^{(1)} &= 2\pi i \begin{pmatrix}
        2 \delta_{23}+c_3(1-z_3) & -c_3(1-z_3)^2 -c_4 z_4(z_4-2)+2 \delta_{23}(z_3+z_4-2)\\
        (z_3-1)(c_3(z_3-1)-2\delta_{23}) & -(2\delta_{23}+c_3(1-z_3)).
    \end{pmatrix} \notag
\end{align}
Fixing
\begin{equation}
    \det[M_{23}] = 1 + 4\pi^2 \delta_{23}^2,\qquad \det[M_{45}] = 1 + 4\pi^2 \delta_{45}^2
\end{equation}
where $\Delta_{23,45}=\delta_{23,45}\times c/6$ are the conformal dimension of the exchanged operator in the respective channels. Taking $\delta_{23}=\delta_{45}=\delta$, we get the following equations for equations $c_3$ and $c_4$.
\begin{align}
    \delta^2&=(2\delta-c_3(-1+z_3))((6\delta+c_3(z_3-1)(z_3-2)  -2\delta z_3)z_3-2(\delta+c_4)z_3z_4\notag \\
    &+ c_4(-1+z_3)z_4^2+2(-\delta+z_4 \delta+c_4z_4)) \notag\\
    2\delta^2&=(-2\delta+c_4z_4)(-4\delta+z_4(9\delta+c_3(z_3-2)(z_3-1)-2z_3 \delta+c_4(z_4-2)(z_4-1)-2z_4 \delta)). \notag
\end{align}
The explicit expressions for $c_3$ and $c_4$ are long and cumbersome. Using {\tt mathematica}, we have checked that they match with $c_3$ and $c_4$ computed from the world-line approach.

\bibliography{LargeDCFT}
\end{document}